\documentclass[%
 reprint,
 superscriptaddress,
 amsmath,amssymb,
 aps,
 prl,
]{revtex4-2}

\usepackage{graphicx}
\usepackage{dcolumn}
\usepackage{bm}
\usepackage[normalem]{ulem}
\usepackage{multirow}
\usepackage[hidelinks]{hyperref}
\usepackage{calrsfs}
\DeclareMathAlphabet{\pazocal}{OMS}{zplm}{m}{n}

\usepackage{wrapfig}
\usepackage{url}
\usepackage{color}
\usepackage[dvipsnames]{xcolor}
\usepackage{cancel}
\usepackage[raggedright]{titlesec}
\usepackage{filecontents}

\usepackage{hyperref}
\usepackage[raggedright]{titlesec}

\makeatletter
 \DeclareRobustCommand\ref{%
    \@ifstar\@refstar\T@ref
  }%
  \DeclareRobustCommand\pageref{%
    \@ifstar\@pagerefstar\T@pageref
  }%
 \makeatother

\begin{document}


\pagestyle{empty}

\title{Optical turbulence retrieval of heterogeneous media}

\author{Masaki Watabe}
\altaffiliation{Corresponding author:  m-watabe@nibb.ac.jp}
\affiliation{Cell Modeling and Simulation Research Group, The Exploratory Research Center on Life and Living Systems, Okazaki, Aichi 444-8787, Japan}
\affiliation{Interdisciplinary Research Unit, National Institute for Basic Biology, Okazaki, Aichi 444-8585, Japan}
\author{Joe Sakamoto}
\altaffiliation{Contributed equally with the first author}
\altaffiliation{This author performed experiments.}
\affiliation{Biophotonics Research Group, The Exploratory Research Center on Life and Living Systems, Okazaki, Aichi 444-8787, Japan}
\affiliation{Division of Biophotonics, National Institute for Physiological Sciences, Okazaki, Aichi 444-8585, Japan}
\author{Hideaki Yoshimura}
\altaffiliation{This author performed experiments.}
\affiliation{School of Science, The University of Tokyo, 7-3-1 Hongo, Bunkyo-ku, Tokyo 113-0033, Japan}
\author{Tomomi Nemoto}
\affiliation{Biophotonics Research Group, The Exploratory Research Center on Life and Living Systems, Okazaki, Aichi 444-8787, Japan}
\affiliation{Division of Biophotonics, National Institute for Physiological Sciences, Okazaki, Aichi 444-8585, Japan}
\author{Kazunari Kaizu}
\affiliation{Cell Modeling and Simulation Research Group, The Exploratory Research Center on Life and Living Systems, Okazaki, Aichi 444-8787, Japan}
\affiliation{Laboratory for Biologically Inspired Computing, RIKEN Center for Biosystems Dynamics Research, Kobe, Hyogo 650-0047, Japan}
\date{\today}

\begin{abstract}
The transport of intensity equation (TIE) has revolutionized phase retrieval in optical microscopy, yet its application to complex media with absorption/scattering remains challenging. Here, we present a coupled TIE-TPE (transport of phase equation) framework derived directly from the paraxial wave equation with complex optical potential. By decomposing the refractive index field into a spatially uniform mean field and local fluctuation field, our approach enables simultaneous reconstruction of refractive-index fluctuations and attenuation coefficients without linearization assumptions. We establish reconstruction validity bounds that define the measurable parameter region where reconstruction remains physically consistent. Experimental demonstration with microlens arrays and HeLa cells shows robust recovery of optical properties even in the transparent-limit regime where attenuation signals approach detection thresholds. Furthermore, we provide the first experimental verification of attenuation symmetry---a fundamental property of wave propagation that characterizes reciprocity in light-matter interactions.
\end{abstract}


\maketitle

\textit{Introduction.}
The transport of intensity equation (TIE), introduced by Teague in 1983~\cite{teague1983, teague1982}, provides a deterministic solution to the phase retrieval problem by relating phase gradients to intensity variations. This non-interferometric approach has become essential for label-free imaging of biological specimens~\cite{ruppel2025, yoneda2024, zuo2020}, enabling quantitative measurement of cellular dry mass, protein concentration, and organelle structures. However, conventional TIE assumes purely refractive objects---an assumption that becomes invalid for real biological samples exhibiting both refractive index variations and attenuation (scattering and absorption). The coupled evolution of amplitude and phase in complex or absorbing media has remained largely inaccessible to conventional wave-transport theory.

Recent approaches to overcome this issue include transport of intensity diffraction tomography (TIDT)~\cite{li2022, jenkins2015}, which extends TIE to three-dimensional reconstruction under the first Born approximation---valid only for weakly scattering samples. Within this framework, Bai et al.~\cite{bai2022} demonstrated absorption-phase decoupling using multiple illumination apertures based on Wolf's scattering potential theory~\cite{wolf1969}. However, their approach introduces additional constraints through weak absorption approximations layered on top of the Born approximation. Furthermore, the validity domains for experimental parameters---including defocus distance, numerical aperture, and sample optical properties---remain to be systematically established.

This motivates the fundamental reformulation of the TIE framework to unify amplitude and phase transport within a single non-divergent theoretical scheme. Derived directly from the paraxial wave equation with a complex optical potential, this formulation establishes explicit validity bounds that define the measurable parameter regime of optical turbulence retrieval. Crucially, we demonstrate this framework's robustness through quantitative phase imaging of microlens arrays and living HeLa cells. Furthermore, we experimentally reveal a hidden invariance in wave propagation---attenuation symmetry---that manifests reciprocity as a conserved property in complex systems.

\textit{Theoretical Framework.}
Light propagation through heterogeneous media involves complex interactions between electromagnetic waves and spatial variations in optical properties. To capture these physics systematically, we introduce a complex optical potential that accounts for both real and imaginary components of the refractive index field. The key insight is to express the spatially varying refractive index field as the product of a uniform mean field $n_0$ and a dimensionless function: $n(\mathbf{r}_\perp,z) = n_0[1 + \Delta n(\mathbf{r}_\perp,z)]$, where $n_0$ represents the mean field of the refractive index and $\Delta n(\mathbf{r}_\perp,z)$ is the fluctuation field capturing local deviations from this mean. This decomposition naturally separates the bulk optical properties from structural heterogeneities, allowing us to treat the fluctuation field $\Delta n$ as a small parameter when $|\Delta n| \ll 1$.

Starting from the Helmholtz equation and applying the paraxial approximation, the wave equation becomes:
\begin{equation}
i\frac{\partial A}{\partial z} = \left[-\frac{1}{2kn_0}\nabla_\perp^2 - \frac{\partial}{\partial z}(\beta + i\alpha)\right]A
\label{eqn;pwe}
\end{equation}
Here $A(\mathbf{r}_\perp,z)$ is the complex amplitude of the wavefield, $kn_0 = 2\pi n_0/\lambda$ is the wave number in the mean medium, and $\nabla_\perp^2$ is the transverse Laplacian. The complex optical potential is encoded in two functions: $\alpha(r_\perp,z)$ governing amplitude variations and $\beta(r_\perp,z)$ controlling phase evolution. The amplitude reduction captured by $\alpha(r_\perp,z)$ encompasses both absorption (where photon energy is converted to other forms such as heat) and scattering (where photons are redirected from their original path), collectively contributing to wave attenuation. These are directly related to the physical properties through:
\begin{align}
\frac{\partial\alpha}{\partial z} &= kn_0(1 + \Delta n)\kappa 
\label{eqn;alpha}\\
\frac{\partial\beta}{\partial z} &= kn_0\left[\Delta n - \frac{\kappa^2}{2}\right]
\label{eqn;beta}
\end{align}
where $\kappa$ is the attenuation index. Physically, Eq.~(\ref{eqn;alpha}) describes how local absorption and scattering reduce wave amplitude---the factor $(1 + \Delta n)$ accounts for how refractive index variations modulate the effective absorption cross-section. Eq.~(\ref{eqn;beta}) reveals that phase accumulation arises from two sources: direct contribution from refractive index variations ($\Delta n$ term) and an indirect effect from absorption ($\kappa^2/2$ term). This quadratic absorption term, often overlooked, represents an additional phase shift that emerges from the complex nature of the optical potential when both refraction and absorption are present.

This decomposition strategy---separating the total refractive index field into mean field and fluctuation field components---ensures mathematical convergence and physical interpretability. When $|\Delta n| < 1$, the fluctuation field remains bounded, preventing divergences that plague direct approaches. Moreover, this formulation naturally identifies the mean field $n_0$ with the surrounding medium (e.g., culture medium for cells, mounting medium for tissues), while the fluctuation field $\Delta n$ directly corresponds to measurable structural features like cell boundaries, organelles, or material interfaces.

Expressing the complex amplitude in terms of observable quantities, $A = \sqrt{I}\ e^{i\varphi}$, and separating real and imaginary parts yields two coupled transport equations. From the imaginary part emerges the generalized TIE:
\begin{equation}
\frac{\partial I}{\partial z} = -\frac{1}{kn_0}\nabla_\perp \cdot (I\nabla_\perp\varphi) - 2\frac{\partial\alpha}{\partial z}I
\label{eqn;tie}
\end{equation}
We previously derived this generalized form of TIE in Ref.~\cite{watabe2023}, where the first term represents Teague's original TIE describing intensity redistribution from phase gradients, and the second term $-2(\partial\alpha/\partial z)$ captures direct intensity attenuation from absorption and scattering. This coupling reveals that intensity changes in heterogeneous media arise from both wavefront curvature and absorptive losses.

\begin{figure}[t]
\centering
\includegraphics[width=1.00\linewidth]{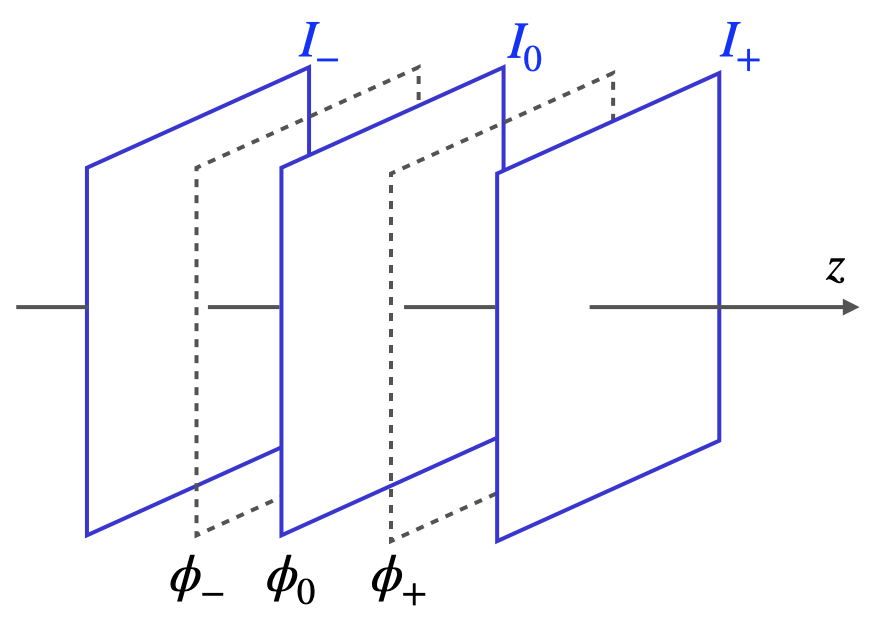}

\caption{Phase retrieval via TIE imaging typically involves a through-focus series of image intensities at three different z-positions, $I(z_0 - \Delta z)$, $I(z_0)$ and $I(z_0 + \Delta z)$. These images are then used to compute the axial intensity derivatives, which, when combined with the intensity distributions at the in-focus and half of the defocus z-positions, solves the TIE and reconstructs the phase distributions $\phi(z_0 - \Delta z/2)$, $\phi(z_0)$ and $\phi(z_0 + \Delta z/2)$.}
\label{fig01;scheme}
\end{figure}

From the real part emerges the transport of phase equation (TPE):
\begin{equation}
\frac{\partial\varphi}{\partial z} = \frac{\partial\beta}{\partial z} - \frac{1}{2kn_0}(\nabla_\perp\varphi)^2 + \frac{1}{2kn_0}\frac{\nabla_\perp^2\sqrt{I}}{\sqrt{I}}
\label{eqn;tpe}
\end{equation}
Each term carries distinct physical meaning: $\partial\beta/\partial z$ represents the direct phase accumulation rate from optical path length variations through heterogeneous media. The nonlinear term $-(\nabla_\perp\varphi)^2$ describes wavefront self-correction---regions with steep phase gradients experience additional phase retardation, analogous to self-focusing in nonlinear optics but arising here purely from propagation geometry. The final term $\nabla_\perp^2\sqrt{I}$ represents amplitude-phase coupling: intensity variations induce phase modulations through diffractive spreading, ensuring energy conservation during propagation. The derivation of the TPE is fully discussed in Sec.~A of the Supplementary Material (SM)~\cite{sm}.

Together, Eqs.~(\ref{eqn;tie}-\ref{eqn;tpe}) form a non-divergent system that rigorously describes wave propagation through complex media without assuming pure phase or weak absorption---a fundamental advance over conventional approaches.

\textit{Reconstruction Algorithm.}
The practical implementation of our framework requires extracting optical parameters from measurable intensity distributions. Consider through-focus intensity measurements at three planes: in-focus ($I_0$ at $z_0$) and symmetric defocus positions ($I_\pm$ at $z_0\pm\Delta z$). For small propagation distances where the fluctuation field induces only linear changes in intensity (see Fig.~\ref{fig01;scheme}), the intensity evolution follows:
\begin{equation}
I_\pm \approx I_0 \mp \frac{\Delta z}{kn_0}\nabla\cdot[I_0\nabla\varphi_0] - 2I_0\left.\frac{\partial\alpha}{\partial z}\right|_{z_0}\Delta z
\end{equation}
This equation reveals how defocused intensities encode both phase and absorption information. The antisymmetric component $(I_+ - I_-)$ isolates phase contributions, enabling retrieval through standard TIE solvers like the geometric multigrid method (GMGM)~\cite{mazumder2016,xue2011,pinhasi2010,press1992}. This yields phase distributions at $z_0$ and half-defocus positions $z_0\pm\Delta z/2$ (see Fig.~\ref{fig01;scheme}).

The symmetric component $(I_+ + I_- - 2I_0)$ directly provides the intensity reduction parameter:
\begin{equation}
\left.\frac{\partial\alpha}{\partial z}\right|_{z_0} \approx \frac{I_+ + I_- - 2I_0}{-4\Delta z I_0}
\end{equation}
Physically, this symmetric difference isolates pure attenuation effects by canceling phase-induced intensity redistributions. Since phase gradients cause equal but opposite intensity changes at symmetric defocus positions, their contribution vanishes in the sum, leaving only the monotonic intensity decay from absorption.

With phase distributions at multiple planes, we compute axial phase derivatives $(\varphi_+ - \varphi_-)/\Delta z$ and apply the TPE to extract the phase-coupling parameter $\partial\beta/\partial z$. This quantifies how local refractive index variations and attenuation jointly influence phase evolution. 

Finally, these intermediate parameters convert to physically meaningful quantities through:
\begin{align}
\Delta n &\approx \frac{\partial_z\beta/kn_0 + (\partial_z\alpha/kn_0)^2/2}{1 - 2(\partial_z\beta/kn_0)}
\label{eqn;dn}\\
\mu &= kn_0\kappa = \frac{\partial_z\alpha}{1 + \Delta n}
\label{eqn;mu}
\end{align}
where we assume $|\Delta n|^2\ll 1$ for media with weak fluctuation fields. The refractive index fluctuation $\Delta n$ directly relates to sample structure and composition, while the attenuation coefficient $\mu$ (units: 1/length) quantifies the total optical losses arising from both absorption and scattering processes in the medium. This reconstruction requires only three intensity measurements, which is much more efficient compared to tomographic approaches. The complete four-step reconstruction procedure, including the detailed implementation of the GMGM solver and intermediate parameter calculations, is provided in SM Sec. B.

\begin{figure}[t]
\leftline{\bf (a)}
\centering
\includegraphics[width=0.99\linewidth]{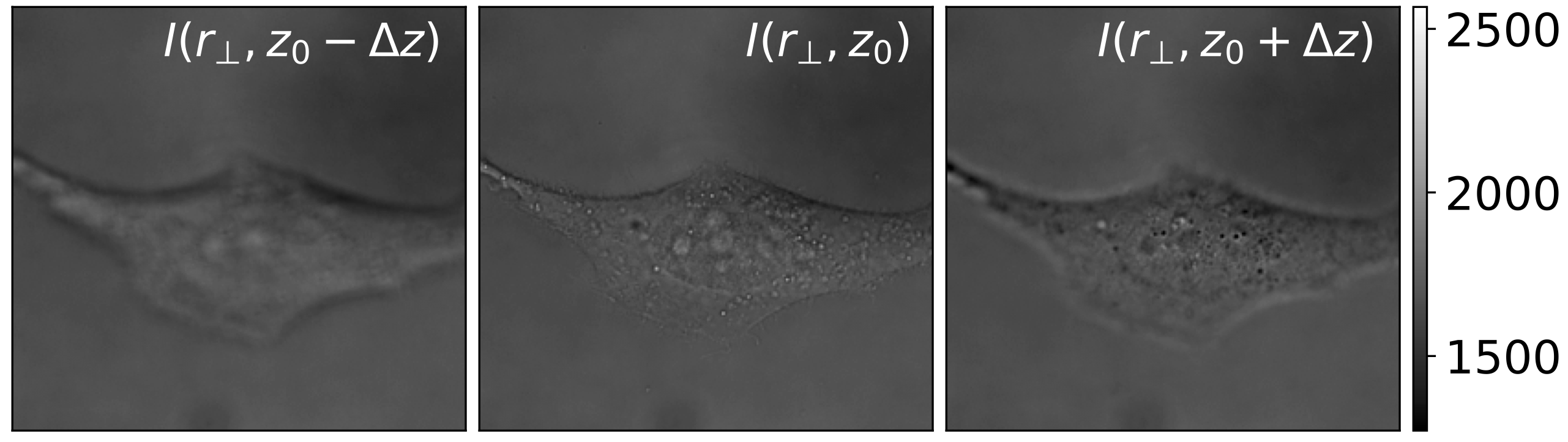}

\leftline{\bf (b)}
\centering
\includegraphics[width=0.99\linewidth]{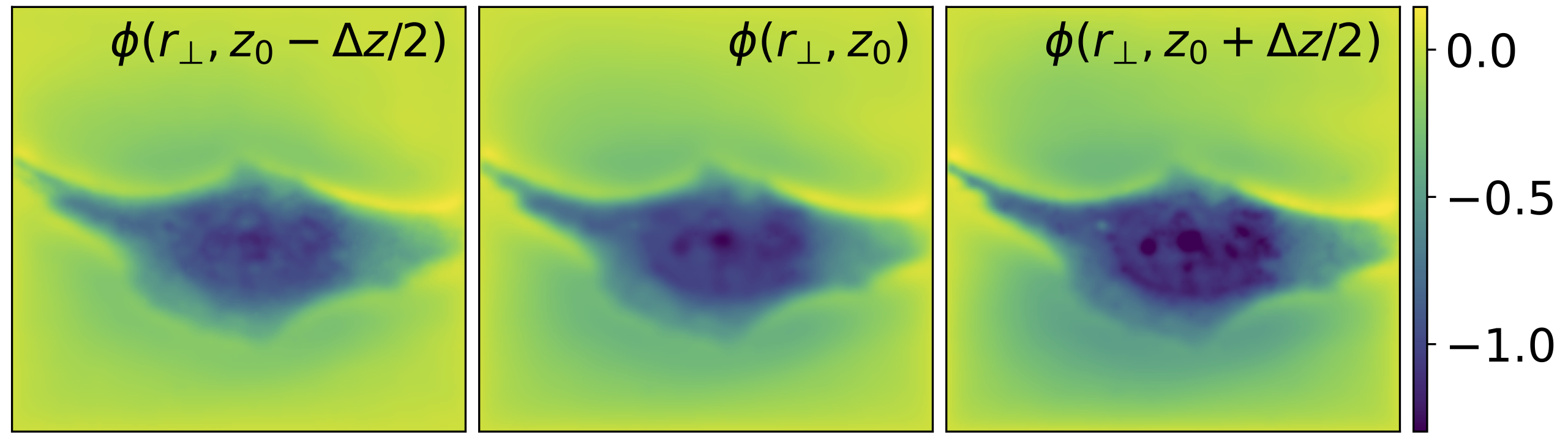}

\caption{The computational inputs of HeLa cells: (a) the image intensities ($I_{-}$, $I_{0}$ and $I_{+}$), and (b) the phases ($\phi_{-}$, $\phi_0$ and $\phi_{+}$) reconstructed by using the GMGM.}
\label{fig02;inputs}
\end{figure}

\begin{figure*}[t]
\leftline{\bf (a) \hspace{0.17\linewidth} (b) \hspace{0.18\linewidth} (c) \hspace{0.25\linewidth} (d)}
\centering
\includegraphics[width=0.21\linewidth]{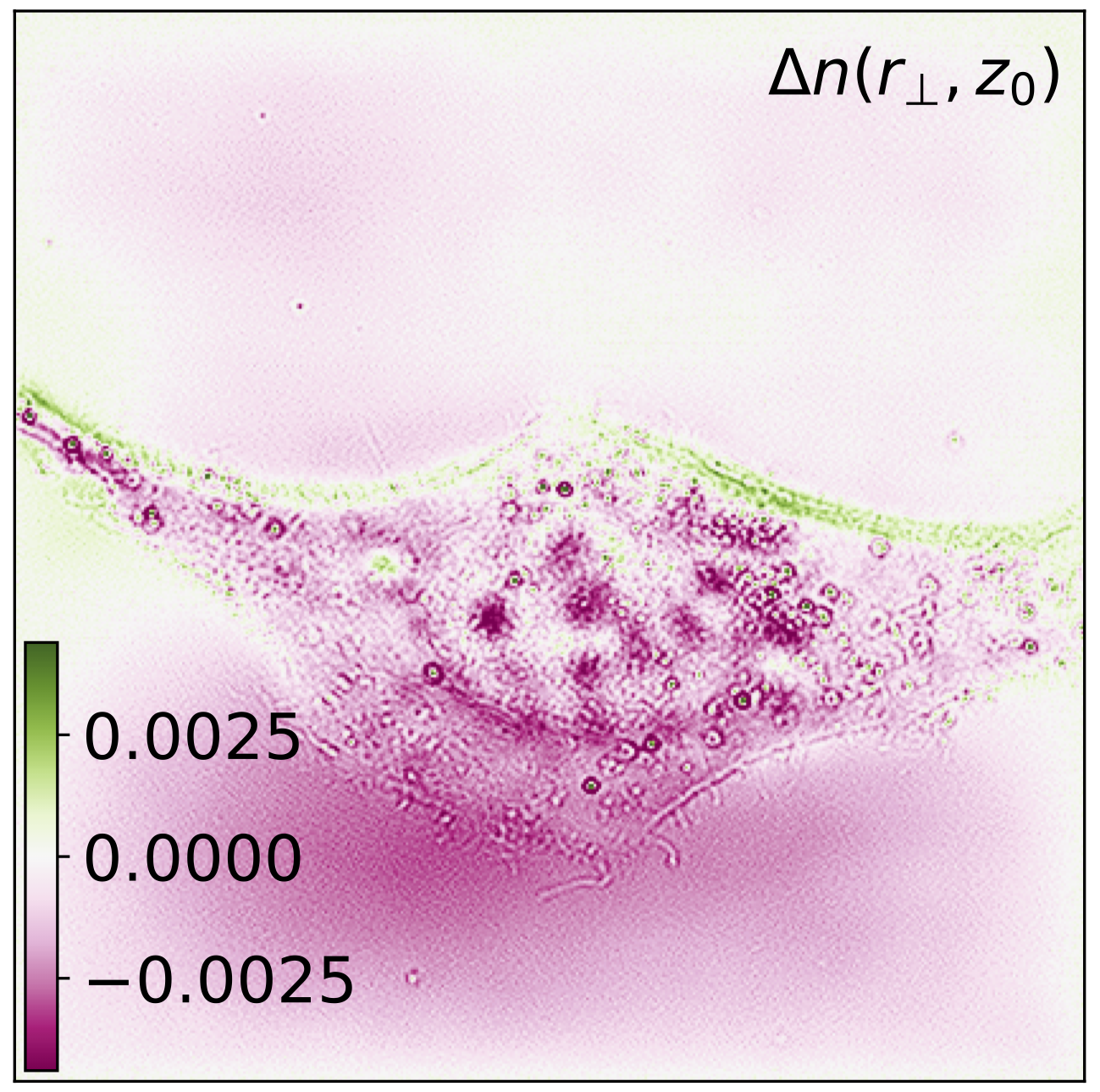}
\includegraphics[width=0.21\linewidth]{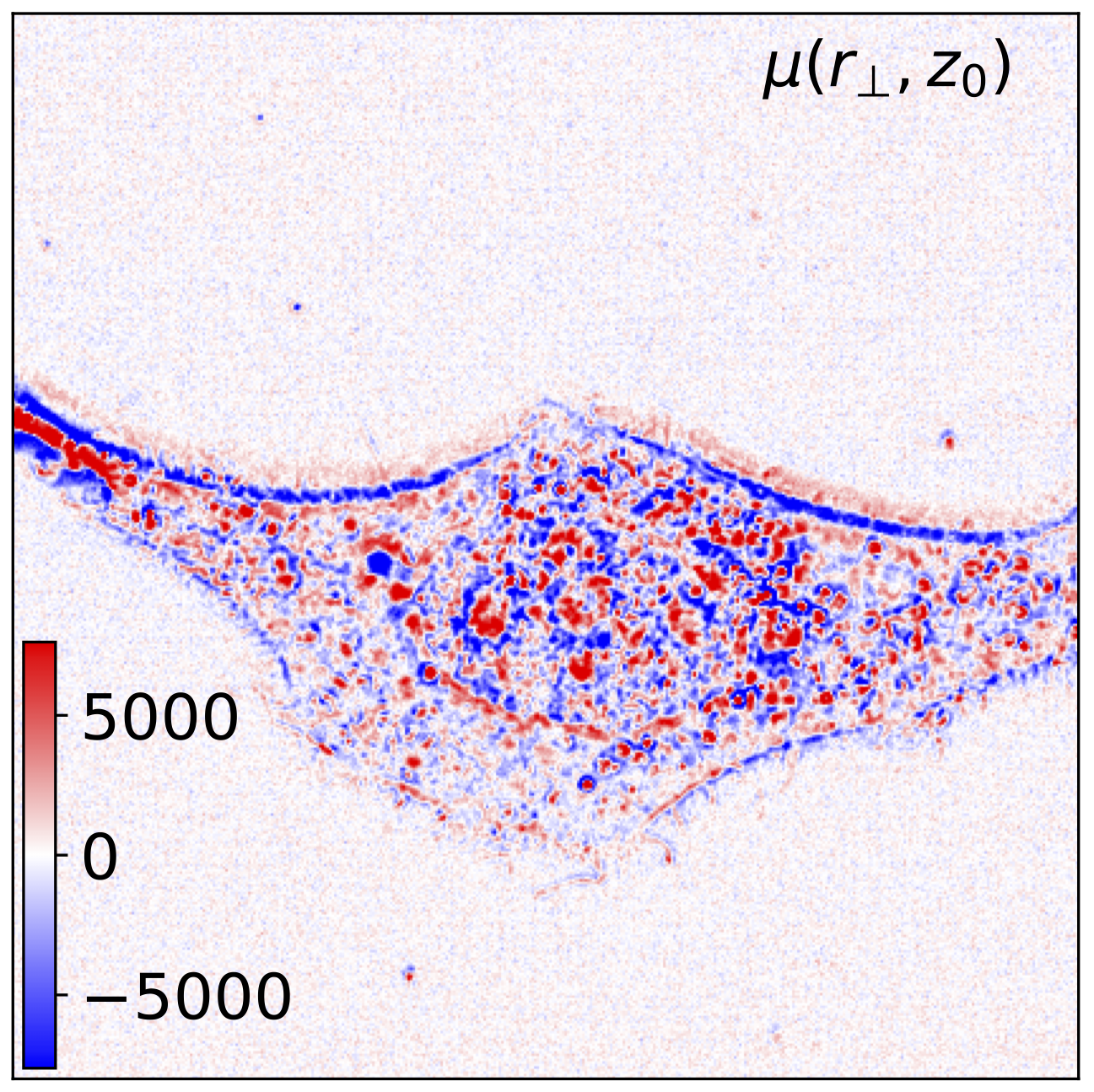}
\includegraphics[width=0.28\linewidth]{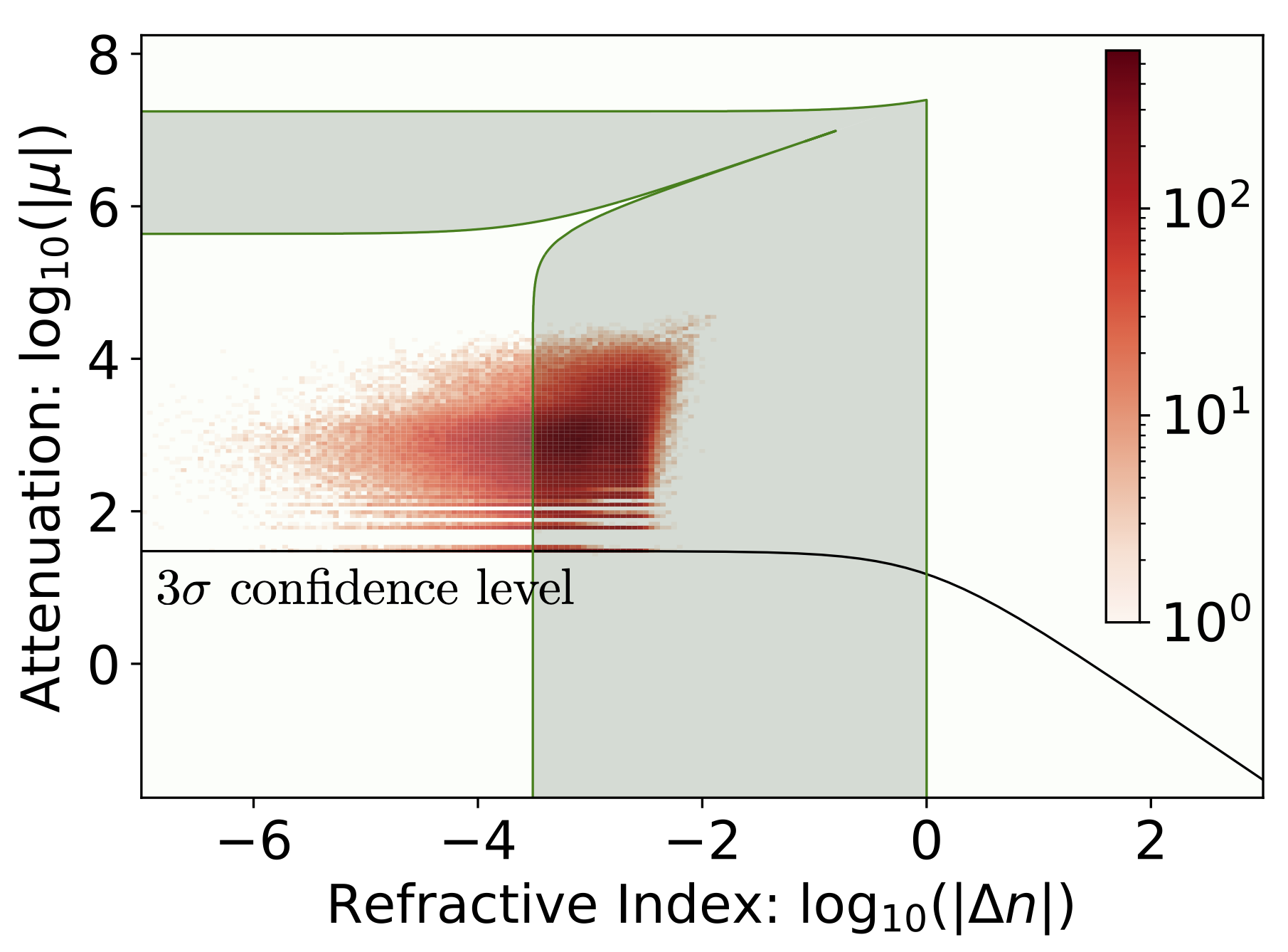}
\includegraphics[width=0.28\linewidth]{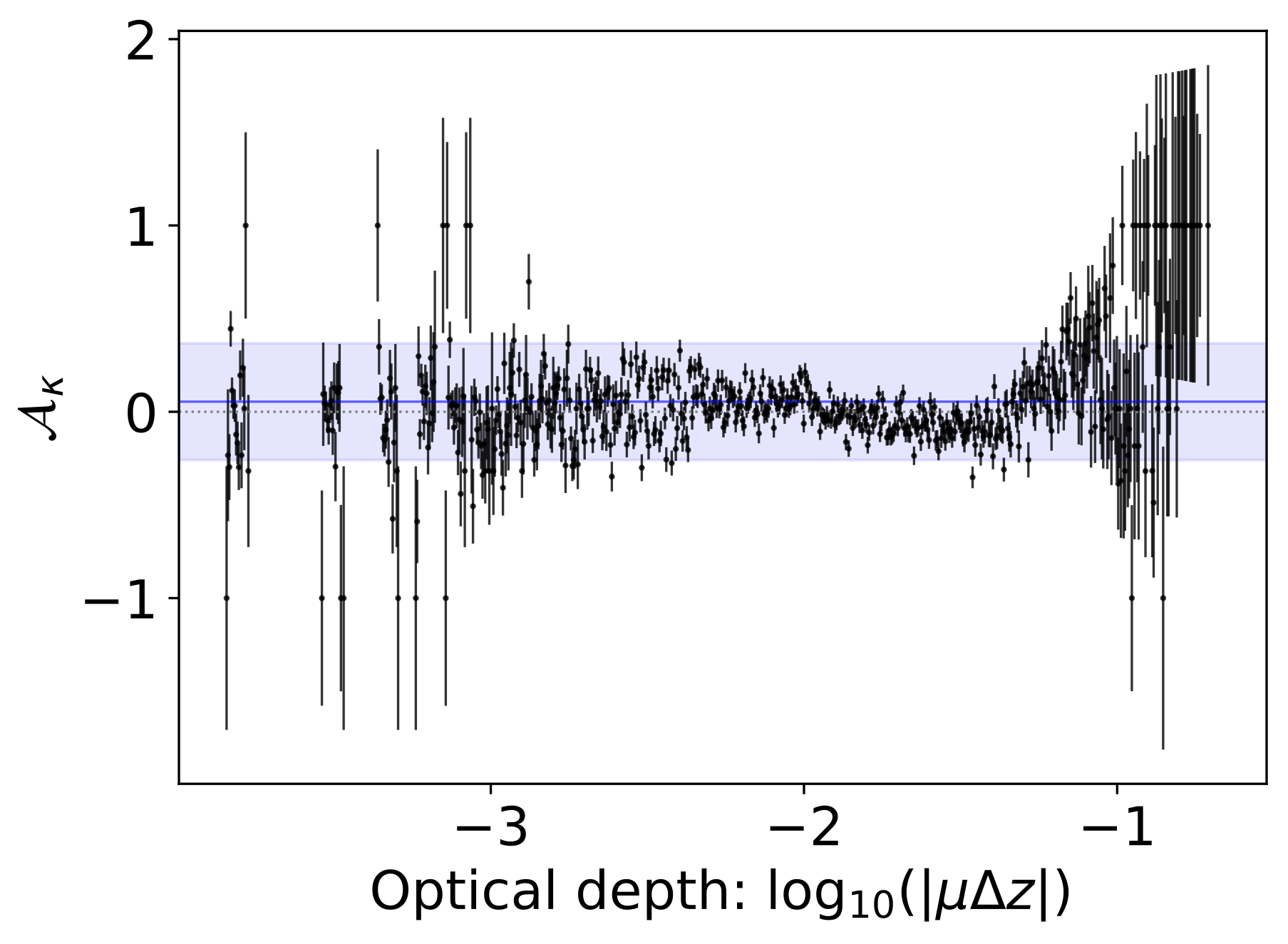}

\caption{Results for HeLa cells: (a) Spatial distribution of refractive-index fluctuations $\Delta n(r_{\bot},z_0)$ and (b) attenuation coefficients $\mu(r_{\bot},z_0)$. The sign of the $\mu(r_{\bot},z_0)$ indicates forward and backward wave propagations along the z-axis. (c) Correlation patterns between $\Delta n(r_{\bot},z_0)$ and $\mu(r_{\bot},z_0)$. The black solid line is the $3\sigma$ confidence level. The green area represents the theoretically measurable region defined by Eq.~(\ref{eqn;limits}), where reconstruction remains physically consistent despite operating in the transparent-limit regime. (d) Measured value of the asymmetry $\mathcal{A}_{\kappa}$ in bins of the optical depth $\log_{10}|\mu \Delta z|$. The blue line and bands are the statistical average and the root mean squared (RMS) value, respectively. Dashed line represents $\mathcal{A}_{\kappa} = 0$.}
\label{fig03;results}
\end{figure*}

\textit{Reconstruction Validity Bounds.}
A critical advancement of our framework is establishing explicit boundaries where reconstruction remains physically meaningful. These constraints emerge from fundamental limits in both wave propagation physics and measurement capabilities.

The statistical reliability of reconstruction depends on the transmission probability through complex media:
\begin{equation}
P(\Delta z) = \int \Phi(k)e^{-2kn_0(1+\Delta n)\kappa\Delta z} dk
\label{eqn;prob}
\end{equation}
where $\Phi(k)$ is the normalized spectral distribution of the incident beam with wavenumber $k$. The exponential term describes attenuation during propagation through absorbing media, with the attenuation factor depending on both the optical properties ($\Delta n$, $\kappa$) and propagation distance ($\Delta z$). From this probability distribution, we establish confidence intervals (e.g., $95\%$) that define the range within which true optical parameters lie, particularly critical near detection limits where intensity reductions or refractive-index variations challenge measurement reliability.

From photon counting statistics and diffraction physics emerge the fundamental validity bounds:
\begin{equation}
\left(\frac{2\Delta x}{\pi\Delta z}\right)^2 < \left|\Delta n - \frac{\kappa^2}{2}\right| < 1 - \frac{\kappa^2}{2}
\label{eqn;limits}
\end{equation}
The lower bound has profound physical meaning: it represents the threshold where phase-induced intensity modulation becomes comparable to shot noise. The factor $(2\Delta x/\pi\Delta z)^2$ quantifies the minimum detectable phase gradient---determined by the ratio of pixel size to propagation distance (see the SM Sec.~C for the derivation). This transparent-limit regime is exemplified by unstained cells where $\mu \to 0$: phase contrast dominates but attenuation signals barely exceed photon shot noise. When phase variations occur over distances smaller than this diffraction length, their intensity signatures blur beyond recognition. This is not a limitation of detector technology but a fundamental consequence of wave physics: information encoded at spatial frequencies beyond the diffraction limit is irretrievably lost during propagation.

The upper bound ensures the paraxial approximation remains valid. As the fluctuation field $|\Delta n|$ approaches or exceeds unity---as seen in thick tissues or highly heterogeneous samples---ray angles become too large, higher-order diffraction becomes significant, thus breaking our framework. In this strong attenuation regime where $|\Delta n|$ approaches $0.5$-$1$ or beyond, multiple scattering dominates. This isn't merely reduced accuracy but fundamental failure of the underlying physics---the mean field decomposition loses meaning when fluctuations become comparable to or larger than the mean field itself.

Understanding these regimes guides experimental design: choosing appropriate defocus distances, optimizing illumination conditions, and predicting reconstruction feasibility for a given experimental configuration.

\textit{Experimental Validation.}
We validated our framework using two complementary systems: microlens arrays (MLA) as calibrated optical standards~\cite{mla} and living HeLa cells as complex biological specimens. Experiments employed laser-scanning confocal microscopy and wide-field microscopy with various numerical apertures (NA=0.45 for MLA, NA=1.2 for HeLa cells, NA=1.49 for cell membranes). Complete optical configurations and results are in SM Sec. D.

For HeLa cells, we deliberately operated in the transparent-limit regime where attenuation signals approach statistical detection thresholds. This regime, characterized by $\mu$$\to$$0$, provides a stringent test of the framework's ability to distinguish weak absorption signals from noise while maintaining stable reconstruction. Through-focus intensity stacks were acquired at $z_0 \pm 5\ {\rm \mu m}$ with $488\ {\rm nm}$ illumination. The computational pipeline revealed rich optical heterogeneity despite weak signals. The intensity reduction parameter $\partial\alpha/\partial z$, extracted from the symmetric intensity component, mapped regions of absorption despite signals approaching noise levels. The phase-coupling parameter $\partial\beta/\partial z$, computed via TPE with axial phase derivatives, revealed how refractive index variations modulate local wavefront curvature.

Converting to physical parameters yielded spatial maps of refractive-index fluctuation field $\Delta n$ [Fig.~\ref{fig03;results}(a)] and attenuation coefficient field $\mu$ [Fig.~\ref{fig03;results}(b)]. The $\Delta n$ distribution clearly delineates cellular boundaries, nuclei, and organelles through refractive index contrast arising from the fluctuation field. The $\mu$ field, though noisier due to weak attenuation in the transparent-limit regime, still reveals cellular structure, with the sign indicating forward (positive) or backward (negative) wave propagation along the z-axis.

Most revealing is the $\Delta n$-$\mu$ correlation plot shown in Fig.~\ref{fig03;results}(c). The data cluster forms a characteristic pattern bounded by two distinct regions: (i) the green area representing the physical validity bounds from Eq.~(\ref{eqn;limits}), and (ii) the black solid line indicating the $3\sigma$ statistical confidence level.

Analysis of multiple cells revealed that $0.2$-$25\%$ of pixels fall outside the physical validity bounds, predominantly clustering near $\Delta n \approx 0$.. At this point, the lower bound condition $(2\Delta x/\pi\Delta z )^2$ becomes the limiting factor---these pixels represent regions where refractive index variations fall below the detection threshold and should be interpreted as $\Delta n \approx 0$.. Importantly, these outliers remain within the $3\sigma$ statistical confidence interval, confirming that the framework correctly identifies the transition between measurable and sub-threshold regions while maintaining mathematical stability.


To verify the generality of our framework across different sample types and imaging modalities, we extended our validation to additional systems. MLA measurements [Figs. S3-S4] provided complementary validation with well-defined optical elements, showing only $3.2\%$ of pixels outside validity bounds but well within the $3\sigma$  confidence level. Wide-field microscopy with xenon lamp illumination of HeLa cell membranes [Figs. S7-S8] tested versatility across illumination modalities, with membrane-specific optical signatures emerging clearly in both fields and minimal outliers ($0.2\%$). Note that systematic uncertainties from finite numerical aperture effects are quantified in SM Sec.~D.4.

\textit{Symmetry in Wave Propagation.}
A fundamental prediction of our theoretical framework is the existence of attenuation symmetry in wave propagation---a property never before experimentally verified in quantitative phase imaging. The attenuation index $\kappa$ can take positive or negative values, corresponding to forward (transmission) and backward (reflection) propagation modes. When the medium exhibits optical recoprocity, the propagation probability $\mathcal{P}(\kappa) = \exp{[-2kn_0(1+\Delta n)\kappa\Delta z]}$ should remain unchanged under directional inversion $\kappa\rightarrow -\kappa$, which we quantify through the asymmetry parameter:
\begin{equation}
\mathcal{A}_\kappa = \frac{\mathcal{P}(+\kappa) - \mathcal{P}(-\kappa)}{\mathcal{P}(+\kappa) + \mathcal{P}(-\kappa)}
\label{eqn;asym}
\end{equation}
where $\mathcal{P}(+\kappa)$ and $\mathcal{P}(-\kappa)$ represent propagation probabilities for forward and backward modes, respectively. Perfect symmetry yields $\mathcal{A}_\kappa = 0$, while nonzero values indicate reciprocity breaking from factors like magneto-optical effects or structural chirality.

Our experimental measurement of $\mathcal{A}_\kappa$ versus optical depth $\log_{10}|\mu\Delta z|$ for HeLa cells [Fig.~\ref{fig03;results}(d)] demonstrates that $\mathcal{A}_\kappa$ remains statistically consistent with zero across three orders of magnitude. The measured values fluctuate within the statistical uncertainty (RMS bands) around zero, with no systematic deviation observed across the entire optical depth range.

This invariance has profound implications for optical reciprocity. In the weak scattering regime (optical depth $< 10^{-2}$), where our paraxial approximation is most valid, the symmetry confirms that microscopic scattering events preserve time-reversal invariance. Even as optical depth increases toward unity---approaching the limits of our framework---the symmetry persists, suggesting robustness beyond the strict validity regime of Eq.~(\ref{eqn;limits}).

The observed $\mathcal{A}_{\kappa}$ invariance demonstrates that biological tissues maintain optical reciprocity at the wavelength scale, preserving identical light-matter interactions in forward and backward propagation. This reciprocity enables reliable quantitative phase measurements regardless of illumination direction---a crucial foundation for next-generation imaging systems in complex biological environments. The first experimental verification of attenuation symmetry in biological samples thus validates both our theoretical framework and reciprocity-based imaging principles. Measurements on microlens arrays and HeLa cell membranes show similar invariance, with detailed analysis in SM Sec.~D.3.

\textit{Conclusion.}
While recent approaches like Bai et al. (2022)~\cite{bai2022} have advanced absorption-phase decoupling in transport of intensity diffraction tomography through linearization and weak absorption approximations, these methods require specific assumptions about the sample properties. We have presented a fundamental reformulation that provides a more general framework through three key advances: (i) a non-divergent coupled TIE-TPE framework derived directly from the paraxial wave equation with complex optical potential, (ii) explicit validity bounds that define the measurable parameter space for any given experimental configuration, and (iii) the first experimental verification of attenuation symmetry $\mathcal{A}_{\kappa}$ across three orders of magnitude in optical depth. 

%

Beyond these advances, the discovery of attenuation symmetry demonstrates that reciprocity---a fundamental principle of wave physics---remains intact even in complex biological tissues. This finding not only validates our theoretical framework but also suggests new possibilities for imaging: the preservation of this symmetry despite tissue heterogeneity could potentially guide strategies for probing deeper structures and imaging more challenging samples. By establishing rigorous validity bounds and revealing this hidden constraint, our work opens pathways toward quantitative phase imaging of thick tissues and strongly scattering materials that were previously beyond reach.


\textit{Data availability.}
The data that support the findings of this article are openly available \cite{zenodo}.

\textit{Acknowledgments.}
The authors express their sincere gratitude to Masayuki Hattori, Naru Yoneda and Osamu Matoba for their invaluable guidance and support throughout this research. We also acknowledge Eiji Watanabe for helpful discussions on the effective use of AI tools such as Claude (claude.ai, Anthropic) and ChatGPT (chatgpt.ai, OpenAI) for computational and manuscript preparation assistance. This work was supported by the Japan Society for the Promotion of Science (JSPS) KAKENHI Grant Numbers JP20H05891, JP20K21836, JP21H05605, JP22H04926, JP23K17364, and JP23K26795. Additional support was provided by the Asahi Glass Foundation, the Frontier Photonic Sciences Project of National Institutes of Natural Sciences (NINS) Grant number 01212504, and by the Joint Research of the Exploratory Research Center on Life and Living Systems (ExCELLS) through the ExCELLS Encouragement Research Program for Young Scientists.


%

\onecolumngrid

\newpage

\leftline{\bf\Large Supplemental Material}

\maketitle

\newcolumntype{C}[1]{>{\centering\arraybackslash}p{#1}}

\setcounter{page}{0}
\setcounter{section}{0} \renewcommand{\thesection}{\Alph{section}}
\setcounter{subsection}{0} \renewcommand{\thesubsection}{\Alph{section}.\arabic{subsection}}
\setcounter{subsubsection}{0} \renewcommand{\thesubsubsection}{\Alph{section}.\arabic{subsection}.\arabic{subsubsection}}
\setcounter{figure}{0} \renewcommand{\thefigure}{S\arabic{figure}}
\setcounter{table}{0} \renewcommand{\thetable}{S\arabic{table}}
\setcounter{equation}{0} \renewcommand{\theequation}{S\arabic{equation}}

\titleformat*{\section}{\large\bfseries}
\titleformat*{\subsection}{\normalsize\bfseries}
\titleformat*{\subsubsection}{\normalsize\bfseries}



%

\section{A. Derivation of the TPE}
Separating the paraxial wave equation [i.e., Eq.~(1) in the main text] into the real and imaginary sections, the TIE can be derived in the imaginary part (see the Sec.~A.1 of the Supplementary Material in Ref.~\cite{watabe2023} for derivation). Conversely, the TPE can be derived in the real part and its derivation is given as follows.\\

\noindent In the real part of the paraxial wave equation, 
\begin{eqnarray}
Re\left[ \nabla_{\bot}^2 A + 2 k n_0 i \frac{\partial  A}{\partial z} + 2 k n_0 \frac{\partial}{\partial z}(\beta + i \alpha)A \right] = 0\nonumber\\
\nonumber\\
Re\left[ \nabla_{\bot}^2 \sqrt{I} e^{i\phi} + 2 k n_0 i \frac{\partial }{\partial z}\sqrt{I} e^{i\phi} + 2 k n_0 \frac{\partial}{\partial z}(\beta + i \alpha)\sqrt{I} e^{i\phi} \right] = 0\nonumber\\
\nonumber\\
Re\left[ \nabla_{\bot}^2 \sqrt{I} e^{i\phi} + 2 k n_0 i \left( \frac{\partial \sqrt{I}}{\partial z} e^{i\phi} +  \sqrt{I} \frac{\partial e^{i\phi}}{\partial z} \right) \right] + 2 k n_0 \frac{\partial \beta}{\partial z} \sqrt{I} e^{i\phi} = 0\nonumber\\
\nonumber\\
Re\left[ \nabla_{\bot}^2 \sqrt{I} e^{i\phi} + 2 k n_0 i \sqrt{I} \frac{\partial e^{i\phi}}{\partial z} \right] + 2 k n_0 \frac{\partial \beta}{\partial z} \sqrt{I} e^{i\phi} = 0\nonumber\\
\nonumber\\
Re\left[ \nabla_{\bot}^2 \sqrt{I} e^{i\phi}\right] -  2 kn_0 \sqrt{I} \frac{\partial \phi}{\partial z}  e^{i\phi} + 2 kn_0 \frac{\partial \beta}{\partial z} \sqrt{I} e^{i\phi} = 0\nonumber\\
\nonumber\\
Re\left[ ( \nabla_{\bot}^2 \sqrt{I} ) e^{i\phi} + 2 ( \nabla_{\bot} \sqrt{I} ) \left(\nabla_{\bot} e^{i\phi} \right) + \sqrt{I} \left( \nabla_{\bot}^2 e^{i\phi} \right) \right] -  2 kn_0 \sqrt{I} \frac{\partial \phi}{\partial z}  e^{i\phi} + 2 kn_0 \frac{\partial \beta}{\partial z} \sqrt{I} e^{i\phi} = 0\nonumber\\
\nonumber\\
\left[ \nabla_{\bot}^2 \sqrt{I} - \sqrt{I} \left(\nabla_{\bot} \phi \right)^2 \right] -  2 kn_0 \sqrt{I} \frac{\partial \phi}{\partial z} + 2 kn_0 \sqrt{I} \frac{\partial \beta}{\partial z} = 0
\label{eqn;pweA}
\end{eqnarray}

\noindent Divided by $2kn_0 \sqrt{I}$, we can deduce the TPE,
\begin{eqnarray}
\frac{\partial \phi}{\partial z} = \frac{\partial \beta}{\partial z} - \frac{1}{2kn_0}\left(\nabla_{\bot} \phi \right)^2 + \frac{1}{2kn_0}\frac{\nabla_{\bot}^2 \sqrt{I}}{\sqrt{I}}
\label{eqn;pweB}
\end{eqnarray}
where the axial derivative of $\alpha(r_{\bot},z)$ vanishes through this modification.\\

\noindent The last term in the Eq.~\eqref{eqn;pweB} can be rewritten as follows.
\begin{eqnarray}
\frac{\nabla_{\bot}^2 \sqrt{I}}{\sqrt{I}} = \frac{1}{\sqrt{I}}\nabla_{\bot} \cdot \left(\frac{\nabla_{\bot} I}{2\sqrt{I}}\right) = \frac{\sqrt{I} \left(\nabla_{\bot}^2 I \right) - (\nabla_{\bot} \sqrt{I})(\nabla_{\bot} I)}{2 I \sqrt{I}} = \frac{2I \left( \nabla_{\bot}^2I \right) - \left(\nabla_{\bot} I\right)^2}{4 I^2} \nonumber
\end{eqnarray}

\noindent Inserting this back to the Eq.~\eqref{eqn;pweB}, the TPE can be also written in the form of
\begin{eqnarray}
\frac{\partial \phi}{\partial z} = \frac{\partial \beta}{\partial z} - \frac{1}{2 k n_0}\left(\nabla_{\bot} \phi \right)^2 + \frac{1}{4 k n_0} \left(\frac{\nabla^2_{\bot} I}{I}\right) - \frac{1}{8 k n_0} \left(\frac{\nabla_{\bot} I}{I}\right)^2.
\label{eqn;pweC}
\end{eqnarray}

\vspace*{\fill}

\newpage

\section{B. Reconstruction method}
Our reconstruction algorithm, developed to systematically recover optical turbulence and interactions in heterogeneous media, consists of four sequential steps:
\begin{itemize}
\item[(1)] As the first step of our linear perturbation approach along the longitudinal axis, we reconstruct the phase distribution $\phi_0$ at the in-focus position by utilizing the axial asymmetric component of the Eq.~(3) in the main text:
\begin{equation}
\frac{I_{+} - I_{-}}{2\Delta z} \approx -\frac{1}{kn_0} \nabla\cdot[I_{0}\nabla\phi_{0}]
\end{equation}
where $I_{0}$ and $I_{\pm}$ are the inputs of image intensities at the in-focus ($z_0$) and defocus positions ($z_0 \pm \Delta z$). This differential equation represents a Poisson equation that we solve using a geometric multigrid method (GMGM)~\cite{watabe2022, mazumder2016, xue2011, pinhasi2010}. The multigrid approach provides efficient numerical solutions by operating on multiple resolution levels simultaneously, which significantly accelerates convergence compared to single-grid iterative methods. Our implementation employs a V-cycle scheme with Gauss-Seidel relaxation, which effectively handles both low and high spatial frequency components of the phase distribution while maintaining computational efficiency for large datasets. This approach is particularly advantageous for handling the boundary conditions and non-uniformities in intensity that are common in experimental data.

Also, we apply the same GMGM technique to retrieve phase distributions at the half-defocus positions $\phi_{\pm}$ using
\begin{equation}
\frac{I_{\pm} - I_{0}}{\Delta z} \approx -\frac{1}{kn_0} \nabla\cdot[I_{\pm1/2}\nabla\phi_{\pm}]
\end{equation}
where the intensity at half-defocus positions is estimated as $I_{\pm1/2} = (I_{0} + I_{\pm})/2$. 

\item[(2)] The intensity reduction parameter $\partial \alpha/\partial z$ is extracted by leveraging the axial symmetric component of the Eq.~(6) in the main text:
\begin{eqnarray}
I_{+} & \approx & I_0 - \frac{\Delta z}{kn_0}\nabla\cdot[I_0\nabla\varphi_0] - 2I_0 \frac{\partial\alpha}{\partial z}\bigg|_{z_0} \Delta z \nonumber\\
I_{-} & \approx & I_0 + \frac{\Delta z}{kn_0}\nabla\cdot[I_0\nabla\varphi_0] - 2I_0 \frac{\partial\alpha}{\partial z}\bigg|_{z_0} \Delta z \nonumber
\end{eqnarray}
Adding those equations, we obtain
\begin{equation}
I_{+} + I_{-} \approx 2 I_0 - 4 I_0 \frac{\partial \alpha}{\partial z}\bigg|_{z_0} \Delta z \nonumber
\end{equation}
Finally, the reduction parameter can be deduced as follows.
\begin{equation}
\frac{\partial \alpha}{\partial z}\bigg|_{z_0} \approx \frac{I_{+} + I_{-} - 2I_{0}}{-4 I_{0} \Delta z}
\end{equation}
This parameter captures how the wavefield's intensity attenuates during propagation due to absorptive and scattering properties of the medium.

\item[(3)] The phase-coupling parameter $\partial \beta/\partial z$ is computed by utilizing the TPE together with axial phase derivatives $\partial \phi/\partial z \approx (\phi^{+} - \phi^{-})/\Delta z$:
\begin{equation}
\frac{\partial \beta}{\partial z}\bigg|_{z_0} \approx \frac{\phi_{+} - \phi_{-}}{\Delta z} + \frac{1}{2kn_0} \big[\nabla_{\bot}\phi_0\big]^{2} - \frac{1}{2kn_0} \frac{\nabla^2_{\bot}\sqrt{I_0}}{\sqrt{I_0}}
\end{equation}
This intermediary parameter effectively quantifies how local phase variations influence wavefront deformation during propagation.

\item[(4)] Finally, the intermediary parameters $\partial \alpha/\partial z$ and $\partial \beta/\partial z$ are converted into physically meaningful quantities: the spatial distributions of refractive-index fluctuation and attenuation coefficients. The conversion functions are given by the Eq.~(4)-(5) in the main text.
\end{itemize}

These transformations establish a direct link between measurable wavefield properties (intensity and phase) and the fundamental optical properties of the medium (refractive index and attenuation), enabling comprehensive characterization of heterogeneous optical media from a minimal set of through-focus measurements.

\newpage

\section{C. Reconstruction limits}
The physical lower bound of the reconstruction method is fundamentally governed by the signal-to-noise ratio (SNR) of the detector's photon counting capability, which must exceed unity for meaningful reconstruction. This constraint emerges through a systematic two-stage derivation process.

First, we establish the lower bound for TIE-based phase retrieval. The fundamental constraint emerges from the requirement that intensity differences $\Delta I$ between the  defocus image distances $2\Delta z$ must exceed the shot noise limit,
\begin{equation}
\Delta I = \left|-\frac{1}{k n_0} \int^{z_0 + \Delta z}_{z_0 - \Delta z} \nabla \cdot\left[ I \nabla\phi \right] dz \right| > \sqrt{I(z_0) + I_{bg}}
\label{eqn;diff_int}
\end{equation}
where $\Delta z$ and $k n_0$ are the defocus lateral distance and the wave number of signaling photons in the mean refractive index field, respectively. $I_{bg}$ is intensity distributions in background. 

For the sake of simplicity, we assume constant image intensity (i.e., $I_0 = const$) and sinusoidal phase distributions [i.e., $\phi = \phi_0 \sin(k_x x) k_z (z - z_0)$]. This constraint translates directly into a minimum detectable phase amplitude. In this assumption, we modify the Eq.~\eqref{eqn;diff_int} to find that the minimum detectable phase variation,
\begin{equation}
\left| \phi \right| > \phi_{min} = \frac{k n_0 \sqrt{I_0 + I_{bg}}}{\Delta z\ k_x^2\ I_0} = \frac{k n_0 \Delta z}{SNR}\left(\frac{2 \Delta x}{\pi \Delta z}\right)^2
\label{eqn;phi_limit}
\end{equation}
where we assume the minimum spatial frequency $k_x = 2\pi/4 \Delta x$ for phase signal detection. Further details are discussed in Ref. \cite{watabe2022}.


Building upon this foundation, we then derive the lower limit for the $\beta$-derivative intermediary function. Substituting our sinusoidal phase assumption and constant intensity to the TPE [i.e., Eq.~(5) in the main text], we obtain 
\begin{eqnarray}
\frac{\partial \beta}{\partial z} & = & \frac{\partial \phi}{\partial z}  + \frac{1}{2kn_0}\left(\nabla_{\bot} \phi \right)^2 - \frac{1}{2kn_0}\frac{\nabla_{\bot}^2 \sqrt{I}}{\sqrt{I}}\nonumber\\
 & = & \phi_0 \sin\left( k_x x \right) k_z + \frac{1}{2kn_0}\left(k_x \phi_0 \cos(k_x x) k_z \Delta z \right)^2 - \frac{1}{2kn_0}\frac{\nabla_{\bot}^2 \sqrt{I_0}}{\sqrt{I_0}}\nonumber\\
 & = & \frac{\phi_0 \sin\left( k_x x \right) k_z \Delta z}{\Delta z} + \frac{1}{2 kn_0} \left( k_x \phi_0 k_z \Delta z \right)^2 \left( 1 - \sin^2(k_x x) \right) \nonumber\\
 & = & \frac{\phi}{\Delta z} + \frac{1}{2 kn_0} \left( k_x \phi_0 \sin(k_x x) k_z \Delta z \right)^2 \left( \frac{1}{\sin^2(k_x x)} - 1 \right) \nonumber\\
 & = & \frac{\phi}{\Delta z} + \frac{1}{2 kn_0} \left(k_x \phi \right)^2 \left(\frac{1}{\sin^2 (k_x x)}  - 1 \right)
 \label{eqn;beta}
 \end{eqnarray}
Inserting the Eq.~\eqref{eqn;phi_limit} into the Eq.~\eqref{eqn;beta}, the $\beta$-derivative intermediary parameter can be limited as follows.
\begin{equation}
\left| \frac{\partial \beta}{\partial z} \right| > \frac{\phi^{min}}{\Delta z} + \frac{1}{2 kn_0} \left(k_x \phi^{min} \right)^2 \cancelto{\mbox{\ 0}}{\left(\frac{1}{\sin^2 (k_x x)}  - 1 \right)}
\label{eqn;bmin}
\end{equation}
where the nonlinear term vanishes as $\sin^2(k_x x) \to 1$. Dividing by $kn_0$ and setting $SNR = 1$ at observable minimum, we systematically arrive at the final constraint:
\begin{equation}
\left| \Delta n - \frac{1}{2}\kappa^2 \right| > \left(\frac{2 \Delta x}{\pi \Delta z}\right)^2
\end{equation}
This derived physical lower bound characterizes the fundamental limit where spatial frequencies approach the critical diffraction threshold, causing irreversible information loss in the measurements. As the $\Delta x/\Delta z$ ratio increases---either due to poorer spatial resolution or decreased propagation distance---the minimum detectable refractive index variation increases quadratically, thus significantly reducing the method's sensitivity to fine structural features within the sample. 


%
%

\newpage

\section{D. Analysis}
\subsection{D.1 Microscopy configuration}
Configuration details for bright-field imaging systems are described as follows.

\begin{figure*}[!h]
\leftline{\bf \hspace{0.05\linewidth} (a) \hspace{0.40\linewidth} (b)}
\centering
\includegraphics[width= 0.48\linewidth]{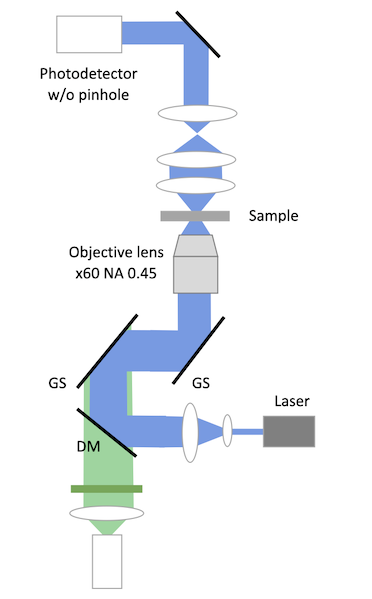}
\includegraphics[width= 0.48\linewidth]{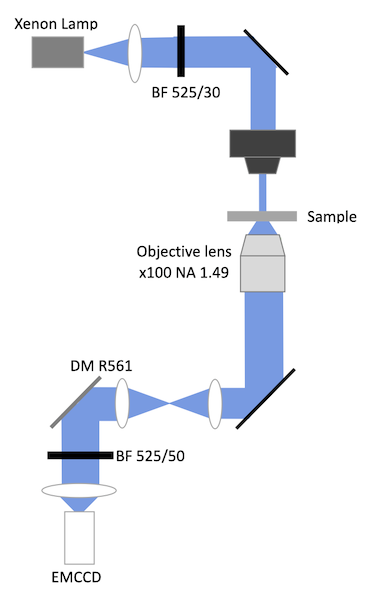}

\caption{Optical path for the bright-field imaging systems: (a) laser-scanning confocal microscope and (b) wide-field microscope.}
\label{figD1;path}
\end{figure*}

\begin{figure*}[!h]
\leftline{\bf (a) \hspace{0.30\linewidth} (b) \hspace{0.30\linewidth} (c)}
\centering
\includegraphics[width= 0.32\linewidth]{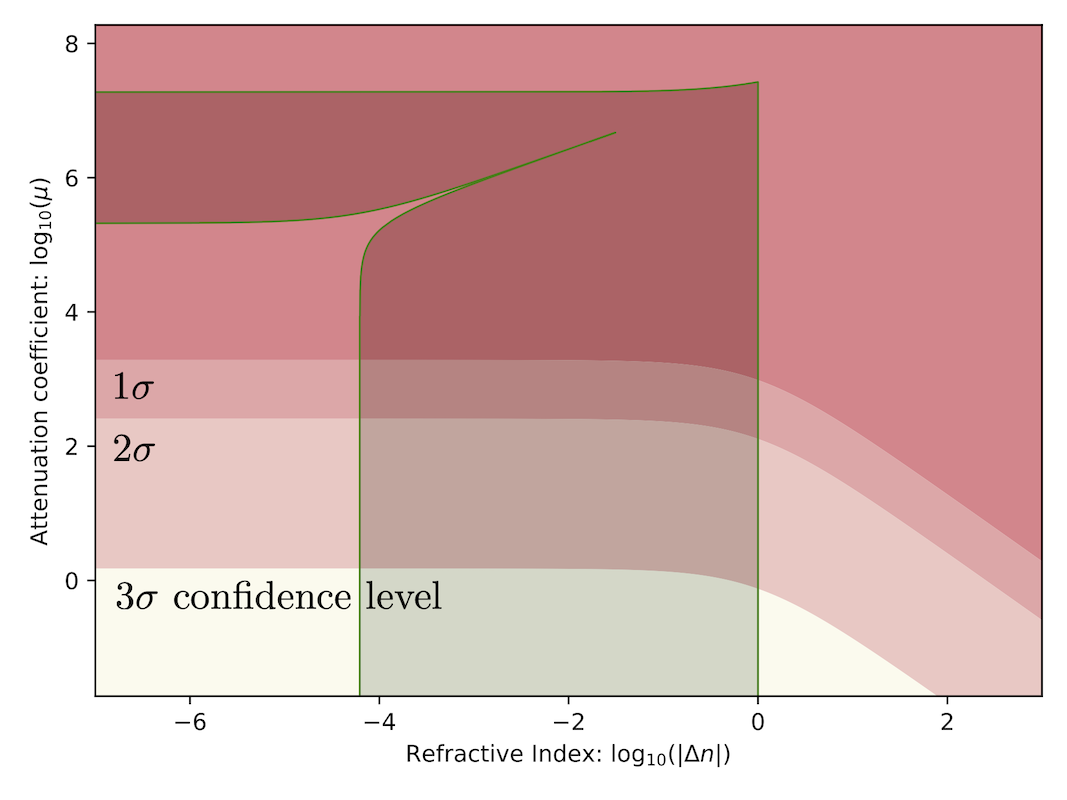}
\includegraphics[width= 0.32\linewidth]{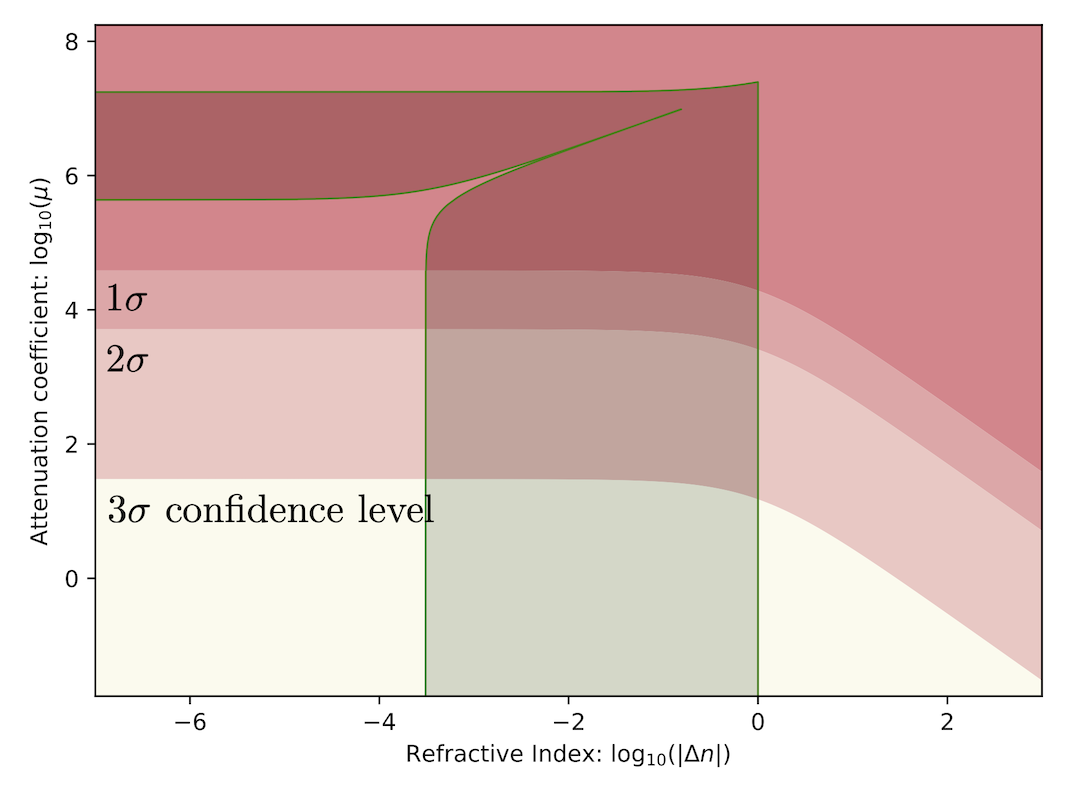}
\includegraphics[width= 0.32\linewidth]{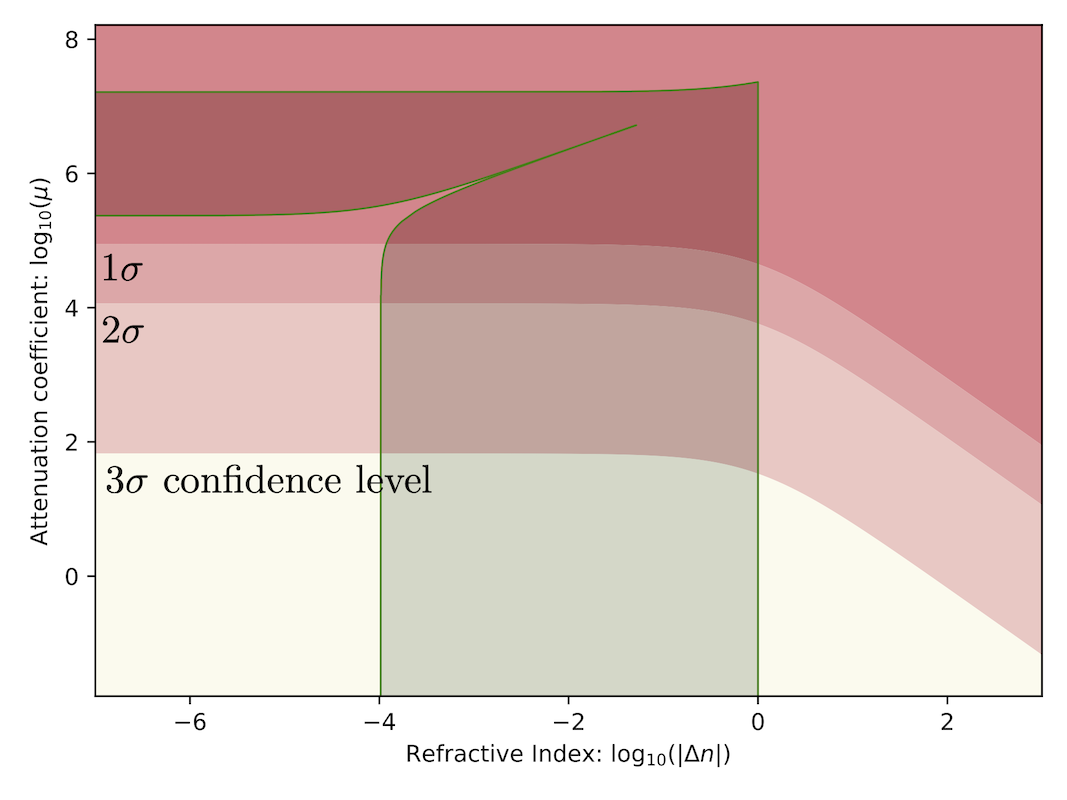}

\leftline{\bf \hspace{0.15\linewidth} MLA \hspace{0.24\linewidth} HeLa cells \hspace{0.18\linewidth} HeLa cell membrane.}

\caption{Parameter allowed regions in the $\Delta n(r_{\bot},z_0)$-$\mu(r_{\bot},z_0)$ spaces: (a) the MLA sample, (b) HeLa cells and (c) HeLa cell membrane. Red contours represent the statistically-allowed regions of the $1\sigma\ (68.3\%)$, $2\sigma\ (95\%)$ and $3\sigma\ (99.7\%)$ confidence intervals. Green area exhibits the physically-allowed region given by the Eq.~(10) in the main text.}
\label{figD1;allow-beta}
\end{figure*}

\begin{itemize}
\item[(1)] Laser-scanning confocal microscope~\cite{yoneda2024}: Bright-field imaging was performed using a Nikon A1 Rsi confocal system equipped with a PlanApo $10\times$ (${\rm NA} = 0.45$; Nikon, Tokyo, Japan) and a PlanApo-VC $60\times$ A water-immersion objective lens (${\rm NA} = 1.2$; Nikon) at room temperature. As illustrated in Figure~\ref{figD1;path}(a), the illumination laser beam is directed to a confocal scanner unit through a series of relay optics, and is reflected by a dichroic mirror (DM). The beam is then scanned by a Galvano scanner (GS) and focused onto the physical and biological samples by the objective lens. The transmitted light is collected by a condenser lens of K{\"o}ller illumination optics, and is then relayed to a photodetector without a pinhole, allowing efficient signal collection for phase retrieval based on TIE.

\begin{itemize}
\item Physical sample: The MLA300-14AR-M, a precision-fabricated microlens array (MLA) from Thorlabs housed in a $\phi 1^{\prime\prime}$ mount ($25.4\ {\rm mm}$ diameter, $3.5\ {\rm mm}$ thickness) compatible with standard optomechanical components~\cite{mla}, was employed for testing the reconstruction performance. This array consists of a $10 \times 10\ {\rm mm^2}$ fused silica substrate patterned with square plano-convex microlenses ($295\ {\rm \mu m}$ side length, $300\ {\rm \mu m}$ pitch) arranged in a regular square grid. The lenslets were fabricated using photolithographic processes to ensure high positional accuracy and shape uniformity. A broadband anti-reflection (AR) coating optimized for the $400$-$900\ {\rm nm}$ wavelength range was applied, reducing surface reflectivity to below $1\%$. The array provides an effective focal length of $14.6 {\rm mm}$ and a fill factor of approximately $\sim 96.7\%$, enabling efficient optical coupling. The MLA was observed using PlanApo $10\times$ objective lens.

Parameters: wavelength of illumination light, $\lambda = 488\ {\rm nm}$; defocus distance, $\Delta z = 100\ {\rm \mu m}$; the scanning pitch, $0.621 {\rm \mu m}$; magnification, $\times 10$; image pixel length, $\Delta x = \Delta y = 1.24\ {\rm \mu m}$; total number of z-stack images, $41$ image took $19 {\rm min}$; image size, $1024 \times 1024\ {\rm pixels}$; beam flux, $\Phi(k) = \delta(k - k_{\lambda})$; and see Figure~\ref{figD1;allow-beta}(a) for parameter allowed region in the $\Delta n(r_{\bot},z_0)$-$\mu(r_{\bot},z_0)$ spaces.

\item Biological sample: HeLa cells (RCB0007, RIKEN Cell Bank, Tsukuba, Japan) were cultured in culture medium: Dulbecco's modified essential medium (11965-092, Thermo Fisher Scientific, Walthum, MA, USA) supplemented with $10\%$ fetal bovine serum (FBS; 10437, Invitrogen) and penicillin/streptomycin (PS; 15140-122, ThermoFisher) at $37^{\circ}C$ under humidified $5\%$ CO2 atmosphere until observation. Before observation, the culture medium to imaging medium; DMEM without phenol red (21063-029, ThermoFisher) supplemented with $10\%$ FBS and PS.

Parameters: wavelength of illumination light, $\lambda = 488\ {\rm nm}$; defocus distance, $\Delta z = 5.00\ {\rm \mu m}$; image pixel length, $\Delta x = \Delta y = 0.138\ {\rm \mu m}$; total number of z-stack images, $81$ frames image took $4.7 {\rm min}$; image size, $512 \times 512\ {\rm pixels}$; beam flux, $\Phi(k) = \delta(k - k_{\lambda})$; and see Figure~\ref{figD1;allow-beta}(b) for parameter allowed region in the $\Delta n(r_{\bot},z_0)$-$\mu(r_{\bot},z_0)$ spaces.

\end{itemize}

\item[(2)] Wide-field microscope: Bright-field observation experiments were conducted using home-built objective-type total internal reflection fluorescence microscopes (TIRFM) based on inverted microscopes (IX81; Olympus Corp.)~\cite{yamada2016}. As illustrated in Fig.~\ref{figD1;path}(b), illumination was provided by a Xenon lamp (SLS401 Light Source with Xenon Arc Lamp for $240$-$2400$ nm, $5800$ K, Thorlabs)~\cite{xenon}, with the beam first collimated and then filtered through a bandpass excitation filter (BF 525/30, Semrock Inc.) to select the desired wavelength range. The light was then reflected by a dichroic mirror (DM) and directed toward the biological sample (i.e., HeLa cells) through a high numerical aperture oil-immersion objective lens ($100\times$, NA $1.49$; Olympus Corp.). After interacting with the sample, the transmitted light was collected by the same objective and passed through an emission optical train that included a second dichroic mirror (DM R561, Semrock Inc.) and an emission bandpass filter (BF 525/50, Semrock Inc.). The filtered emission signal was finally focused onto an electron-multiplying CCD (EMCCD) camera (ImagEM; Hamamatsu Photonics Inc.) for image acquisition. 


\begin{itemize}
\item 
Parameters: wavelength of emission light, $\lambda = 525\ {\rm nm}$; defocus distance, $\Delta z = 5\ {\rm \mu m}$; image pixel length, $\Delta x = \Delta y = 80\ {\rm nm}$; total number of z-stack images, $41$ frames; and image size, $512 \times 512\ {\rm pixels}$; beam flux, $\Phi(k) = \Phi_{Xe}$ \cite{xenon}; and see Figure~\ref{figD1;allow-beta}(c) for parameter allowed region in the $\Delta n(r_{\bot},z_0)$-$\mu(r_{\bot},z_0)$ spaces.

\end{itemize}

\end{itemize}

\newpage


\newpage

\subsection{D.2 Reconstruction results}

\leftline{\bf - MLA sample}

\begin{figure}[!h]

\leftline{\bf (a) Intensity distributions at $z_0-\Delta z$, $z_0$ and $z_0+\Delta z$}
\centering
\includegraphics[width=0.32\linewidth]{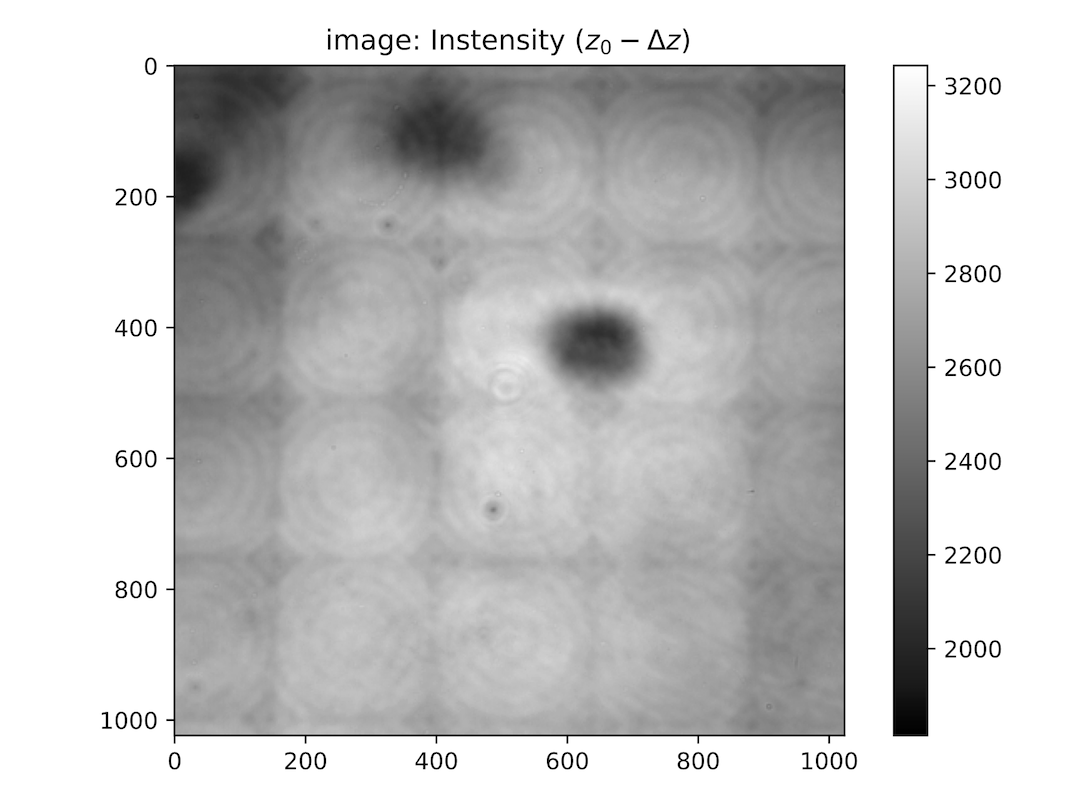}
\includegraphics[width=0.32\linewidth]{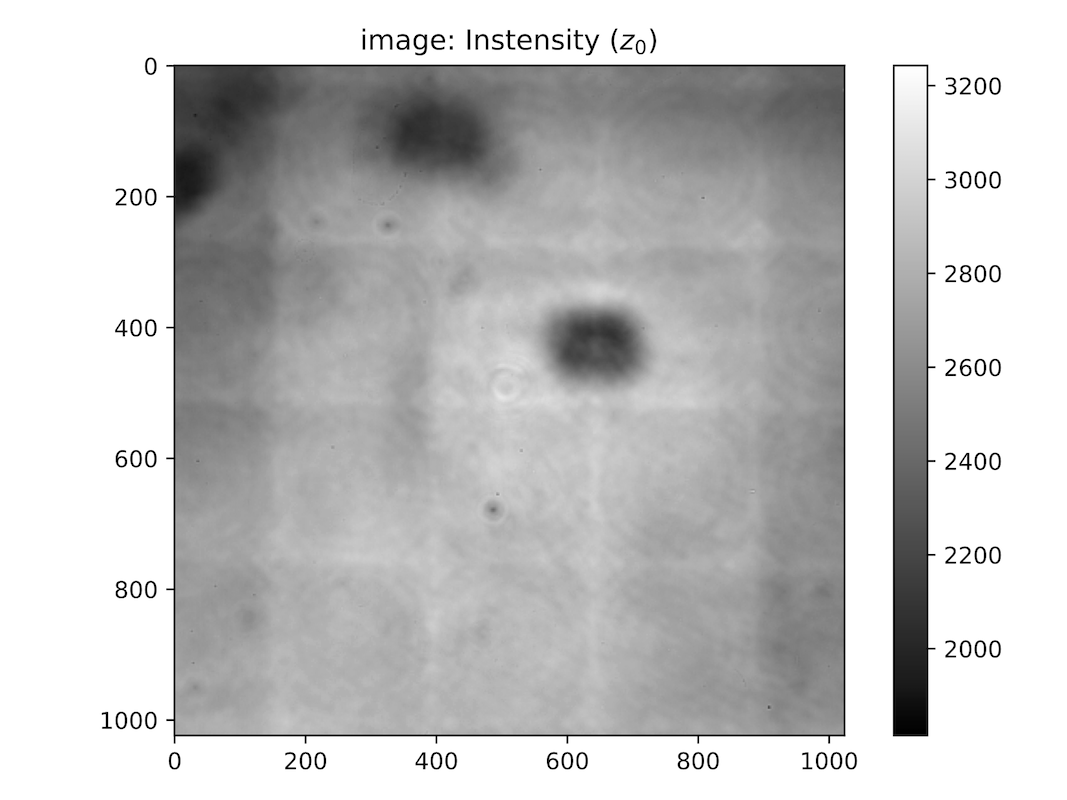}
\includegraphics[width=0.32\linewidth]{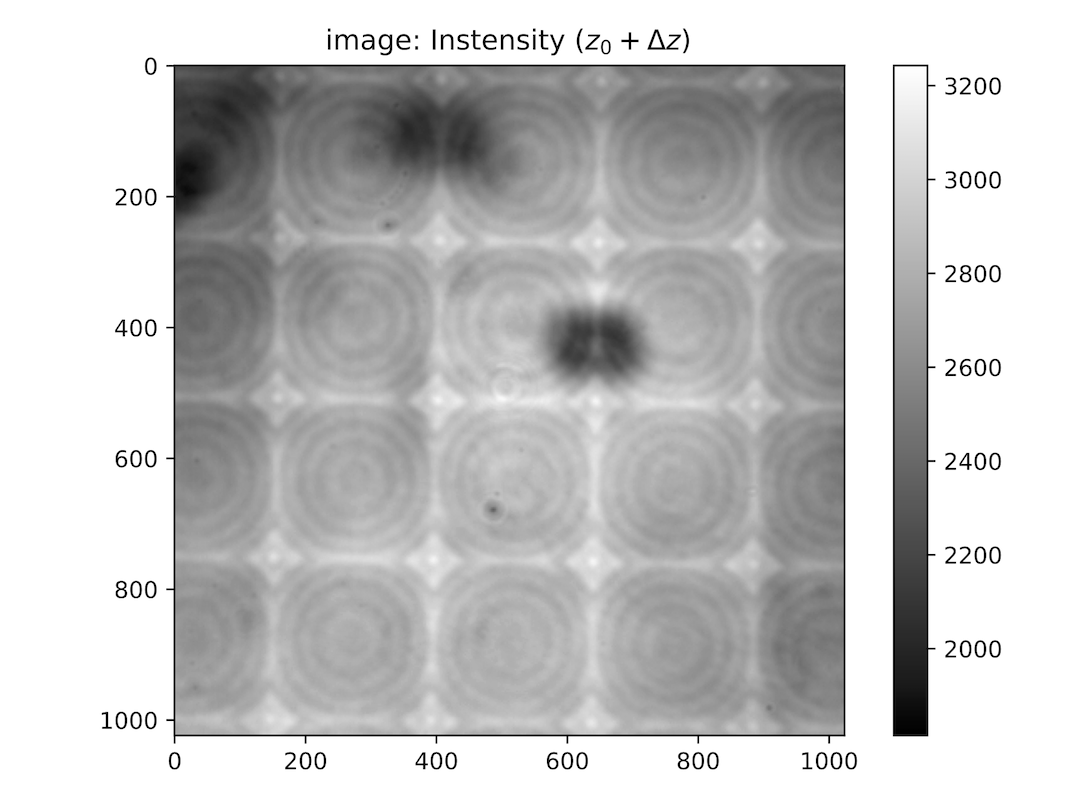}

\leftline{\bf (b) Phase distributions at $z_0-\Delta z/2$, $z_0$ and $z_0+\Delta z/2$}
\centering
\includegraphics[width=0.32\linewidth]{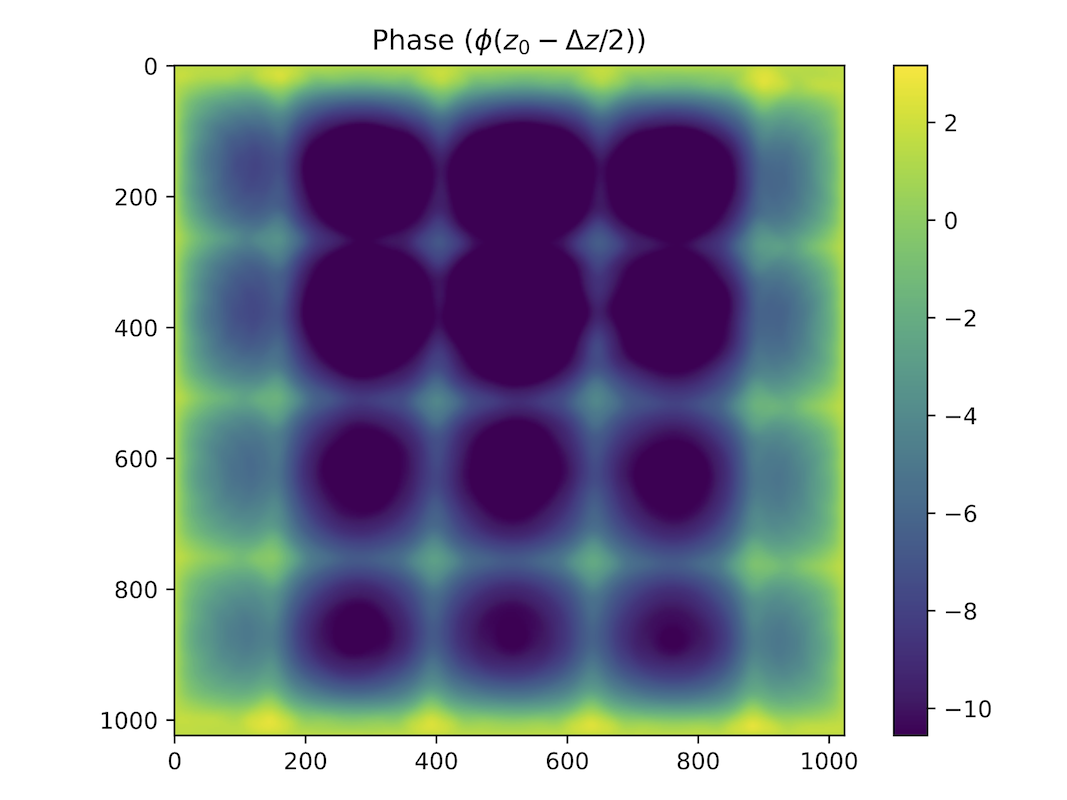}
\includegraphics[width=0.32\linewidth]{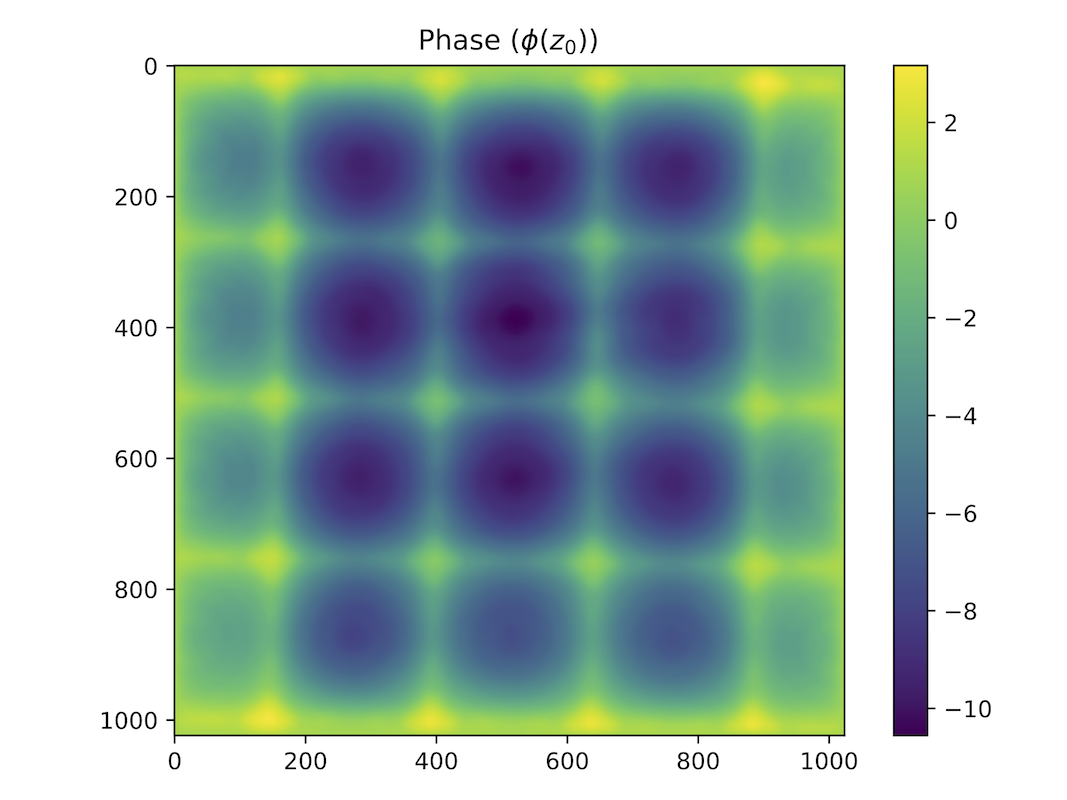}
\includegraphics[width=0.32\linewidth]{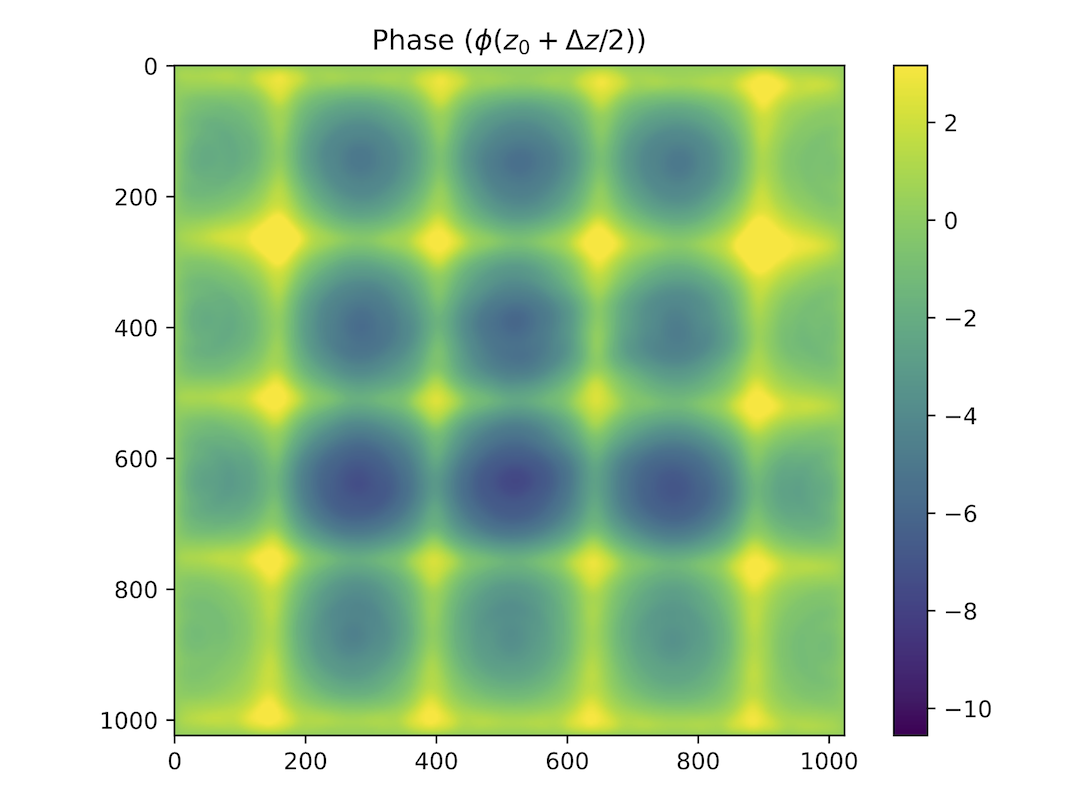}

\leftline{\bf (c) Reconstructed distributions of the $\alpha(r_{\bot}, z_0)$-derivative function}
\centering
\includegraphics[width=0.32\linewidth]{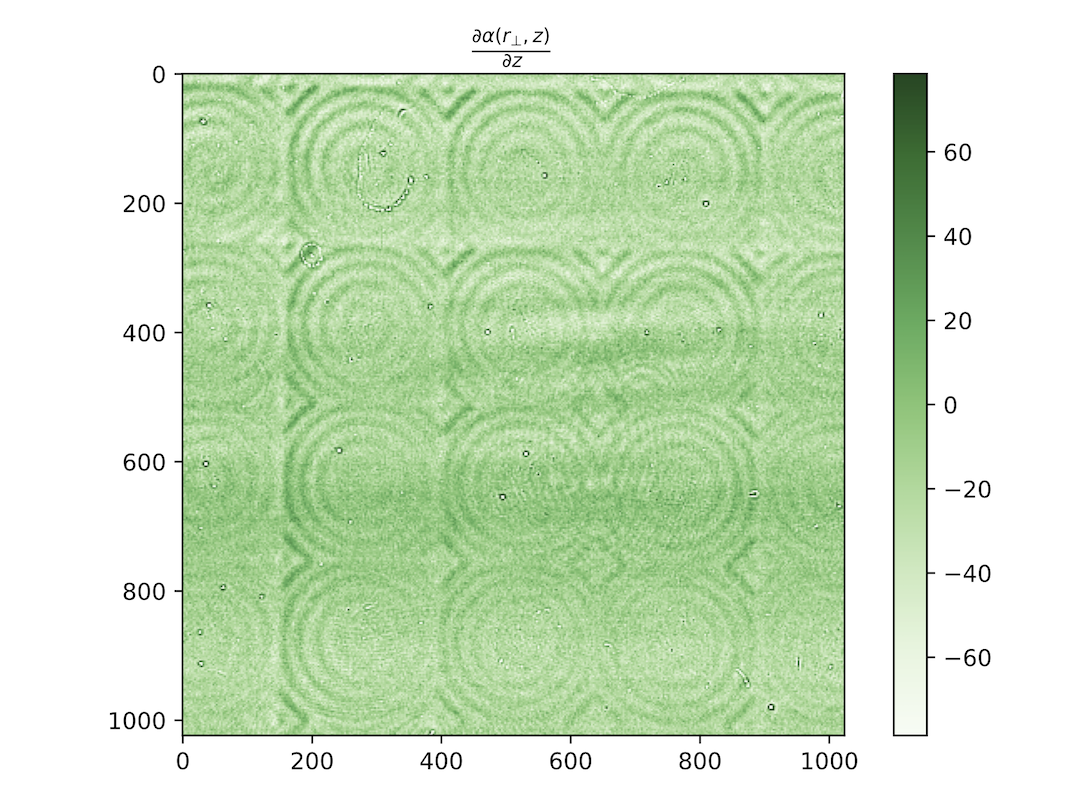}
\includegraphics[width=0.32\linewidth]{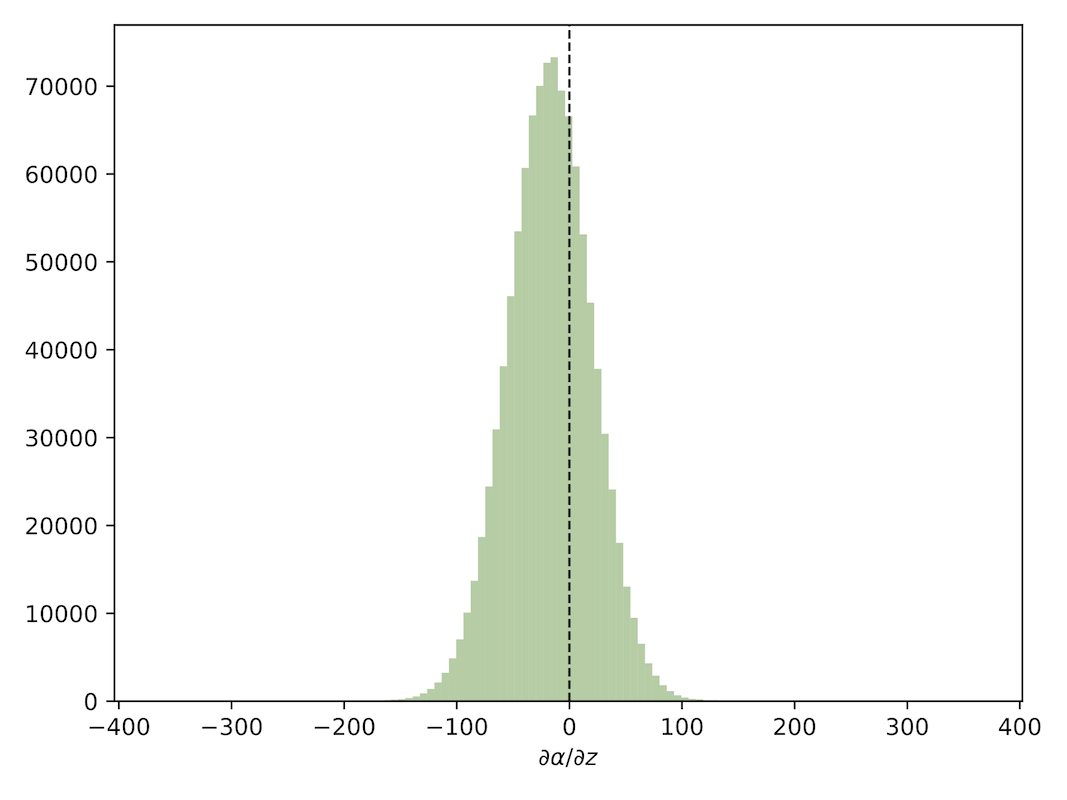}
\hspace{0.34\linewidth}

\leftline{\bf (d) Reconstructed distributions of the $\beta(r_{\bot}, z_0)$-derivative function}
\centering
\includegraphics[width=0.32\linewidth]{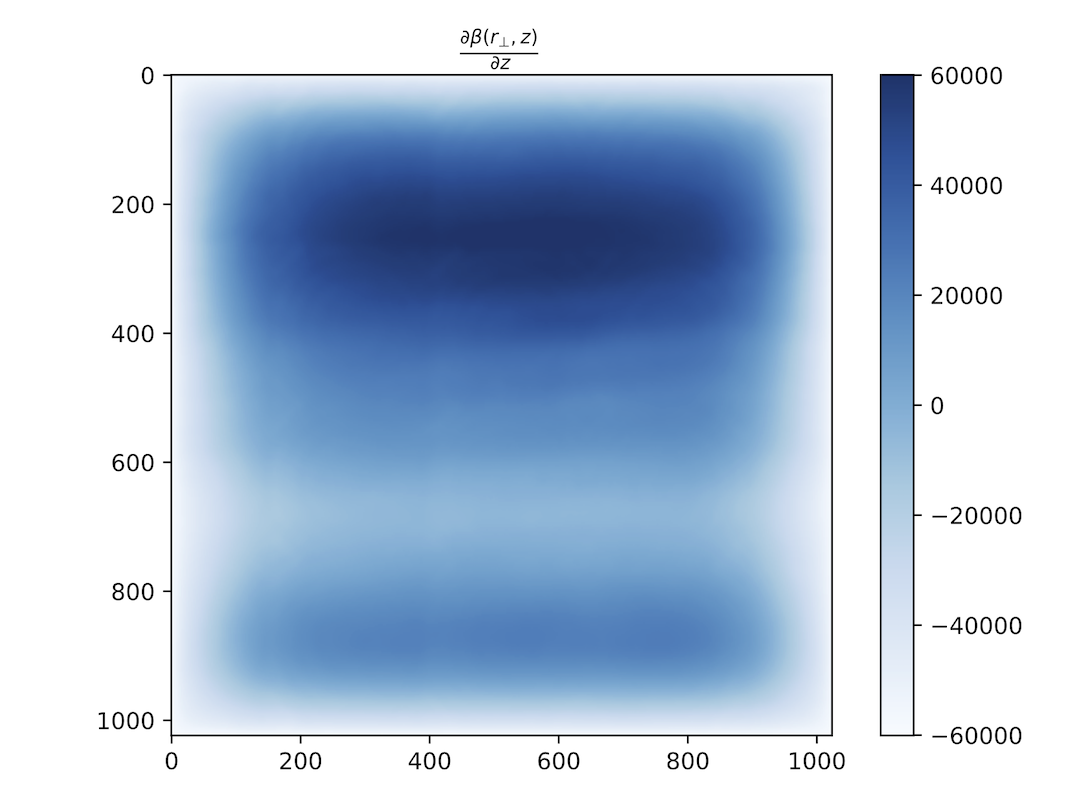}
\includegraphics[width=0.32\linewidth]{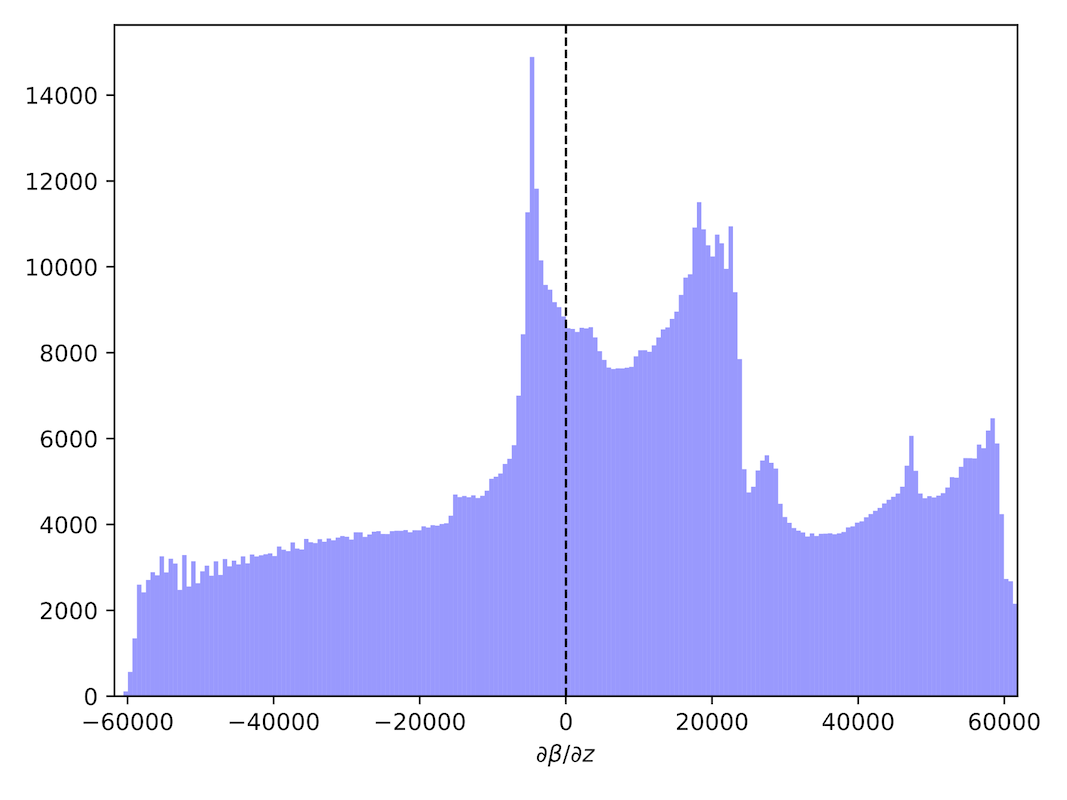}
\hspace{0.33\linewidth}

\caption{Results for the MLA sample 1: (a) Intensity distributions, $I(r_{\bot},z_{0}-\Delta z)$, $I(r_{\bot},z_{0})$, and $I(r_{\bot},z_{0}+\Delta z)$. (b) Phase distributions reconstructed from the three fluorescent cell images. (c) Reconstructed distributions of the intensity reduction parameter $\partial_z \alpha(r_{\bot}, z_{0})$ and (d) the phase-coupling parameter $\partial_z \beta(r_{\bot}, z_{0})$.}
\label{figD2;results01a}
\end{figure}

\newpage

\begin{figure}[!h]
\leftline{\bf (e) Reconstructed distributions of refractive-index fluctuations $\Delta x$}
\centering
\includegraphics[width=0.32\linewidth]{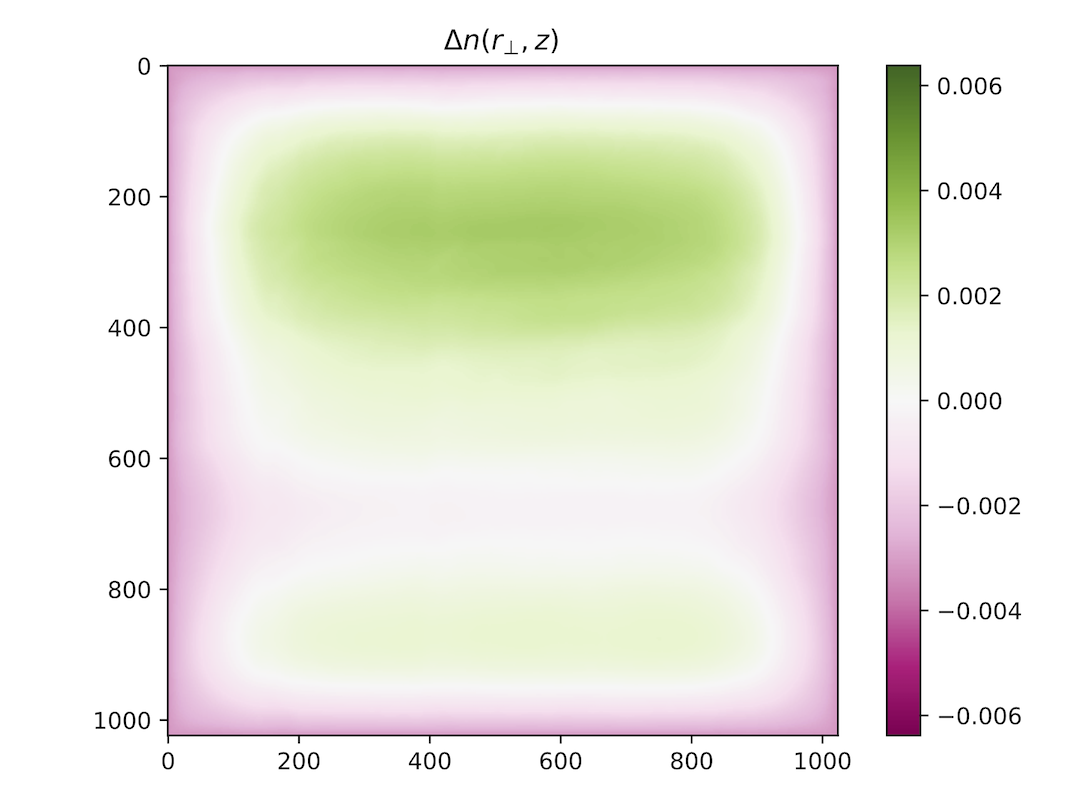}
\includegraphics[width=0.32\linewidth]{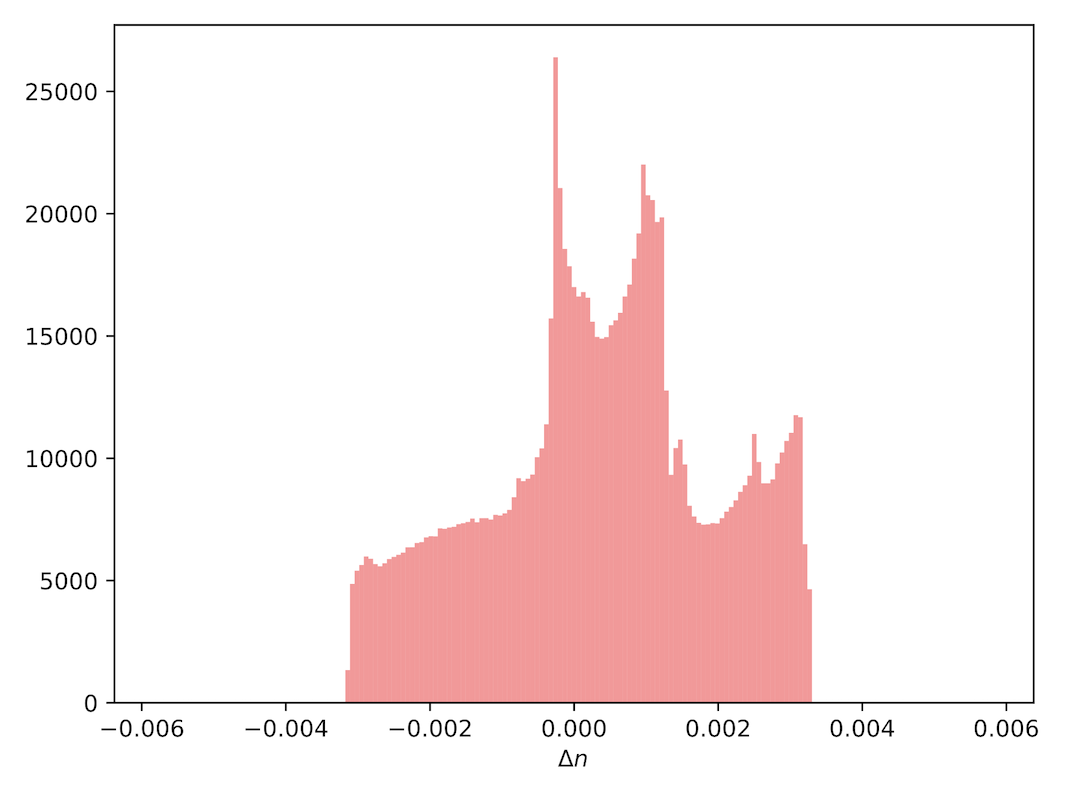}
\includegraphics[width=0.32\linewidth]{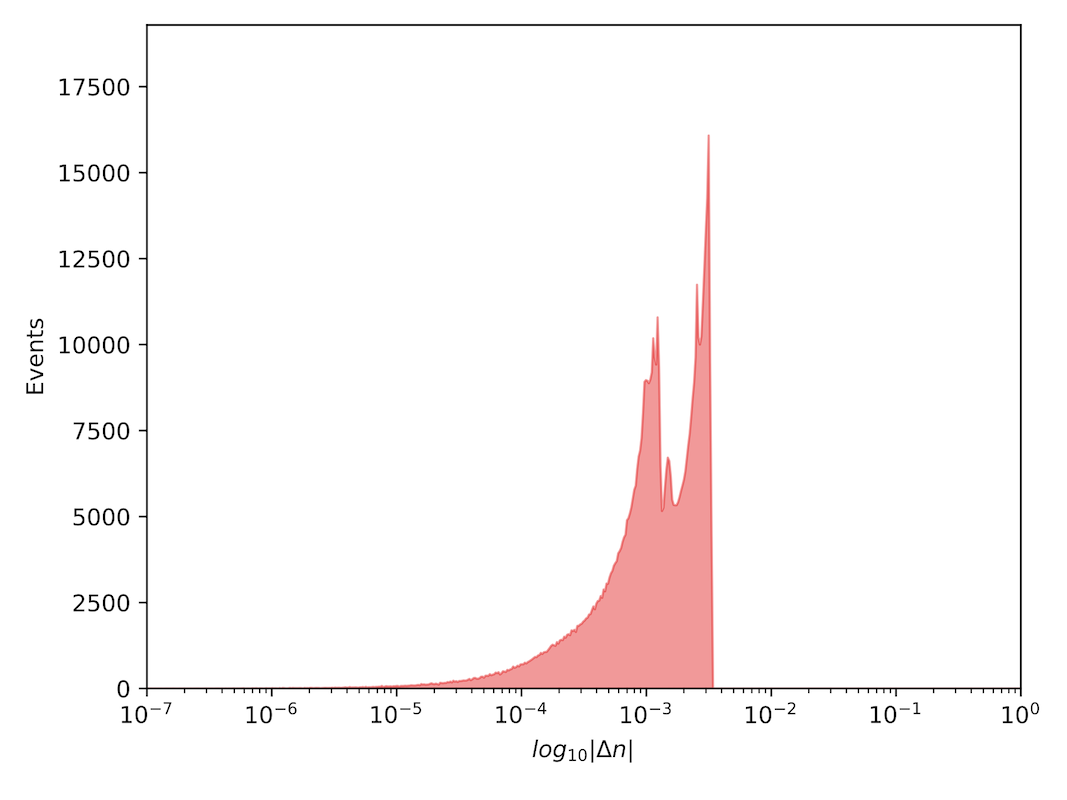}

\leftline{\bf (f) Reconstructed distributions of attenuation coefficients $\mu$}
\centering
\includegraphics[width=0.32\linewidth]{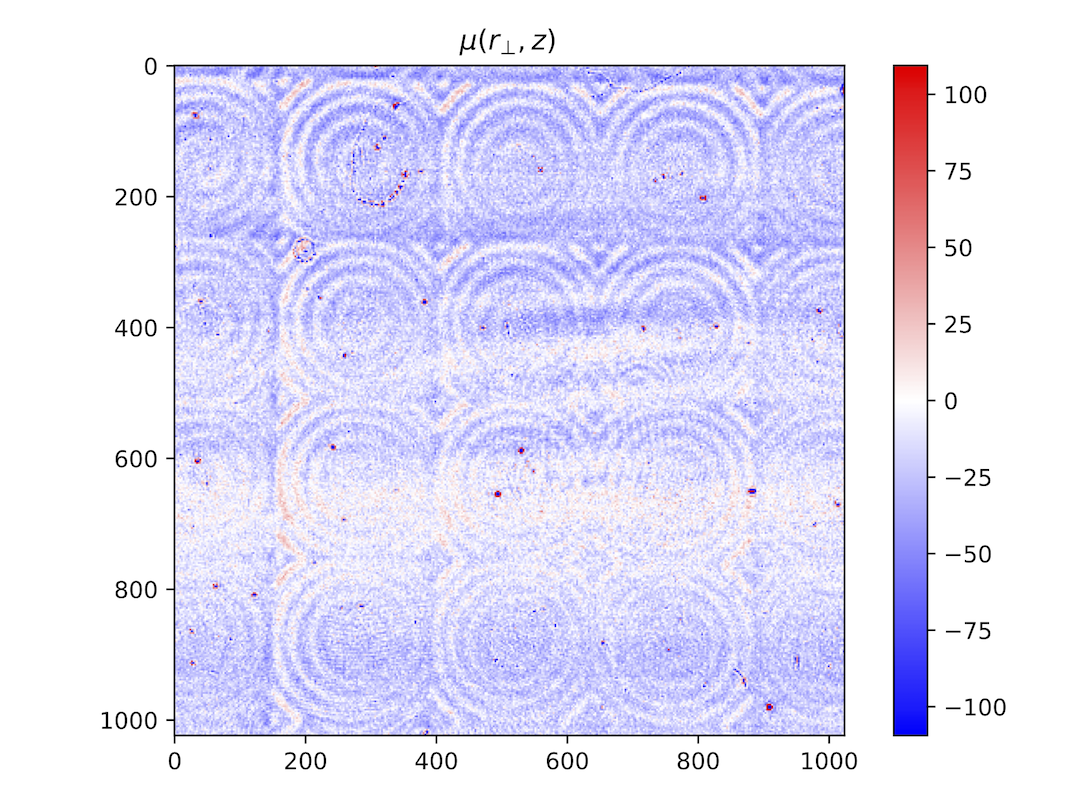}
\includegraphics[width=0.32\linewidth]{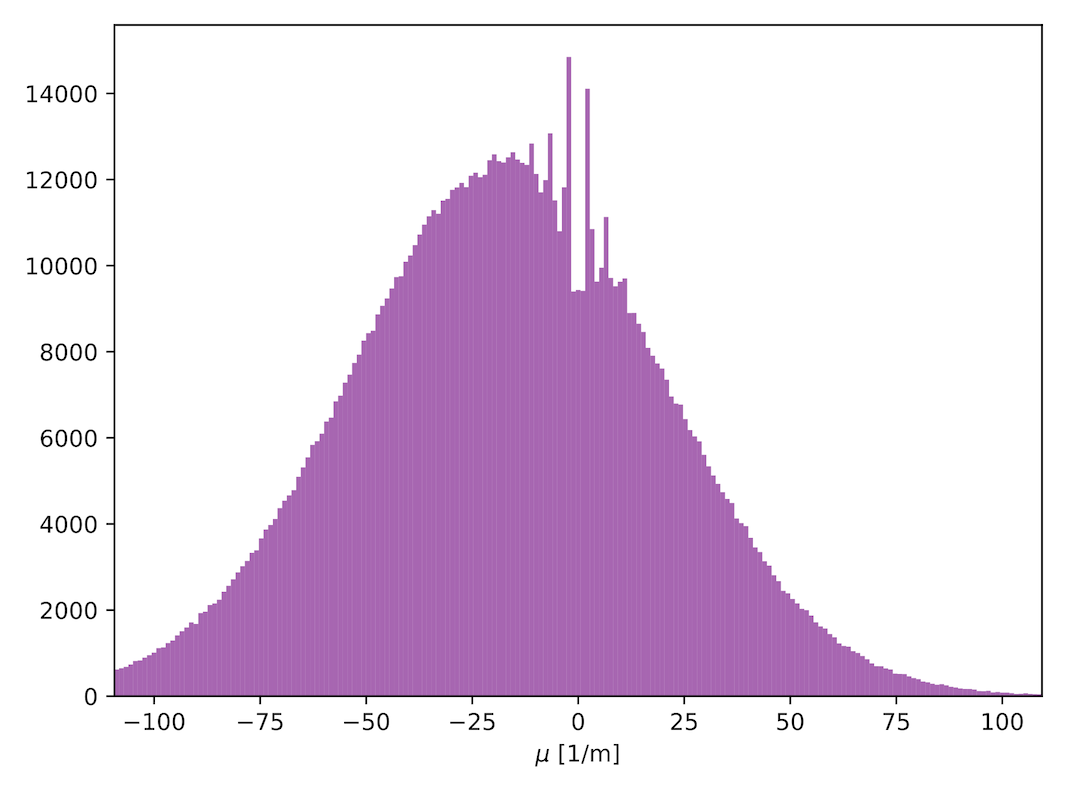}
\includegraphics[width=0.32\linewidth]{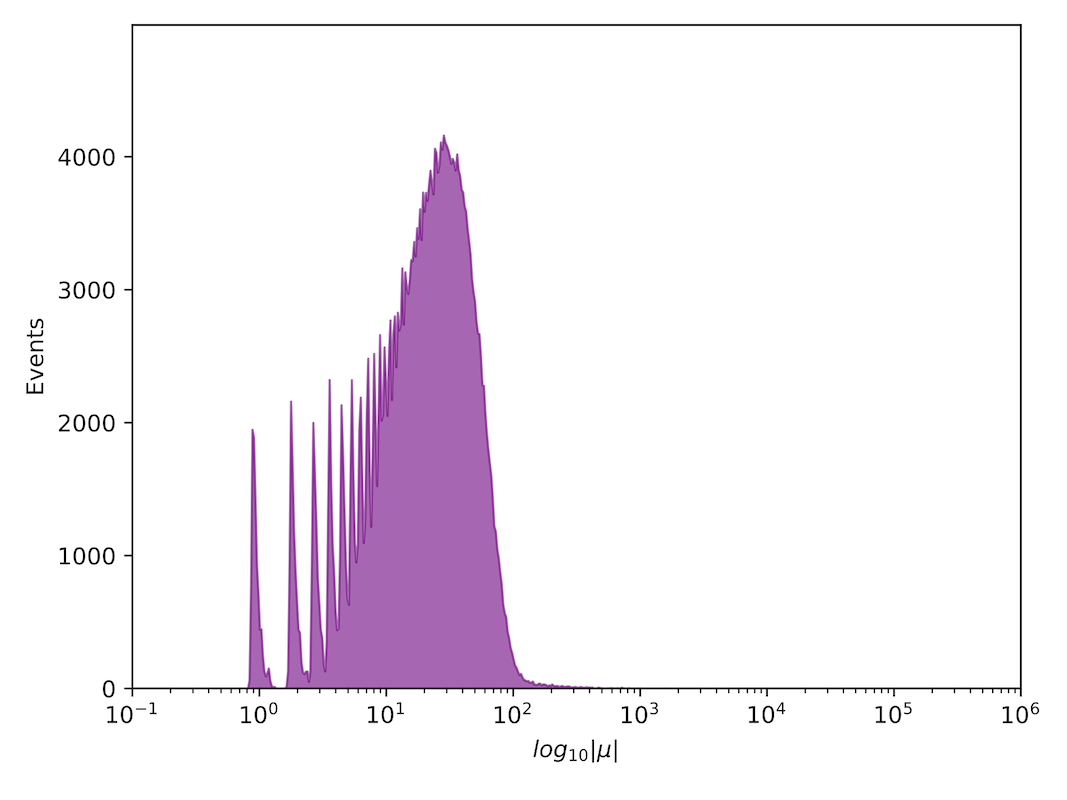}

\leftline{\bf (g) 2D histogram of $\Delta n$ and $\mu$}
\centering
\includegraphics[width=0.63\linewidth]{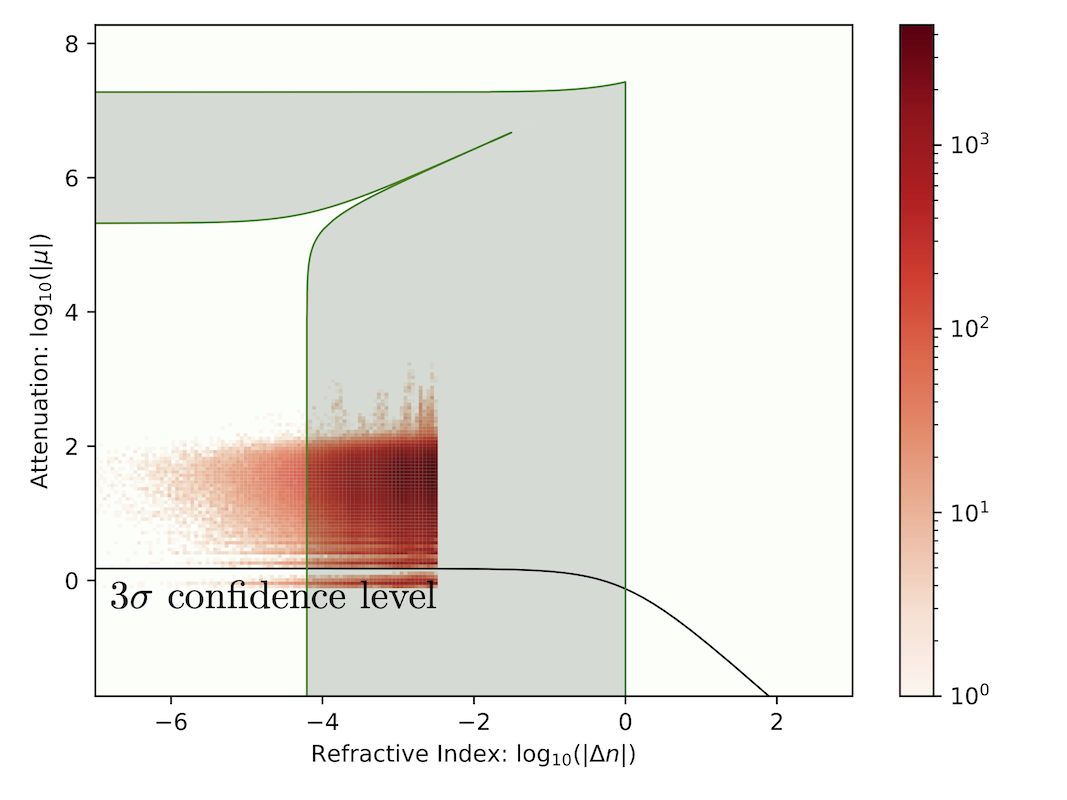}
\hspace{0.30\linewidth}

\caption{Results for the MLA sample: (e) Reconstructed distributions and histogram of the refractive-index fluctuations and (f) attenuation coefficients. (g) Correlation pattern between $\Delta n(r_{\bot},z_0)$ and $\mu(r_{\bot},z_0)$ lies with the $3\sigma$ confidence level (i.e., black solid line). $33,233$ events ($3.169\%$) fall outside the physical parameter boundaries (i.e., green area) given by the Eq.~(10) in the main text.}
\label{figD2;results01b}
\end{figure}

\newpage

\leftline{\bf - HeLa cell}

\begin{figure}[!h]

\leftline{\bf (a) Intensity distributions at $z_0-\Delta z$, $z_0$ and $z_0+\Delta z$}
\centering
\includegraphics[width=0.32\linewidth]{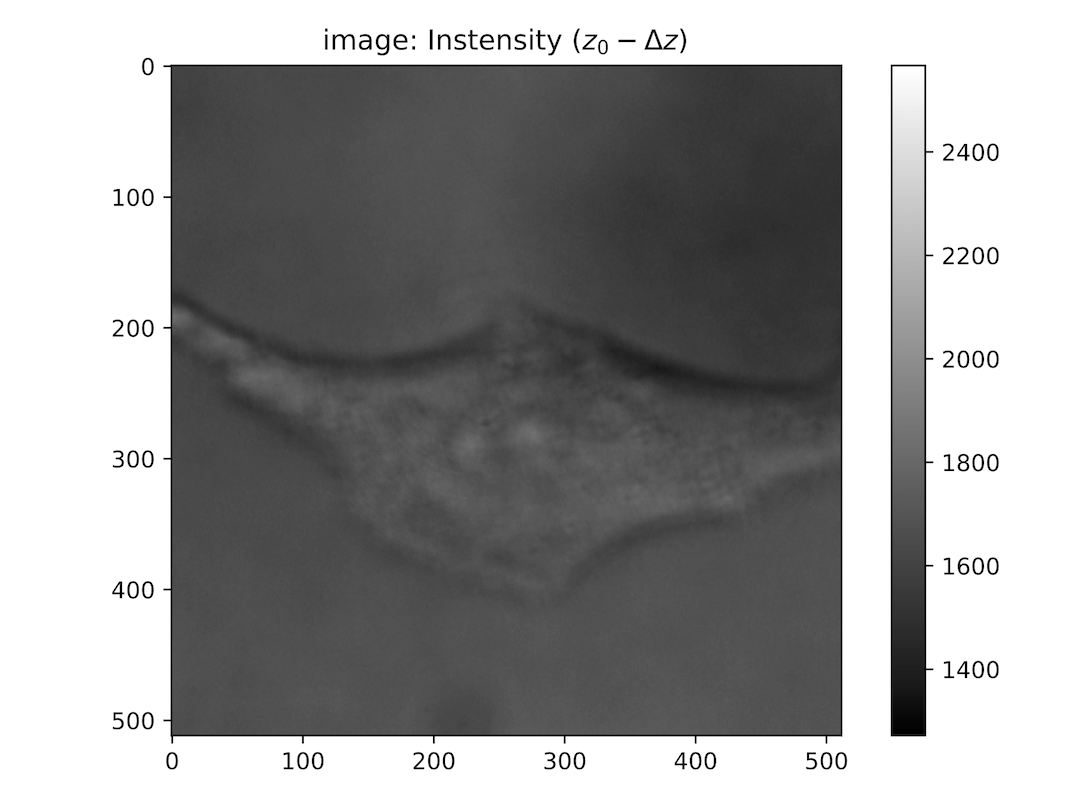}
\includegraphics[width=0.32\linewidth]{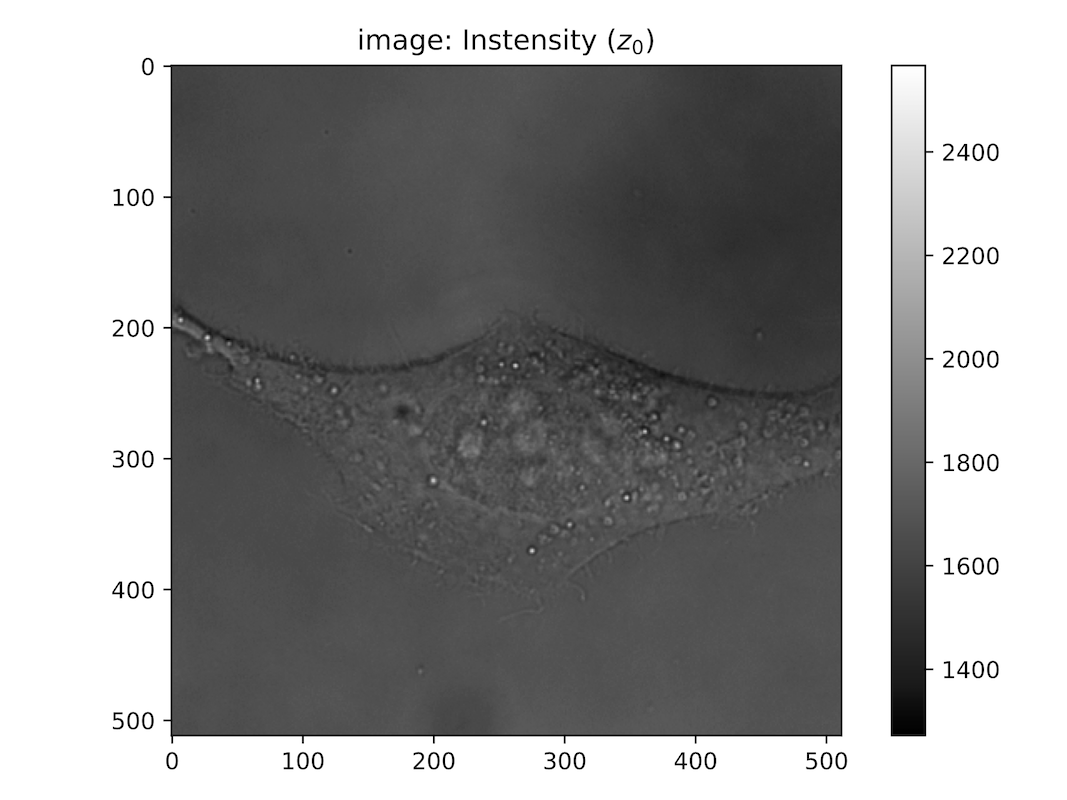}
\includegraphics[width=0.32\linewidth]{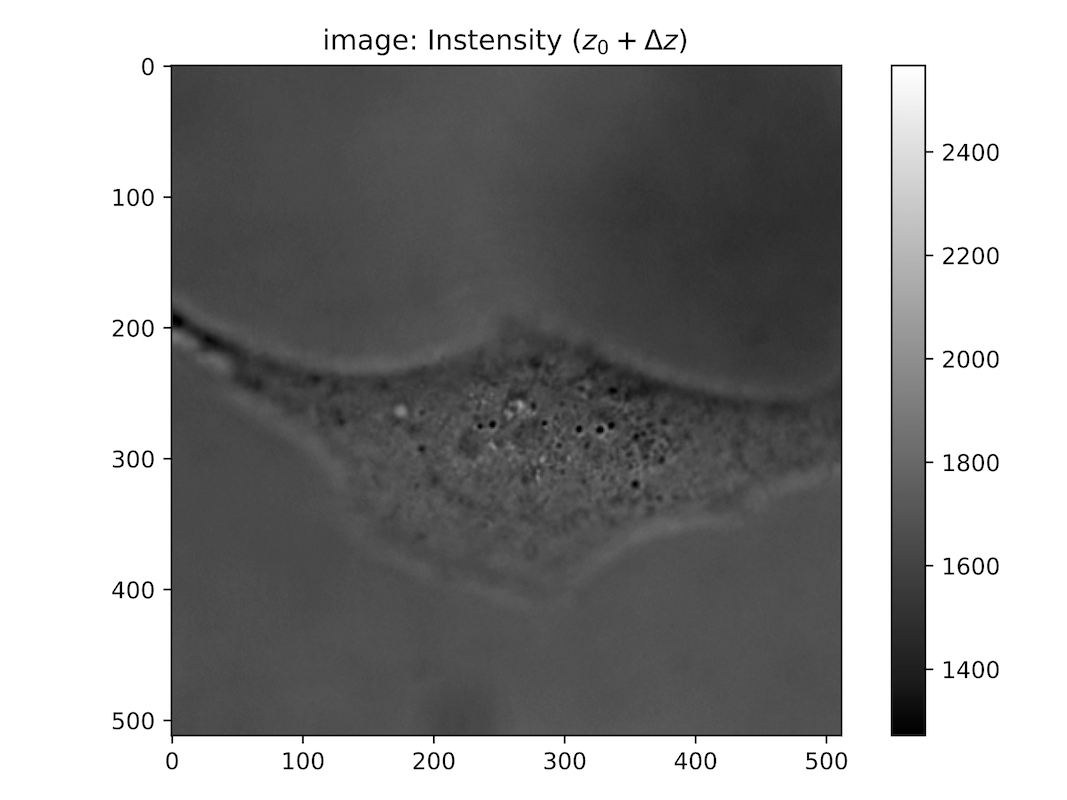}

\leftline{\bf (b) Phase distributions at $z_0-\Delta z/2$, $z_0$ and $z_0+\Delta z/2$}
\centering
\includegraphics[width=0.32\linewidth]{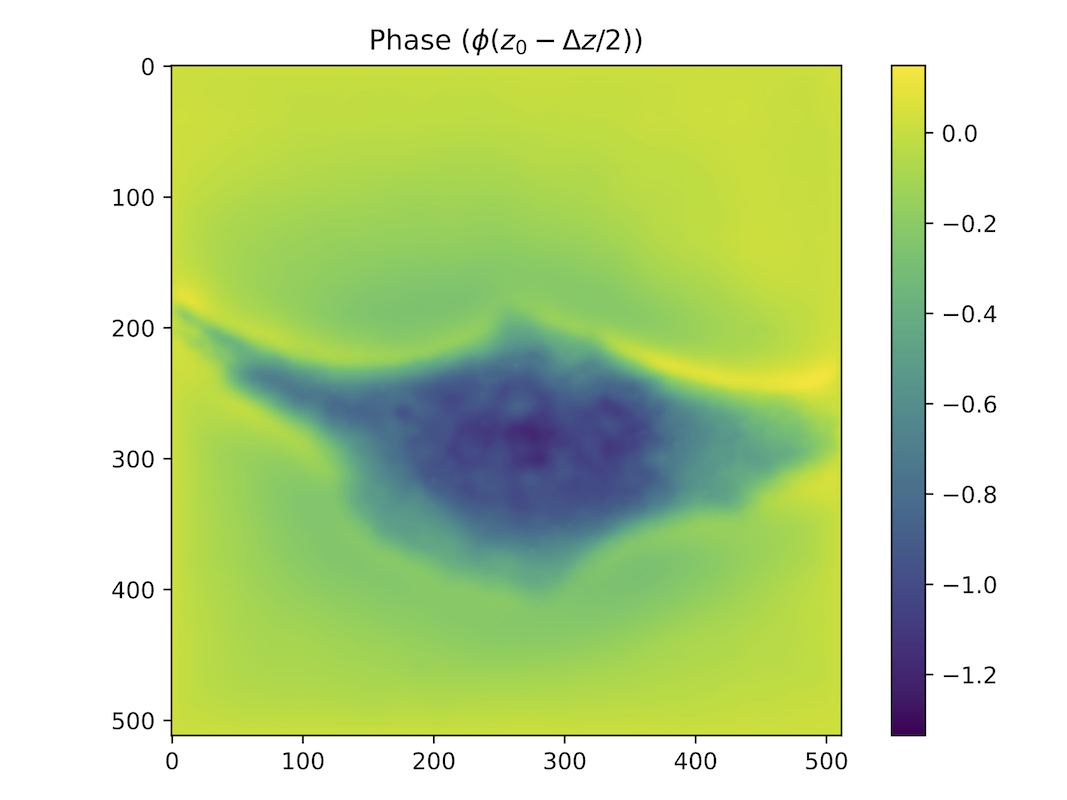}
\includegraphics[width=0.32\linewidth]{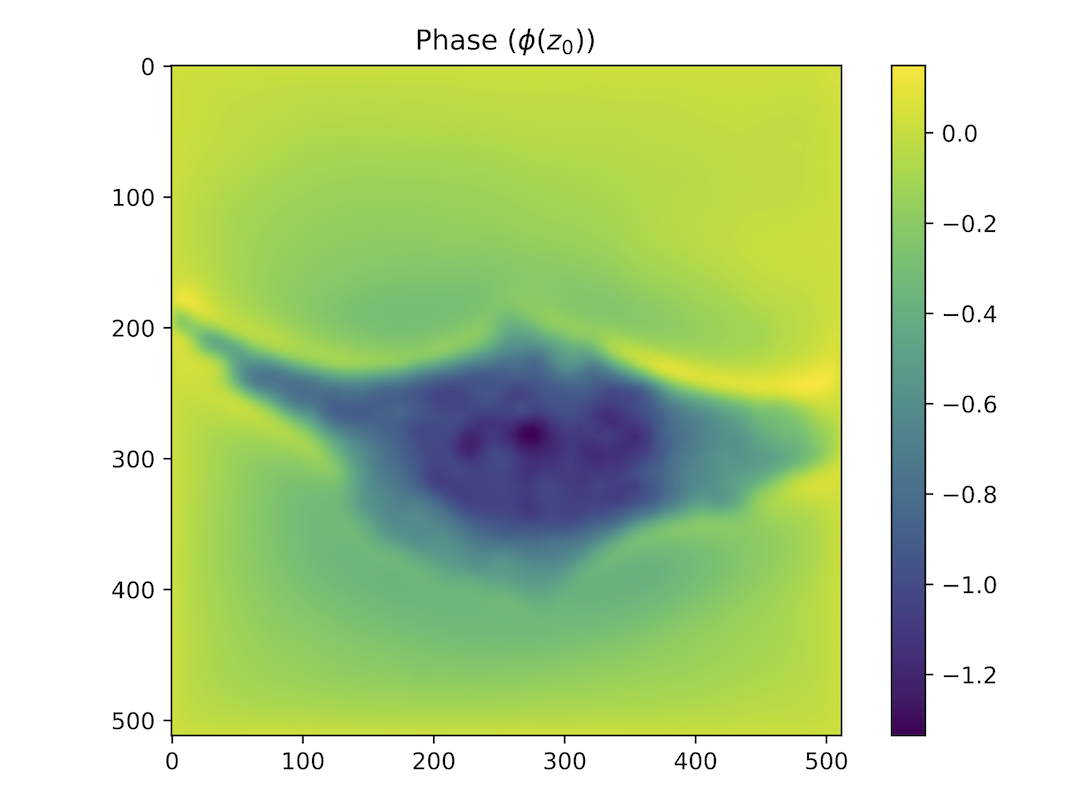}
\includegraphics[width=0.32\linewidth]{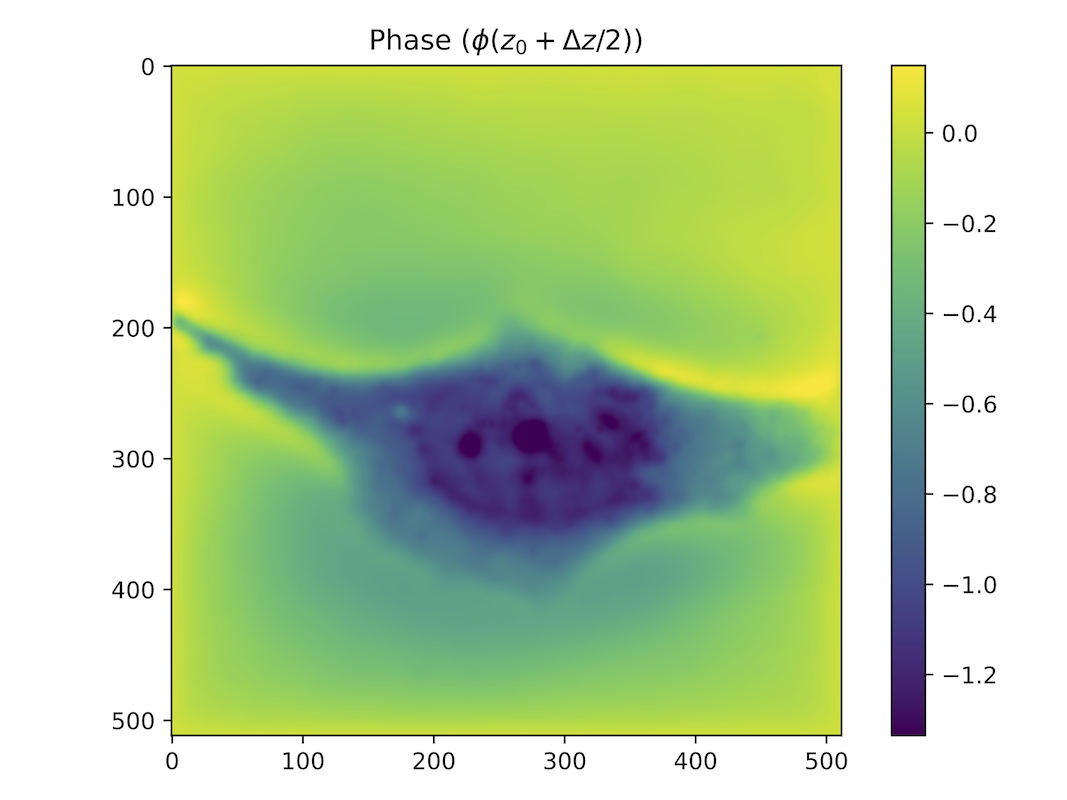}

\leftline{\bf (c) Reconstructed distributions of the $\alpha(r_{\bot}, z_0)$-derivative function}
\centering
\includegraphics[width=0.32\linewidth]{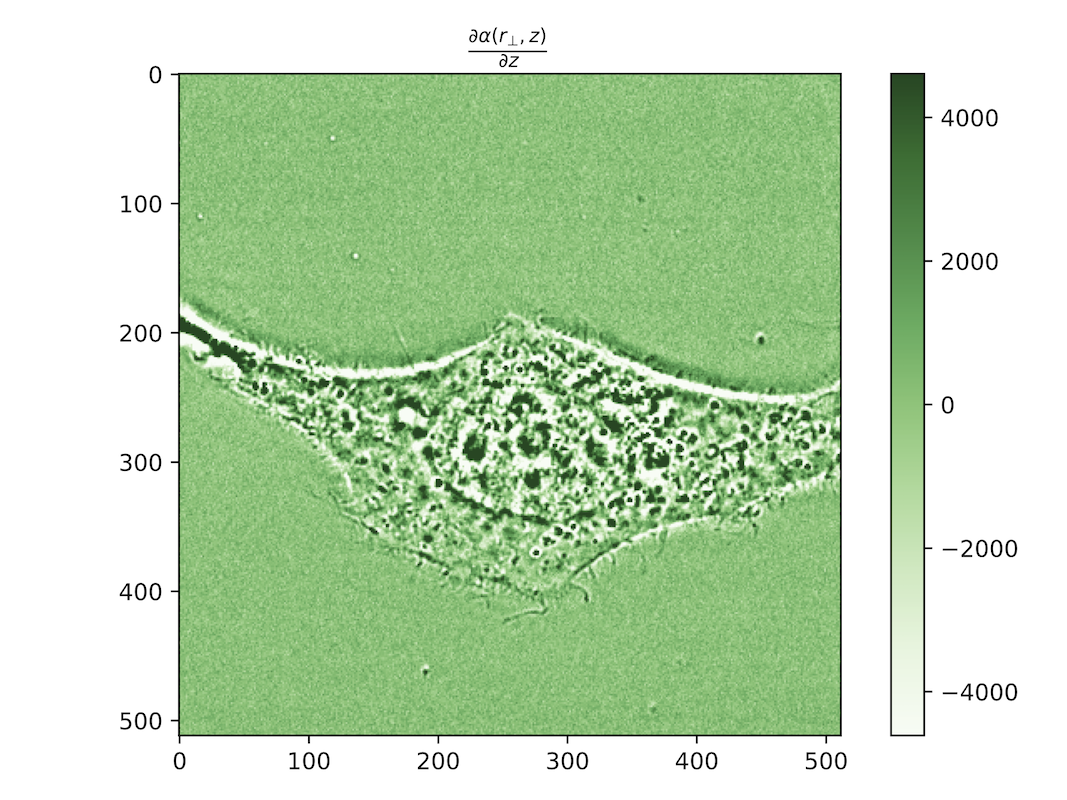}
\includegraphics[width=0.32\linewidth]{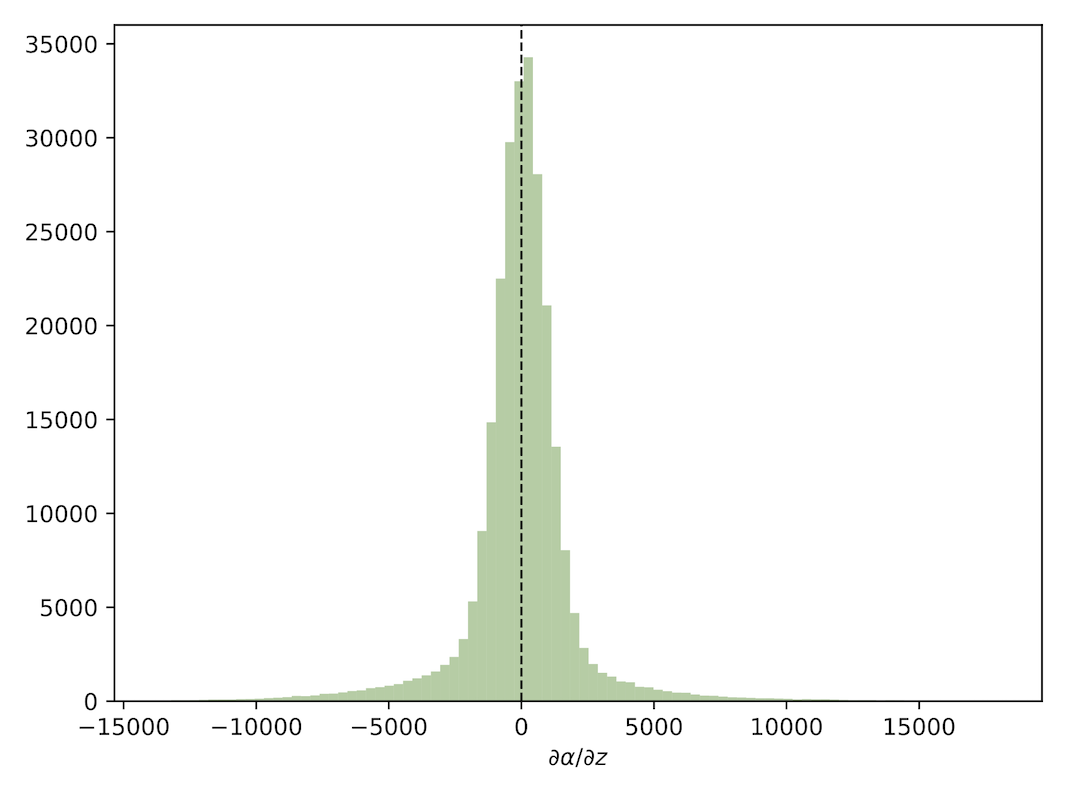}
\hspace{0.34\linewidth}

\leftline{\bf (d) Reconstructed distributions of the $\beta(r_{\bot}, z_0)$-derivative function}
\centering
\includegraphics[width=0.32\linewidth]{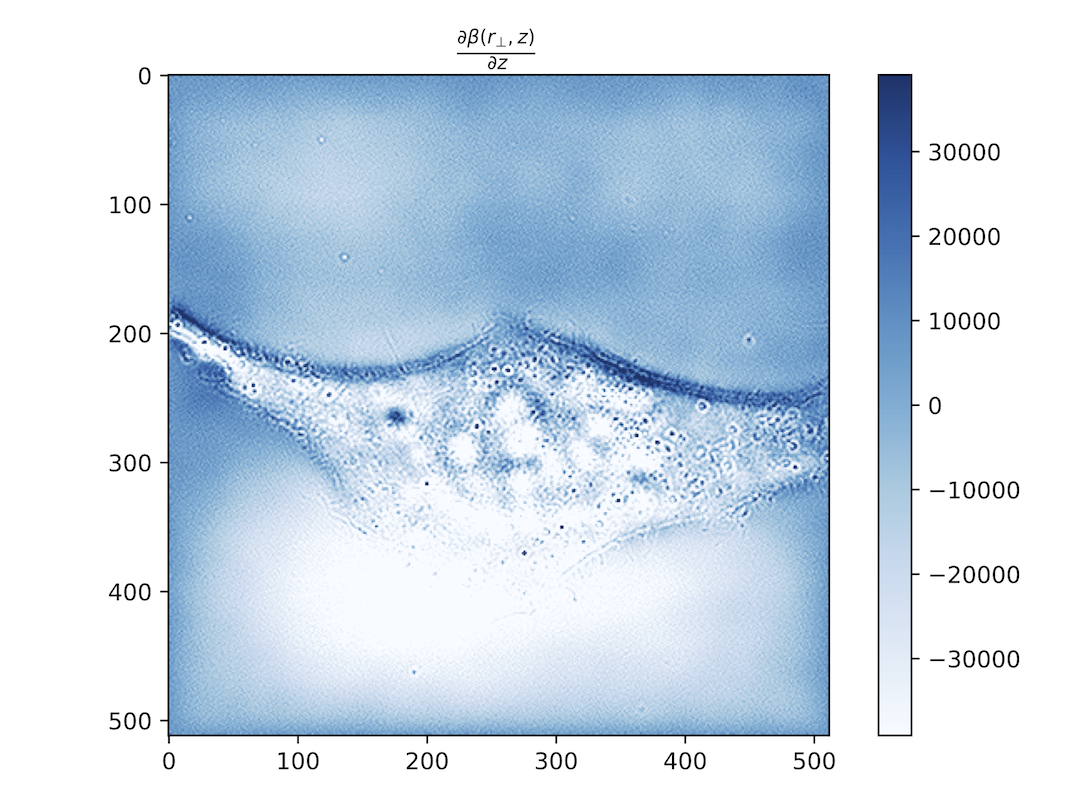}
\includegraphics[width=0.32\linewidth]{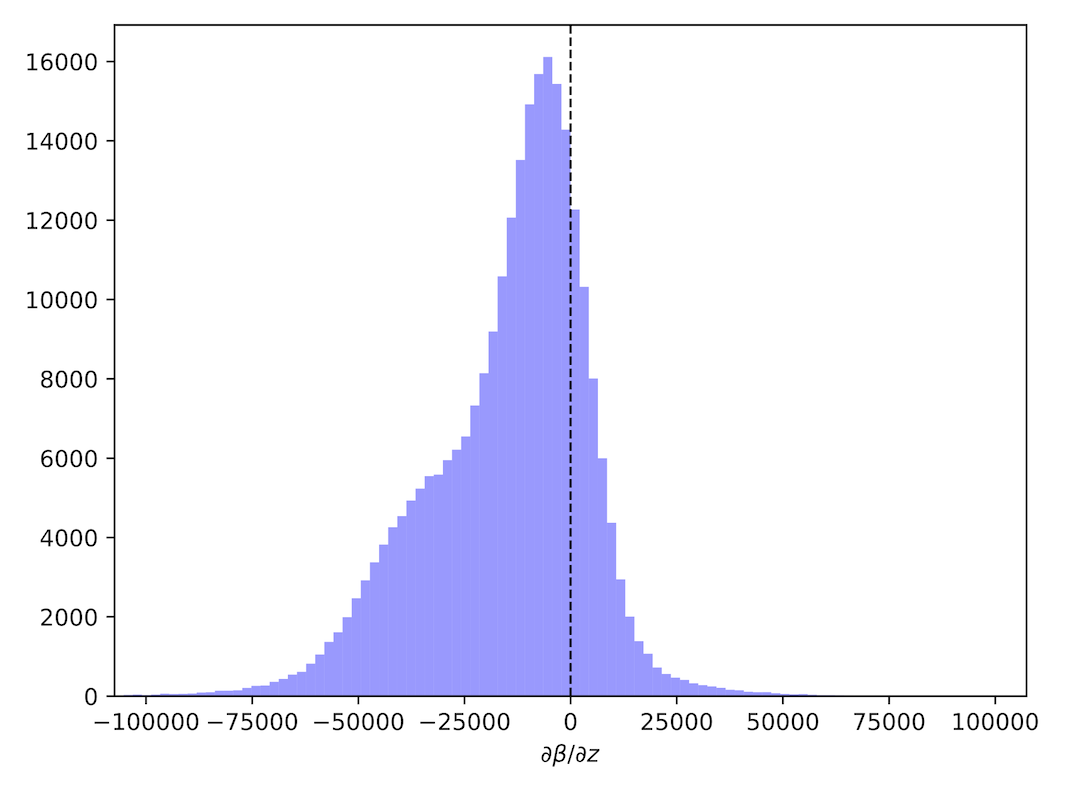}
\hspace{0.33\linewidth}

\caption{Results for HeLa cell 1: (a) Intensity distributions, $I(r_{\bot},z_{0}-\Delta z)$, $I(r_{\bot},z_{0})$, and $I(r_{\bot},z_{0}+\Delta z)$. (b) Phase distributions reconstructed from the three fluorescent cell images. (c) Reconstructed distributions of the intensity reduction parameter $\partial_z \alpha(r_{\bot}, z_{0})$ and (d) the phase-coupling parameter $\partial_z \beta(r_{\bot}, z_{0})$.}
\label{figD2;results02a}
\end{figure}

\newpage

\begin{figure}[!h]
\leftline{\bf (e) Reconstructed distributions of refractive-index fluctuations $\Delta n(r_{\bot}, z_0)$}
\centering
\includegraphics[width=0.32\linewidth]{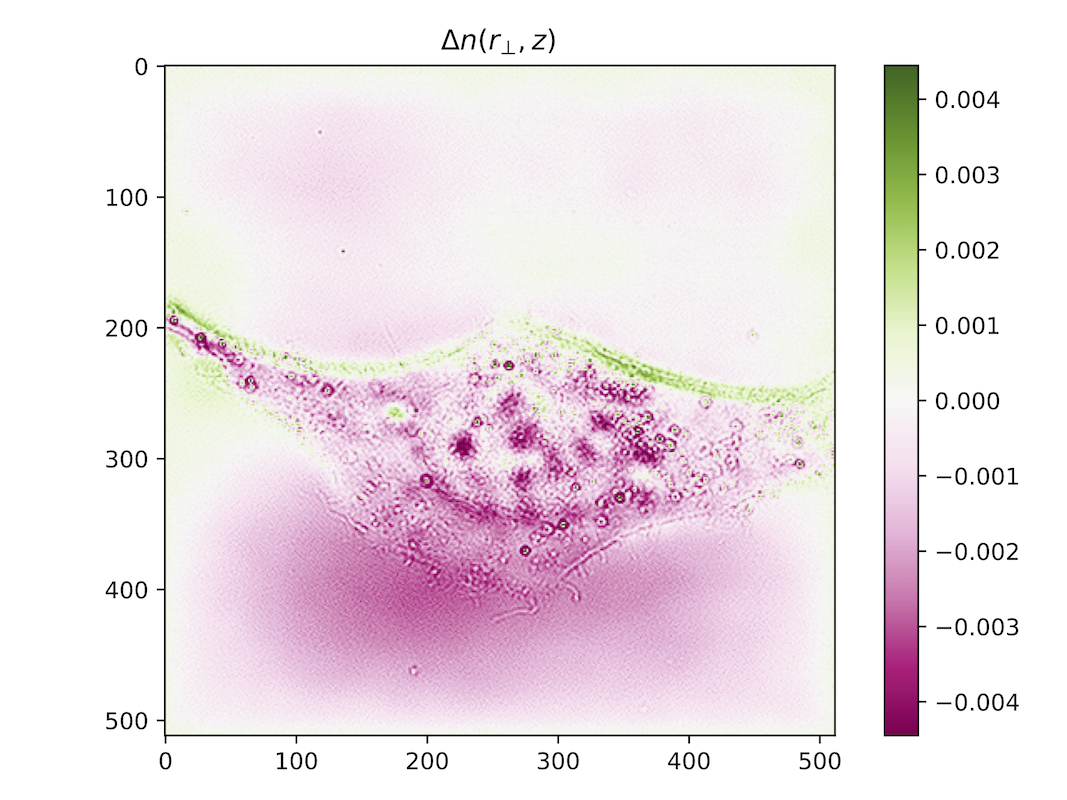}
\includegraphics[width=0.32\linewidth]{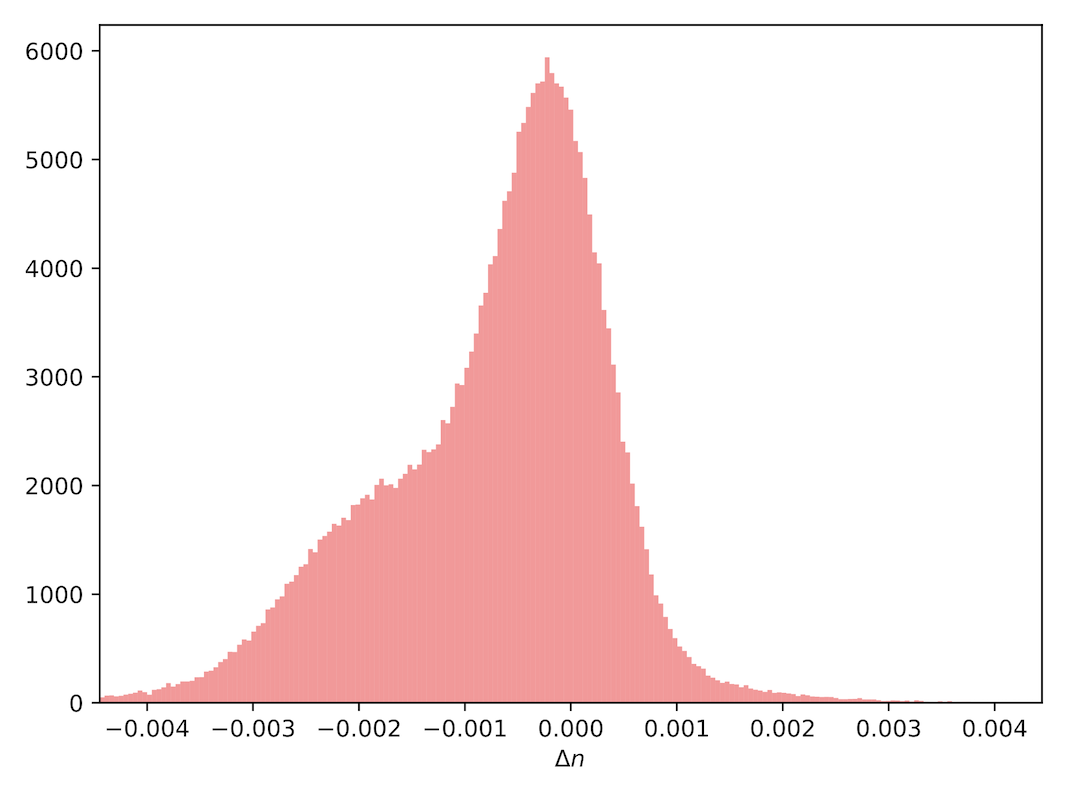}
\includegraphics[width=0.32\linewidth]{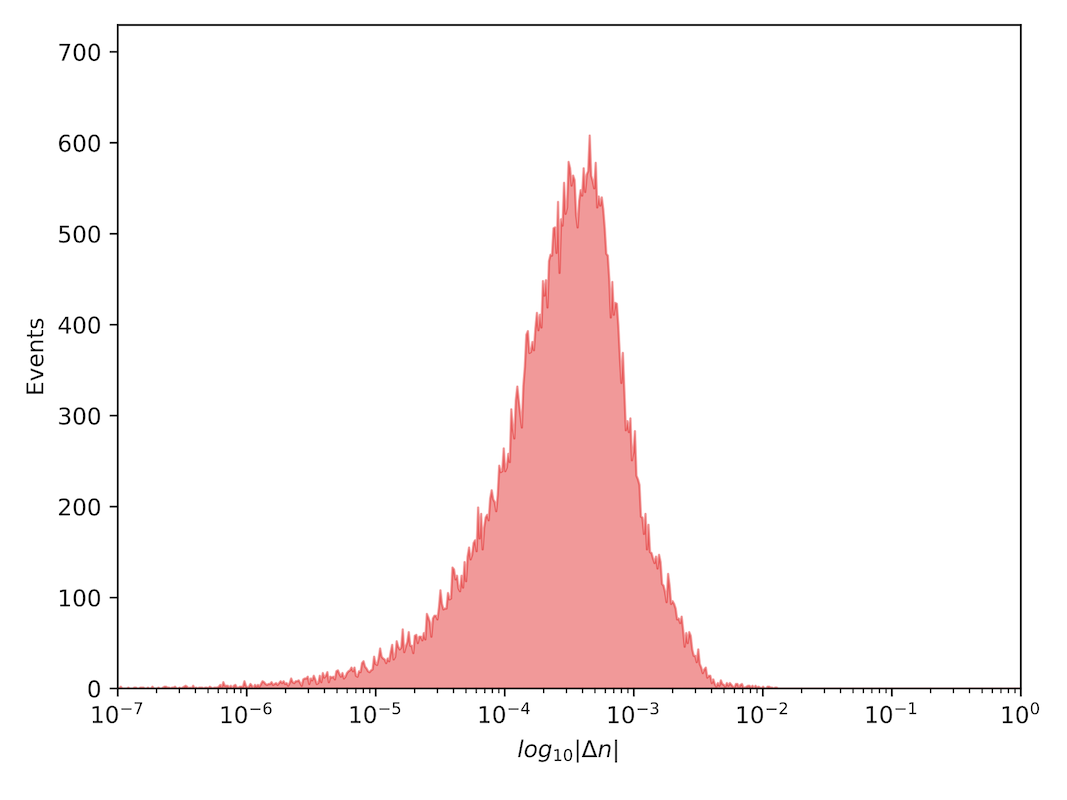}

\leftline{\bf (f) Reconstructed distributions of attenuation coefficients $\mu(r_{\bot}, z_0)$}
\centering
\includegraphics[width=0.32\linewidth]{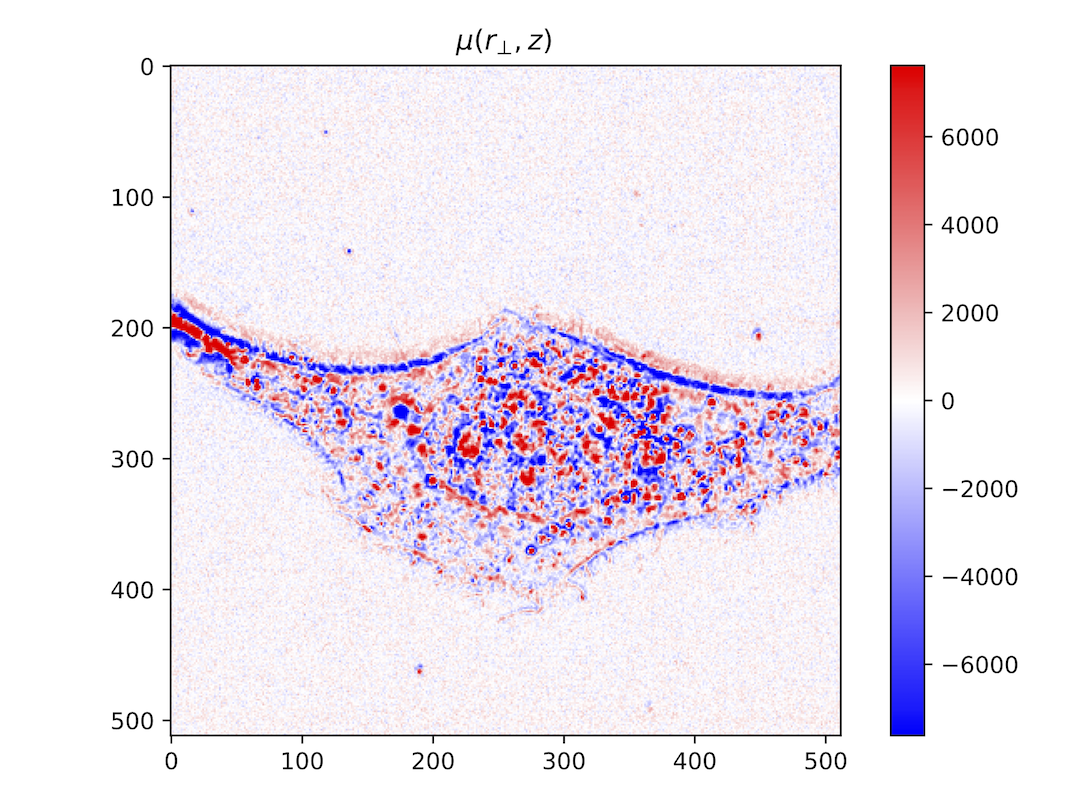}
\includegraphics[width=0.32\linewidth]{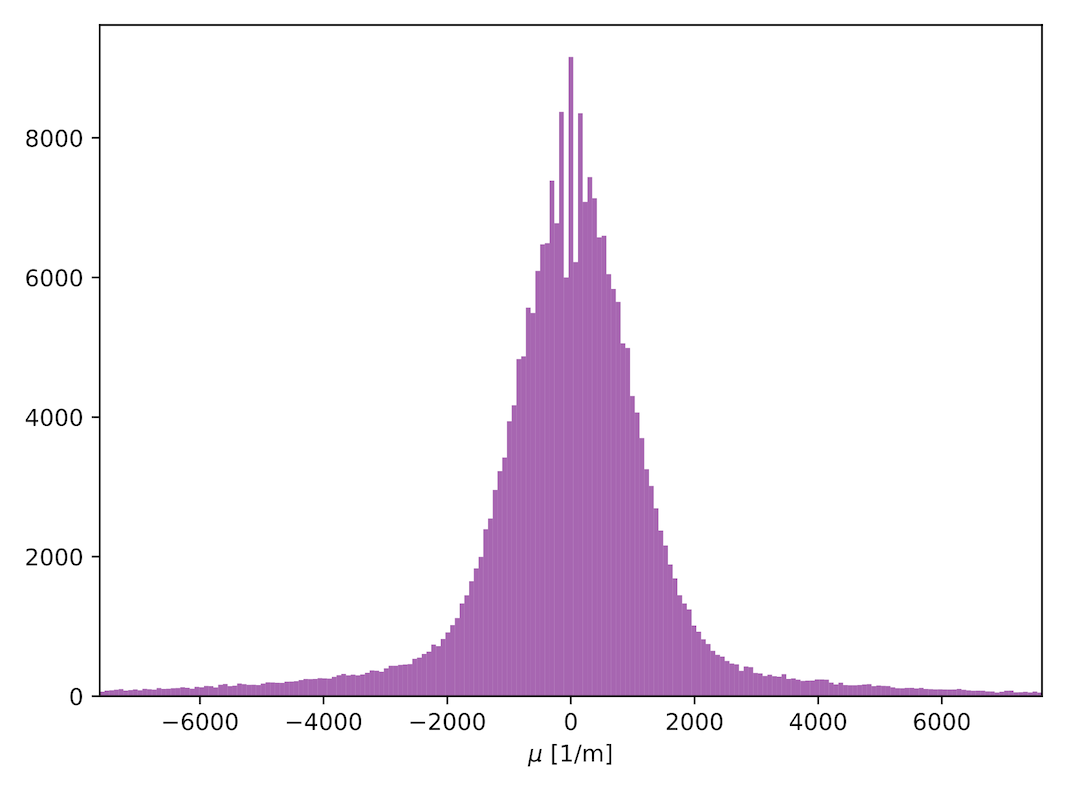}
\includegraphics[width=0.32\linewidth]{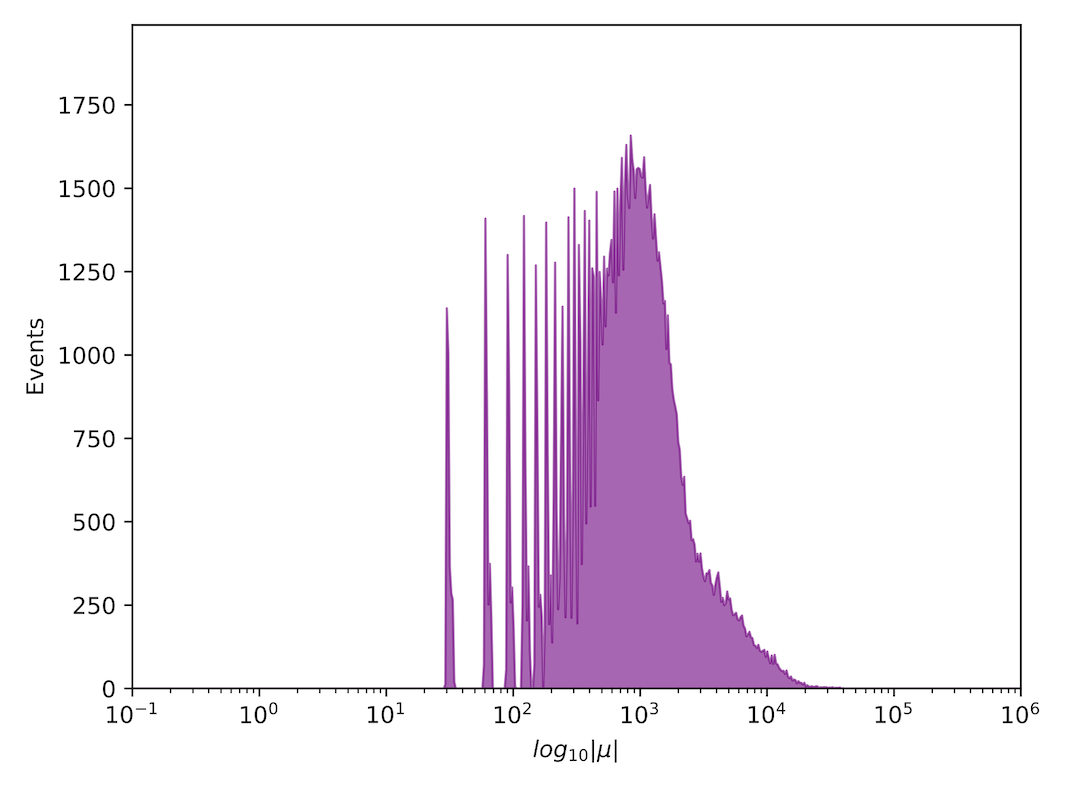}

\leftline{\bf (g) 2D histogram of $\Delta n(r_{\bot}, z_0)$ and $\mu(r_{\bot}, z_0)$}
\centering
\includegraphics[width=0.62\linewidth]{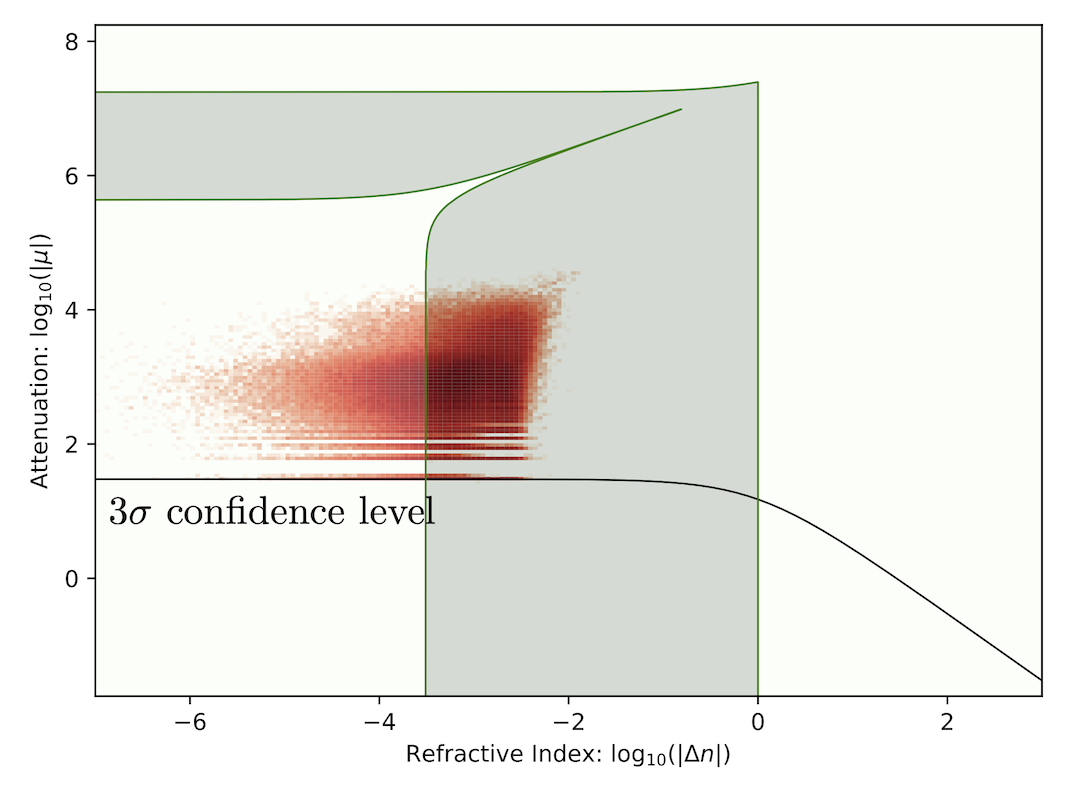}
\hspace{0.30\linewidth}

\caption{Results for HeLa cell 2: (e) Reconstructed distributions and histogram of the refractive-index fluctuations and (f) attenuation coefficients. (g) Correlation pattern between $\Delta n(r_{\bot},z_0)$ and $\mu(r_{\bot},z_0)$ lies with the $3\sigma$ confidence level (i.e., black solid line). $65,055$  events ($24.82\%$) fall outside the physical parameter boundaries (i.e., green area) given by the Eq.~(10) in the main text.}
\label{figD2;results02b}
\end{figure}

\newpage

\leftline{\bf - HeLa cell membrane}

\begin{figure}[!h]

\leftline{\bf (a) Intensity distributions at $z_0-\Delta z$, $z_0$ and $z_0+\Delta z$}
\centering
\includegraphics[width=0.32\linewidth]{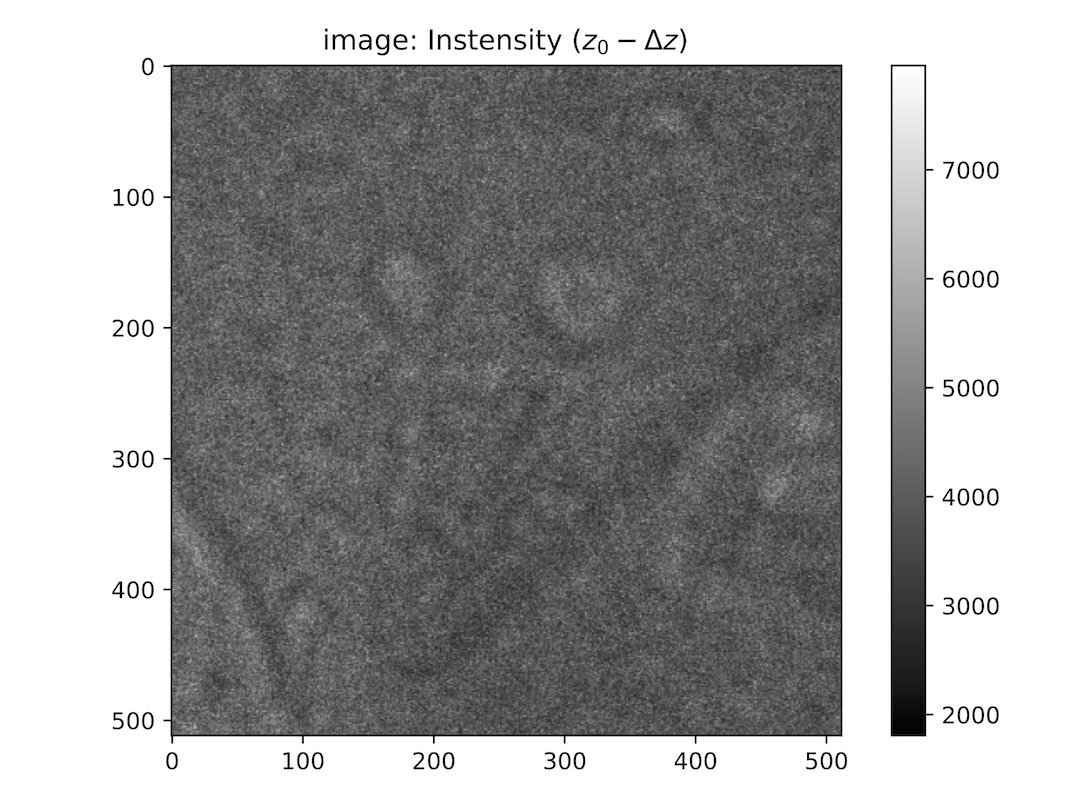}
\includegraphics[width=0.32\linewidth]{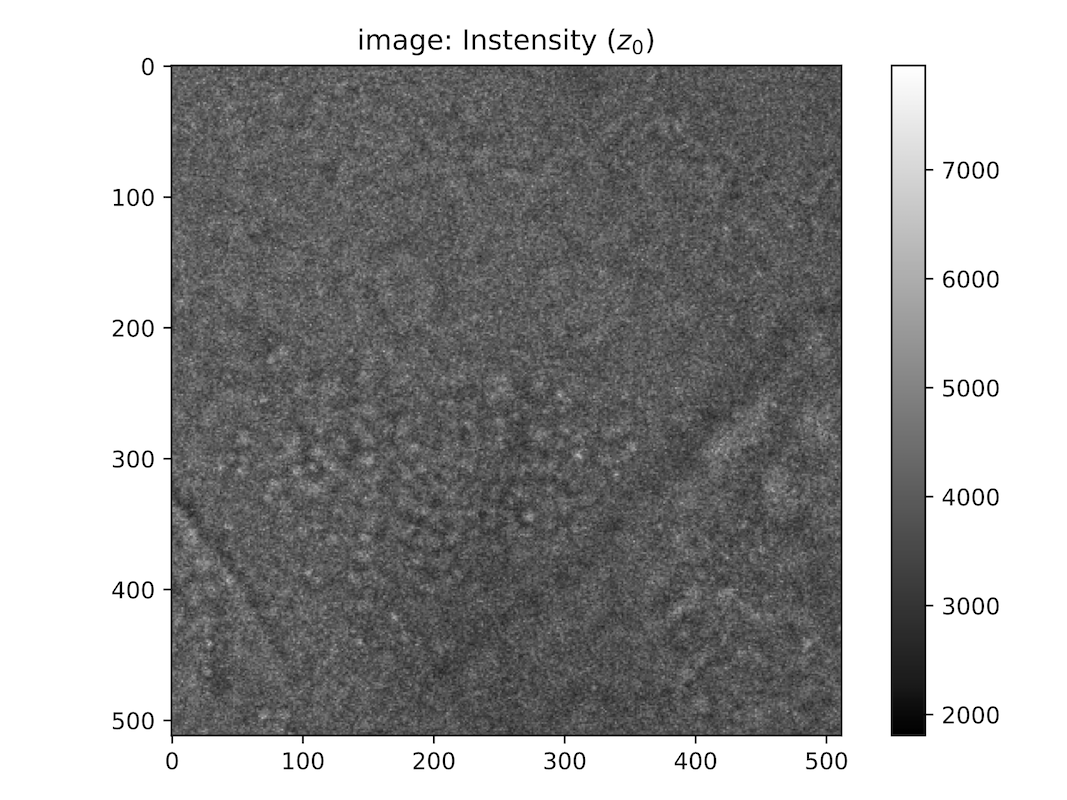}
\includegraphics[width=0.32\linewidth]{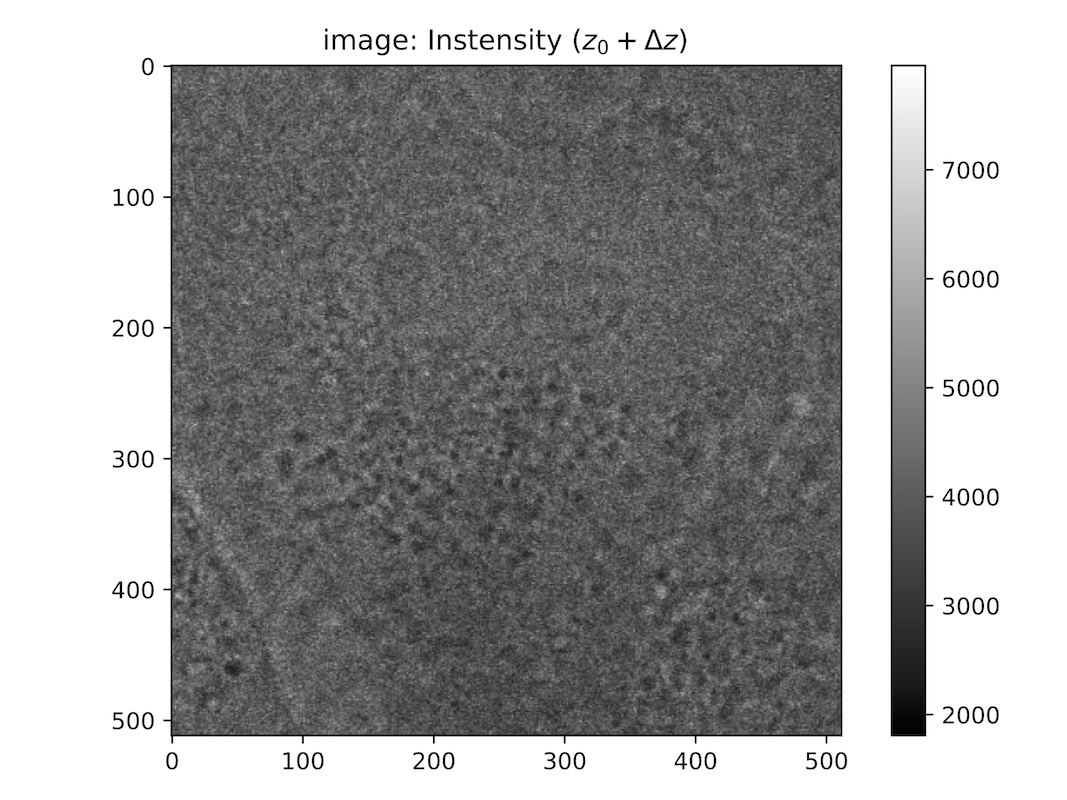}

\leftline{\bf (b) Phase distributions at $z_0-\Delta z/2$, $z_0$ and $z_0+\Delta z/2$}
\centering
\includegraphics[width=0.32\linewidth]{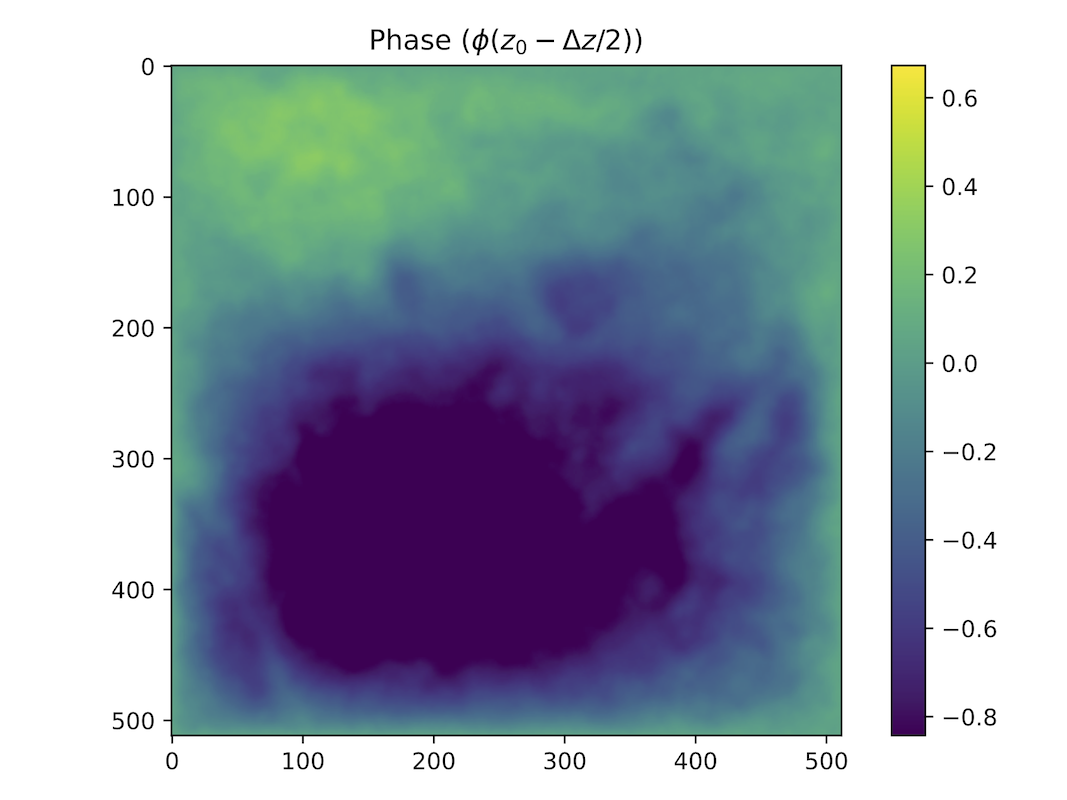}
\includegraphics[width=0.32\linewidth]{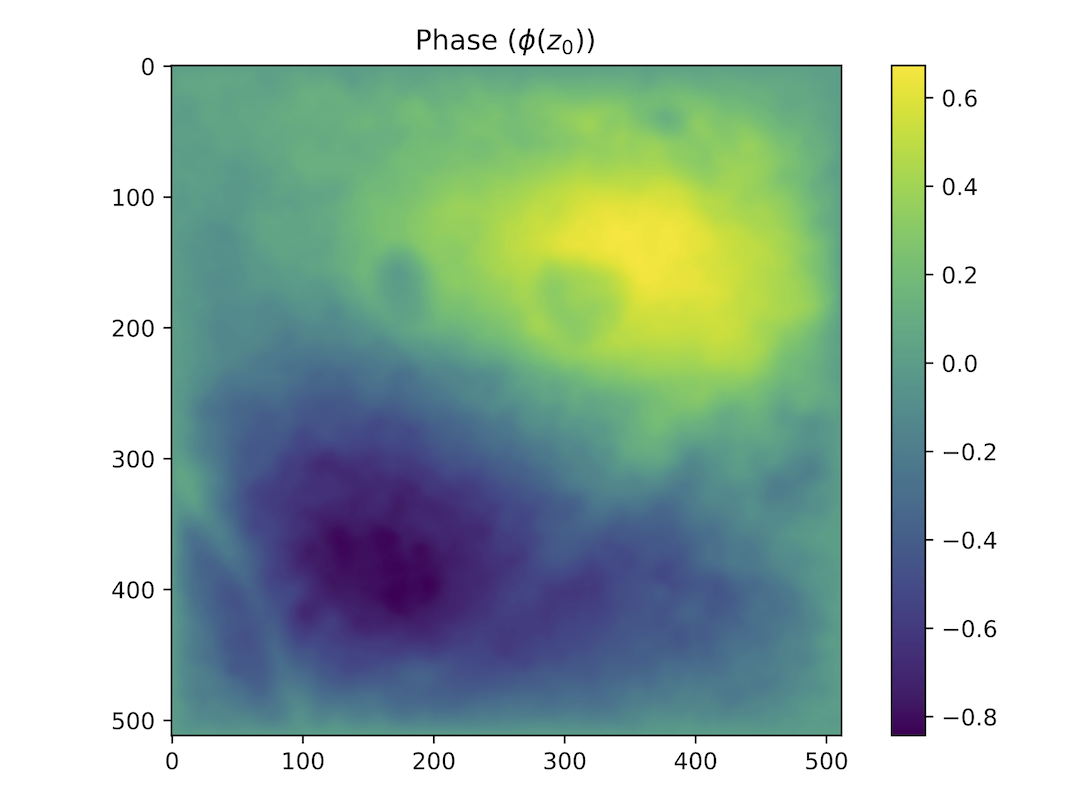}
\includegraphics[width=0.32\linewidth]{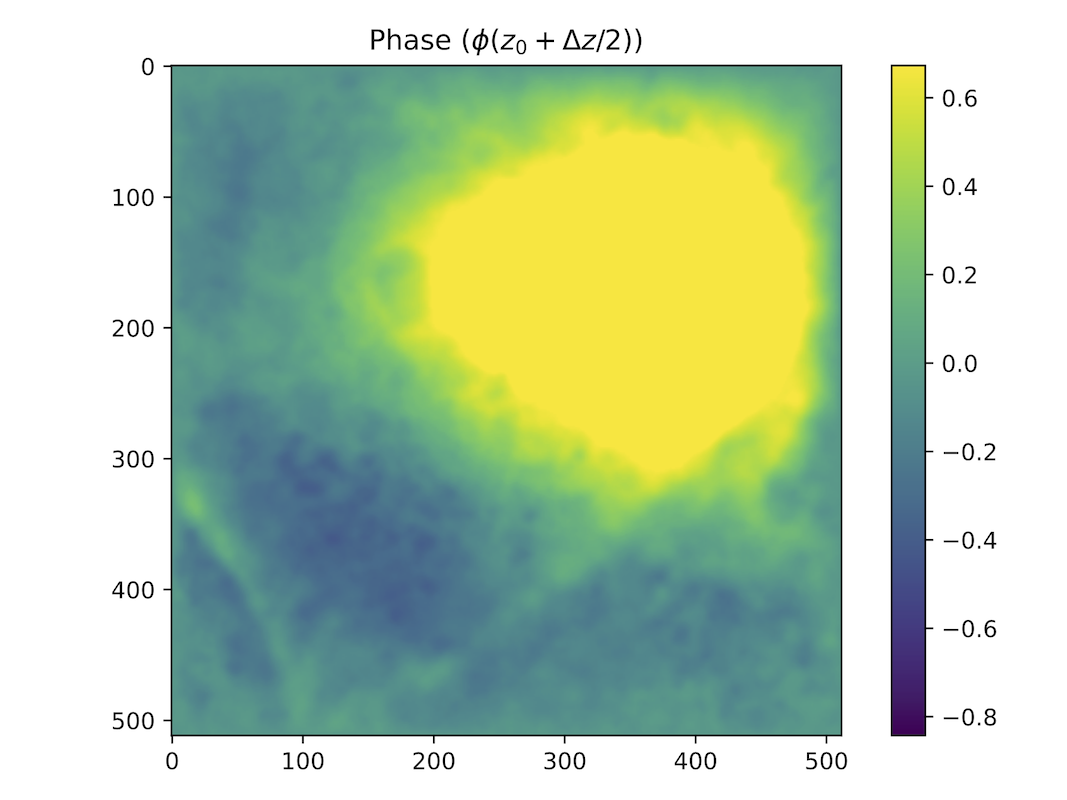}

\leftline{\bf (c) Reconstructed distributions of the $\alpha(r_{\bot}, z_0)$-derivative function}
\centering
\includegraphics[width=0.32\linewidth]{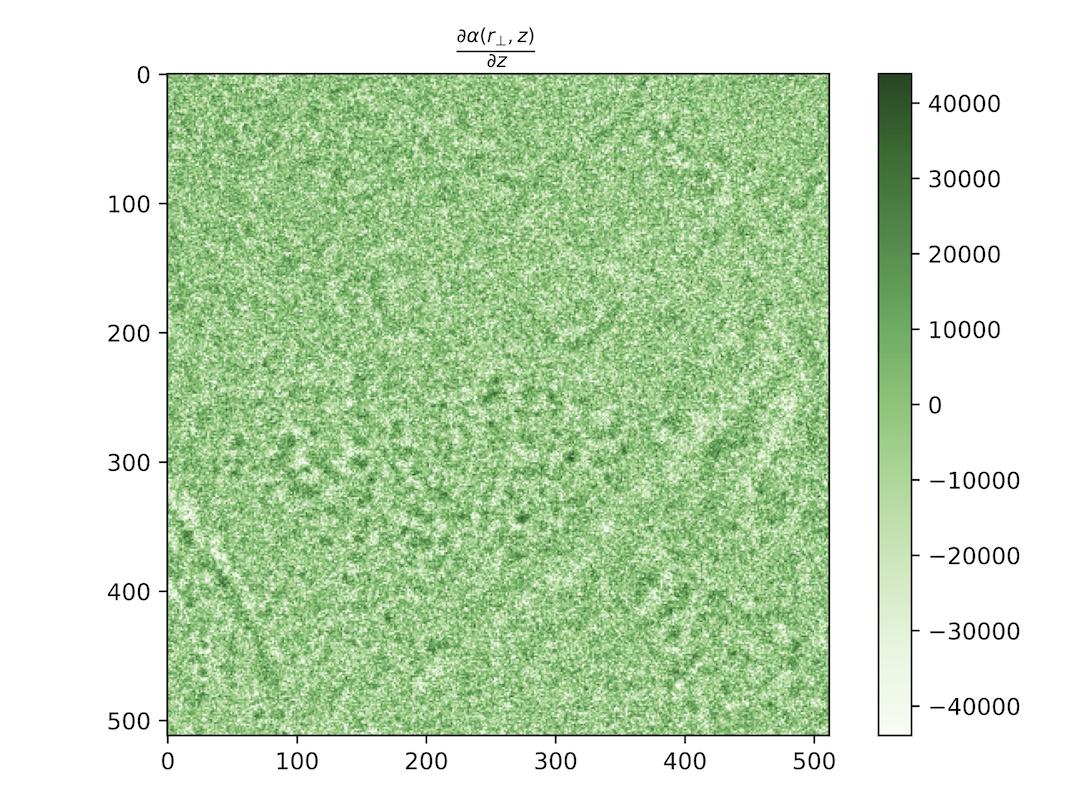}
\includegraphics[width=0.32\linewidth]{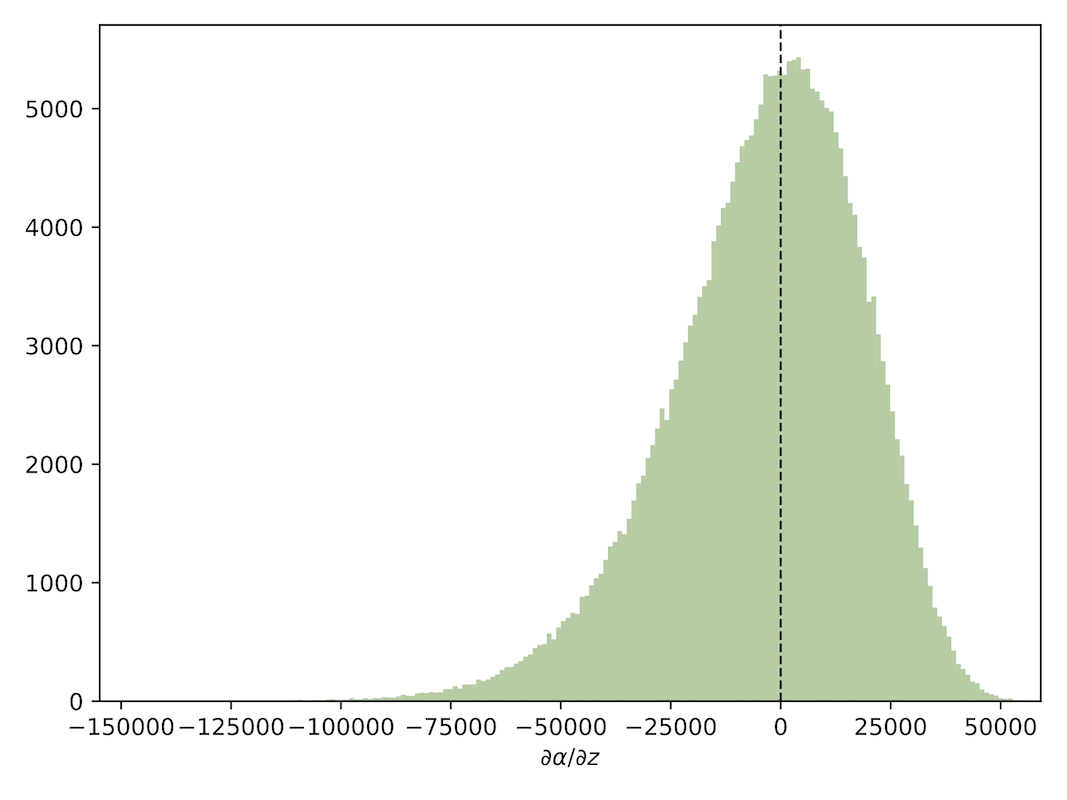}
\hspace{0.34\linewidth}

\leftline{\bf (d) Reconstructed distributions of the $\beta(r_{\bot}, z_0)$-derivative function}
\centering
\includegraphics[width=0.32\linewidth]{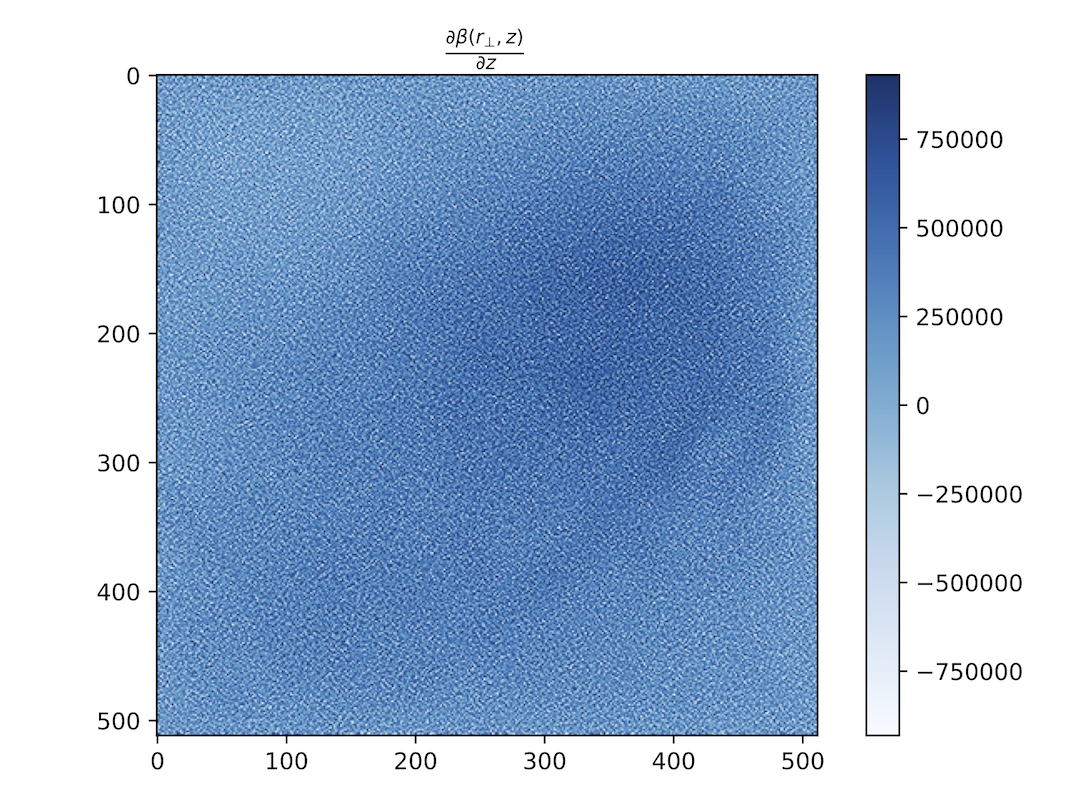}
\includegraphics[width=0.32\linewidth]{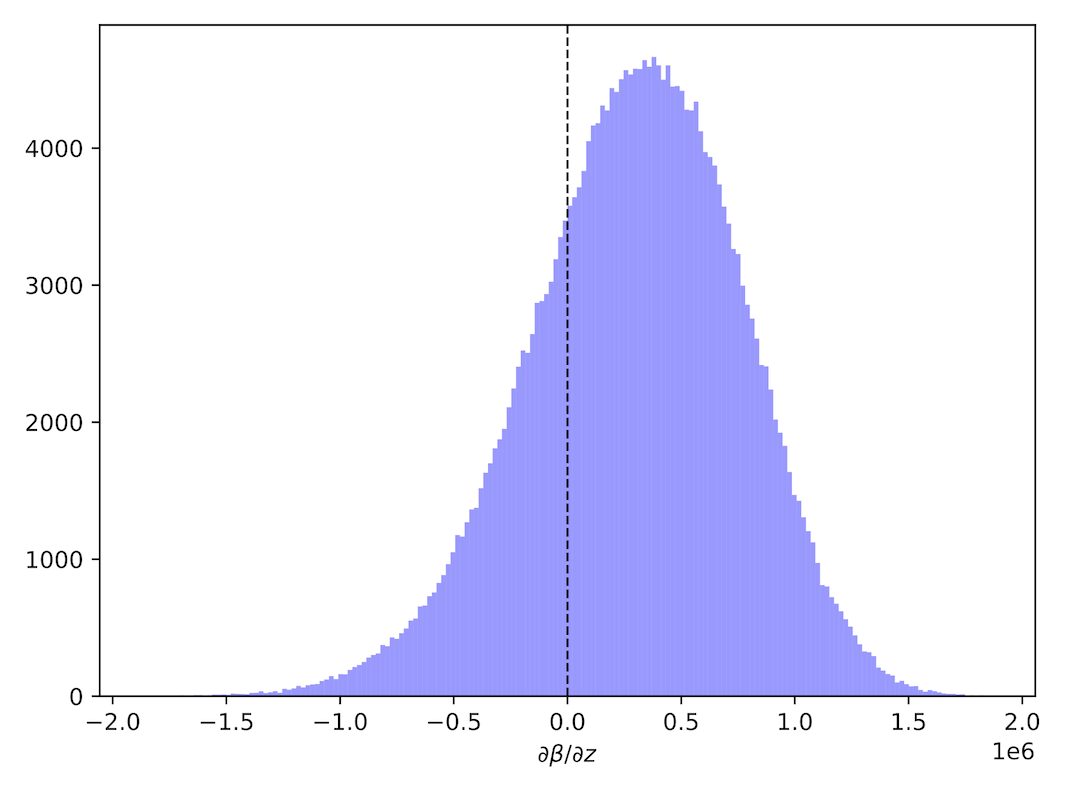}
\hspace{0.33\linewidth}

\caption{Results for HeLa cell membrane 1: (a) Intensity distributions, $I(r_{\bot},z_{0}-\Delta z)$, $I(r_{\bot},z_{0})$, and $I(r_{\bot},z_{0}+\Delta z)$. (b) Phase distributions reconstructed from the three fluorescent cell images. (c) Reconstructed distributions of the intensity reduction parameter $\partial_z \alpha(r_{\bot}, z_{0})$ and (d) the phase-coupling parameter $\partial_z \beta(r_{\bot}, z_{0})$.}
\label{figD2;results03a}
\end{figure}

\newpage

\begin{figure}[!h]
\leftline{\bf (e) Reconstructed distributions of refractive-index fluctuations $\Delta n(r_{\bot}, z_0)$}
\centering
\includegraphics[width=0.32\linewidth]{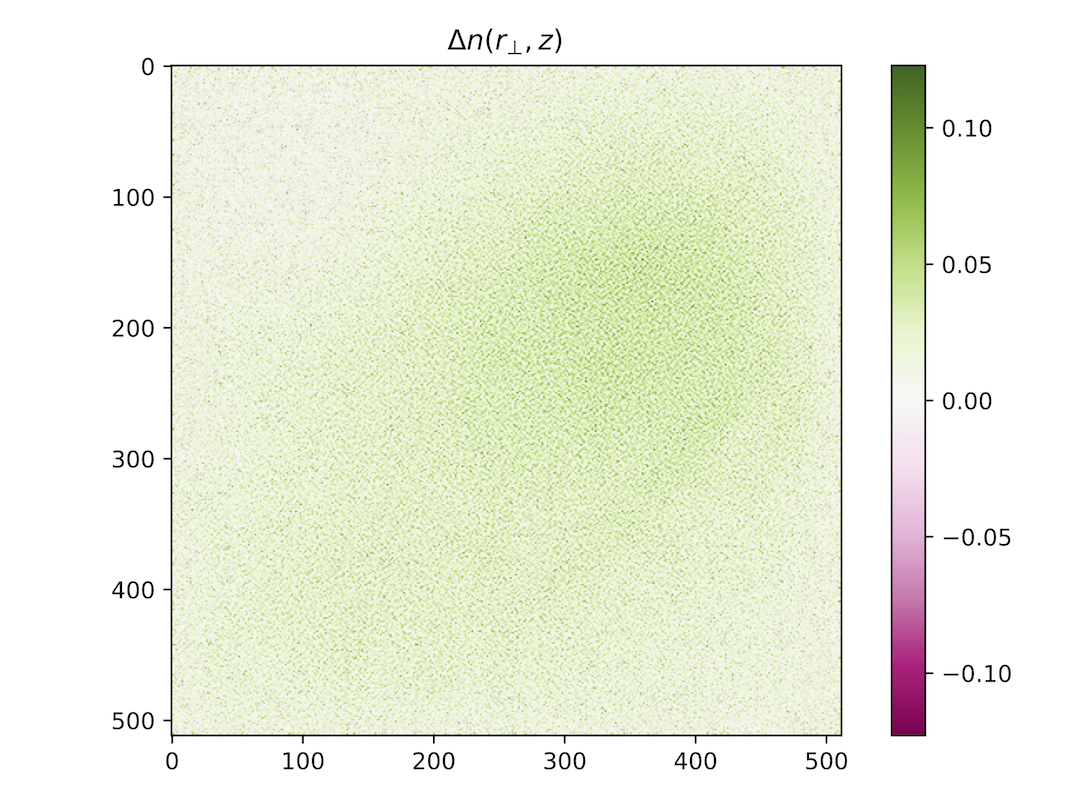}
\includegraphics[width=0.32\linewidth]{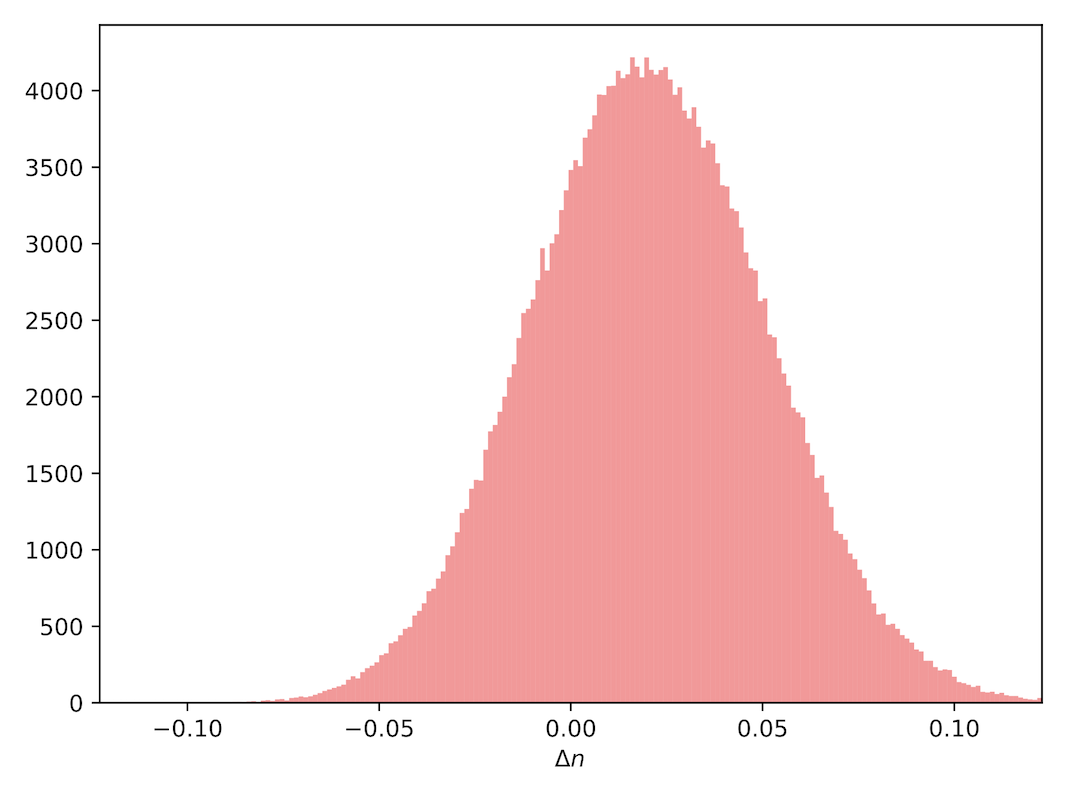}
\includegraphics[width=0.32\linewidth]{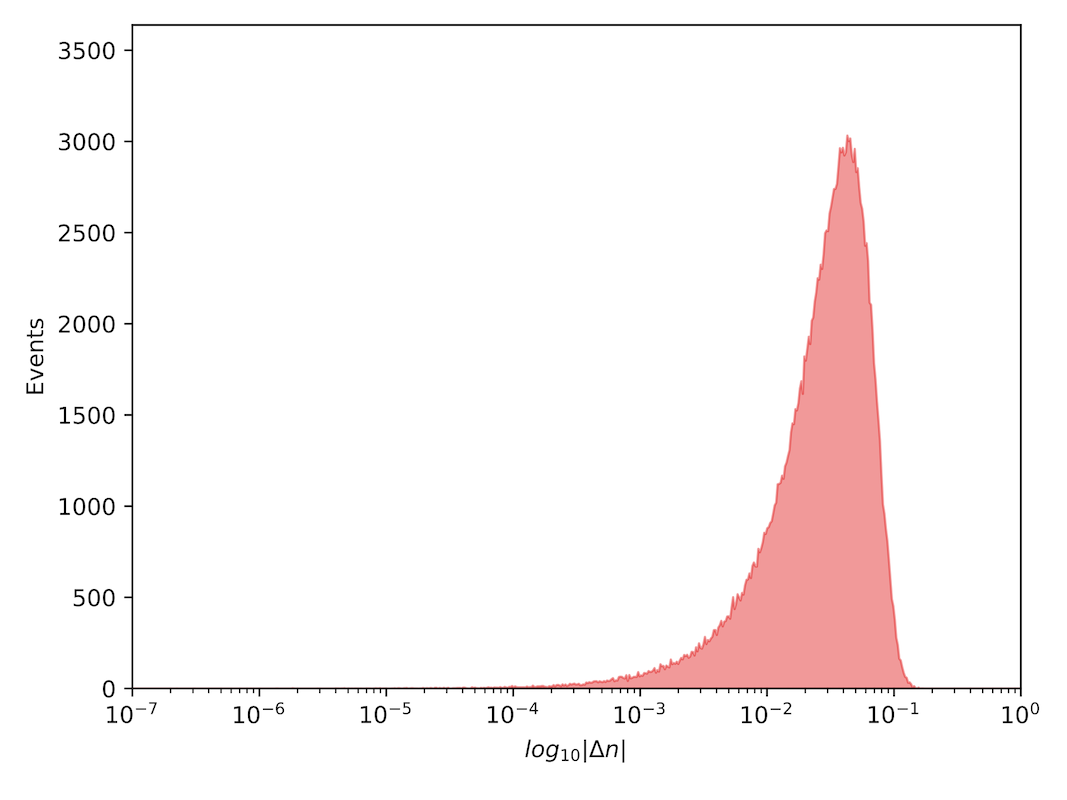}

\leftline{\bf (f) Reconstructed distributions of attenuation coefficients $\mu(r_{\bot}, z_0)$}
\centering
\includegraphics[width=0.32\linewidth]{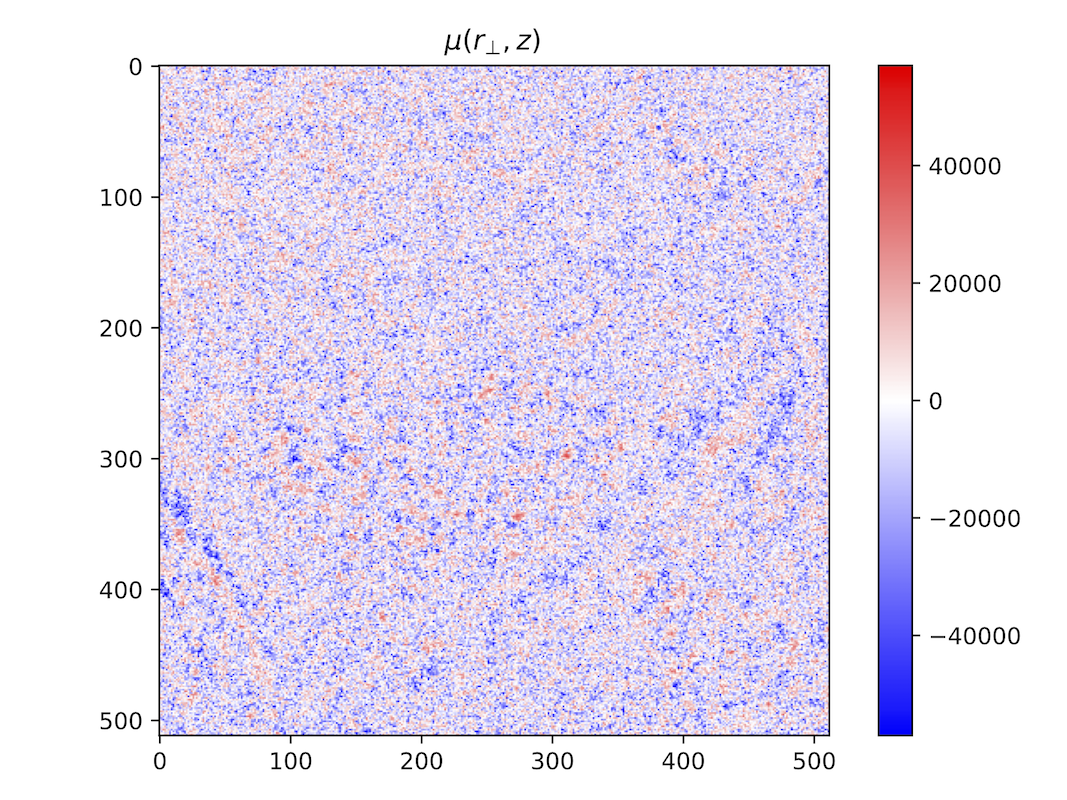}
\includegraphics[width=0.32\linewidth]{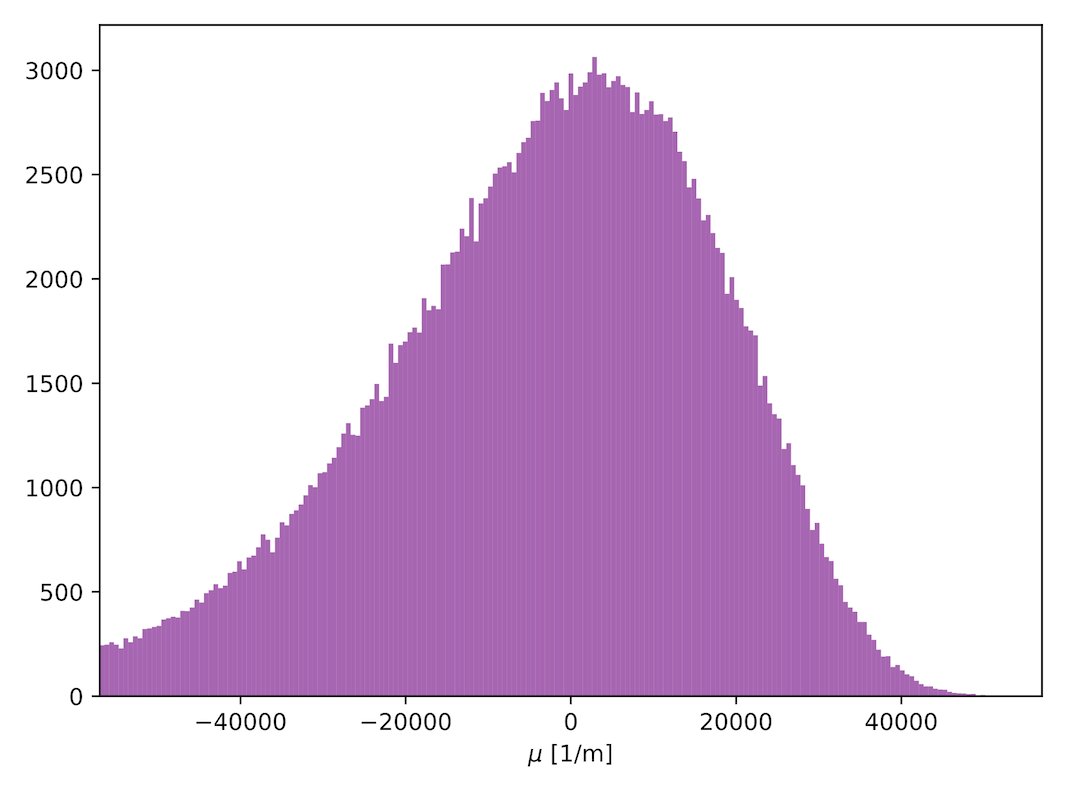}
\includegraphics[width=0.32\linewidth]{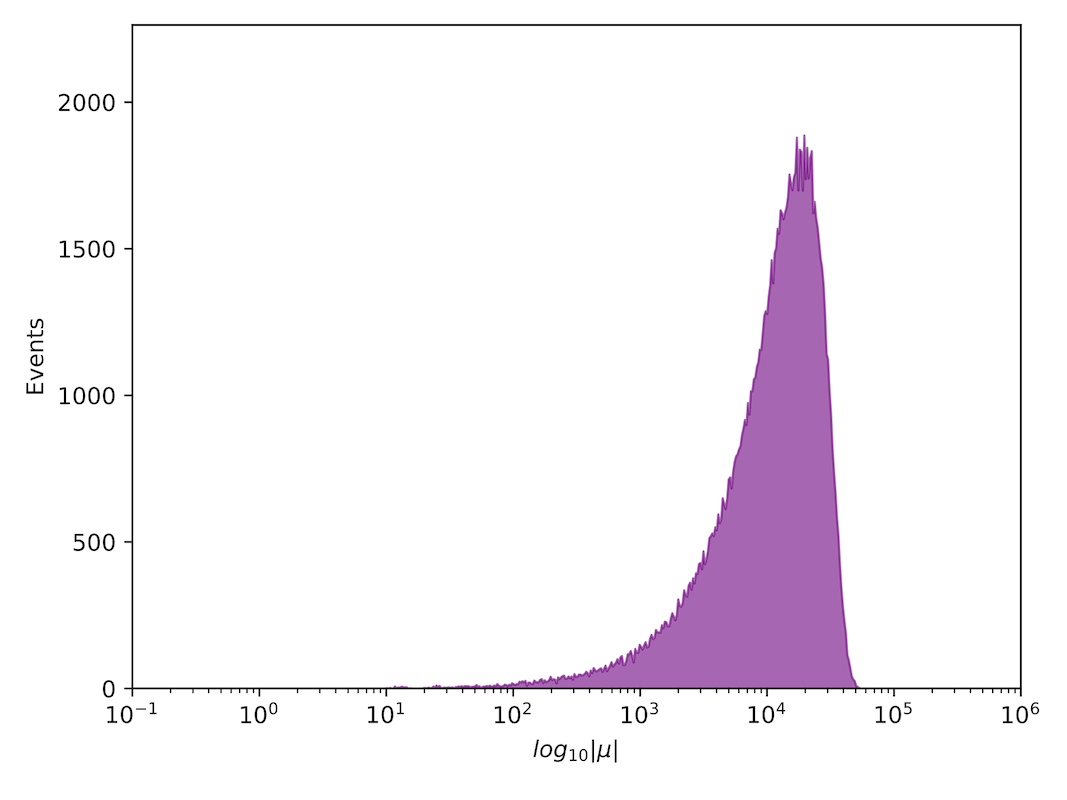}

\leftline{\bf (g) 2D histogram of $\Delta n(r_{\bot}, z_0)$ and $\mu(r_{\bot}, z_0)$}
\centering
\includegraphics[width=0.62\linewidth]{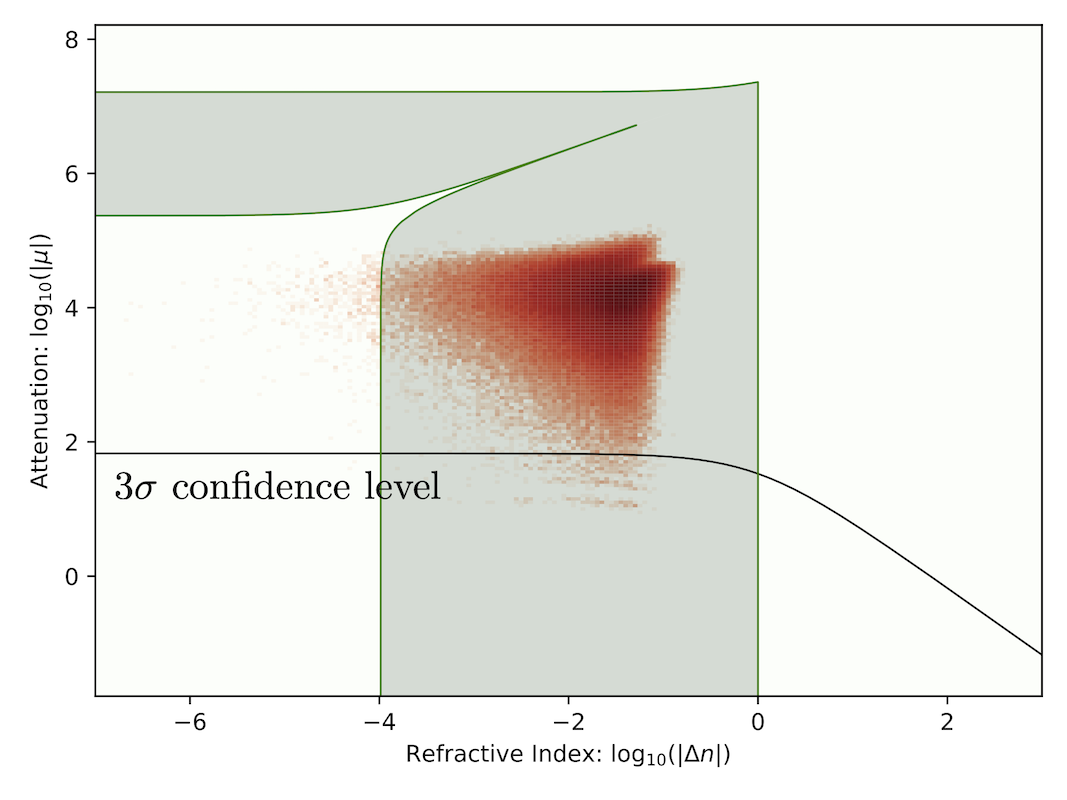}
\hspace{0.30\linewidth}

\caption{Results for HeLa cell membrane 2: (e) Reconstructed distributions and histogram of the refractive-index fluctuations and (f) attenuation coefficients. (g) Correlation pattern between $\Delta n(r_{\bot},z_0)$ and $\mu(r_{\bot},z_0)$ lies with the $3\sigma$ confidence level (i.e., black solid line). $555$ events ($0.217\%$) fall outside the physical parameter boundaries (i.e., green area) given by the Eq.~(10) in the main text.}
\label{figD2;results03b}
\end{figure}

\newpage

\subsection{D.3 Results of symmetry analysis}

\begin{figure}[!h]
\leftline{\bf (a) Propagation probability}
\centering
\includegraphics[width=0.32\linewidth]{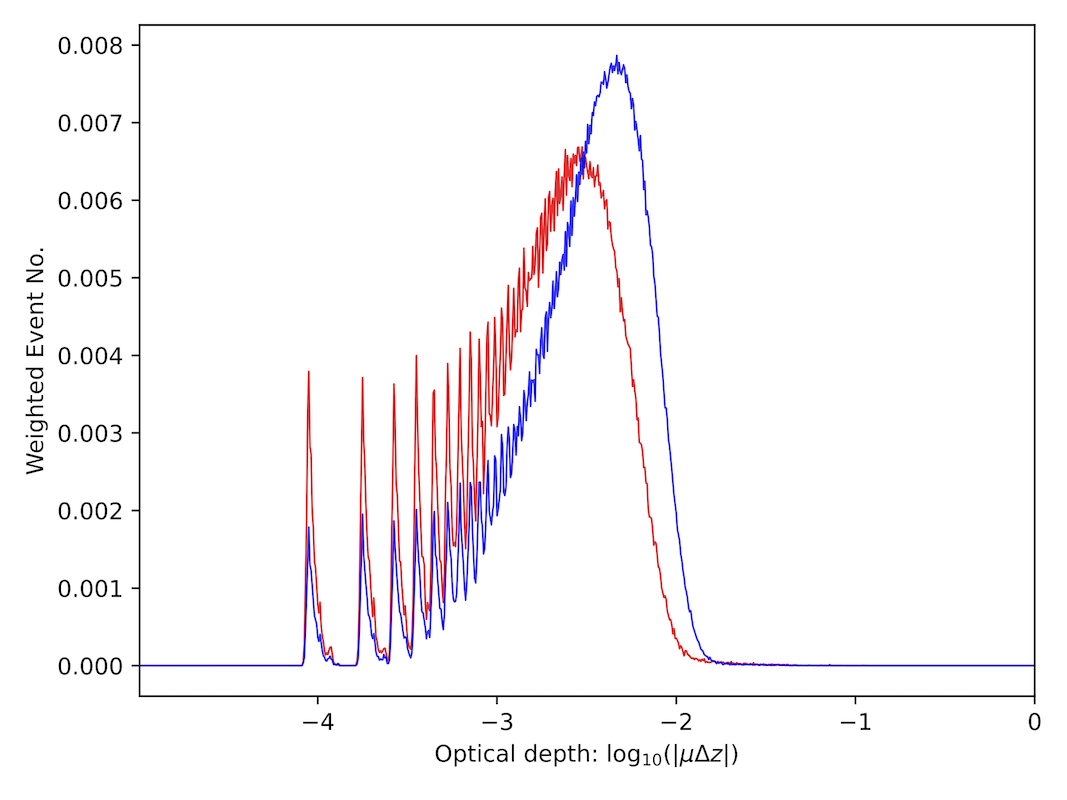}
\includegraphics[width=0.32\linewidth]{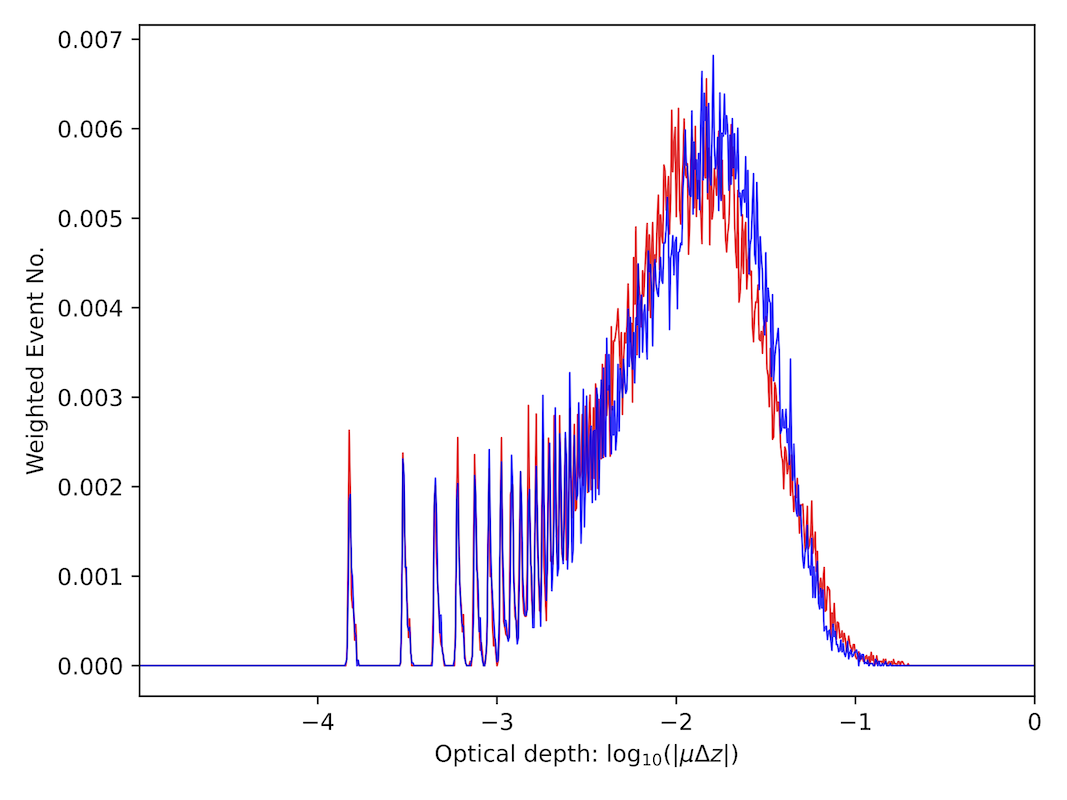}
\includegraphics[width=0.32\linewidth]{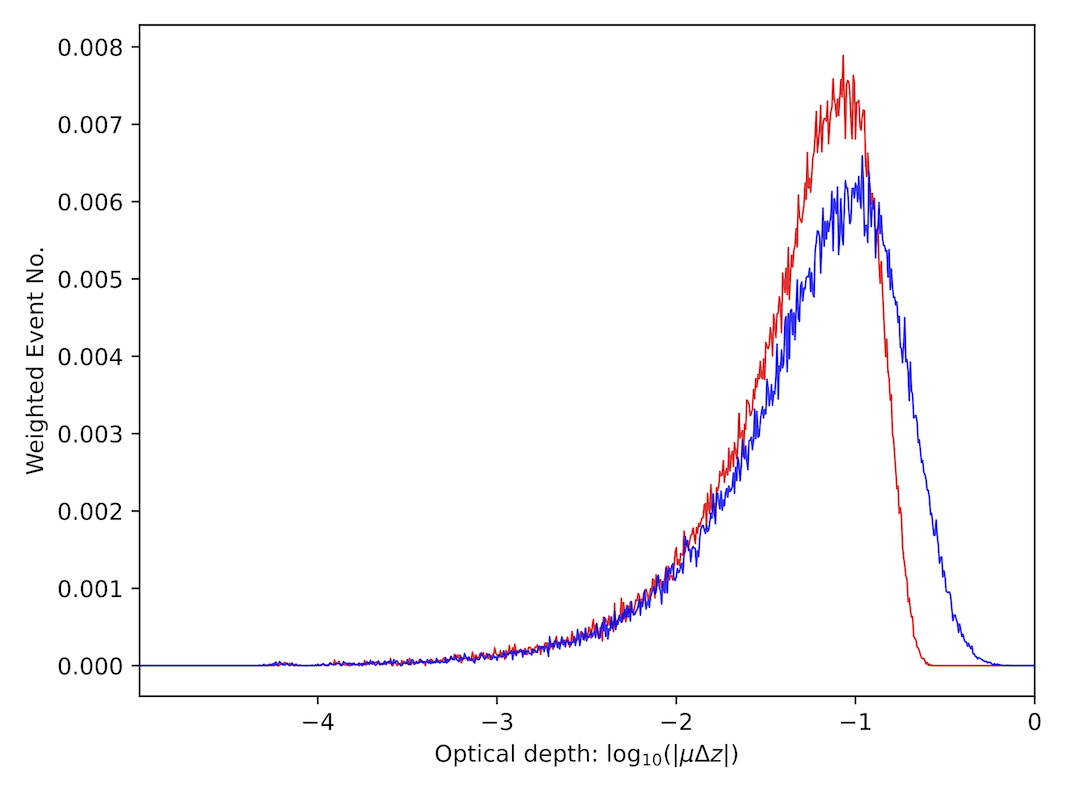}

\leftline{\bf (b) Asymmetry property}
\centering
\includegraphics[width=0.32\linewidth]{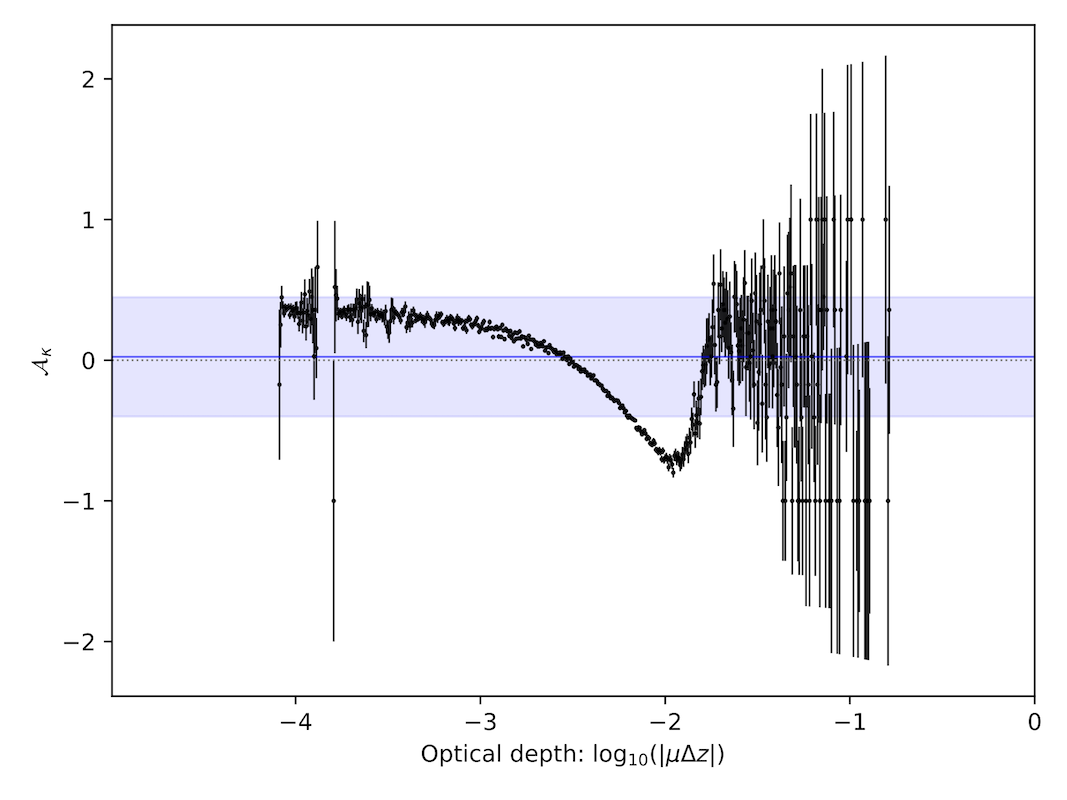}
\includegraphics[width=0.32\linewidth]{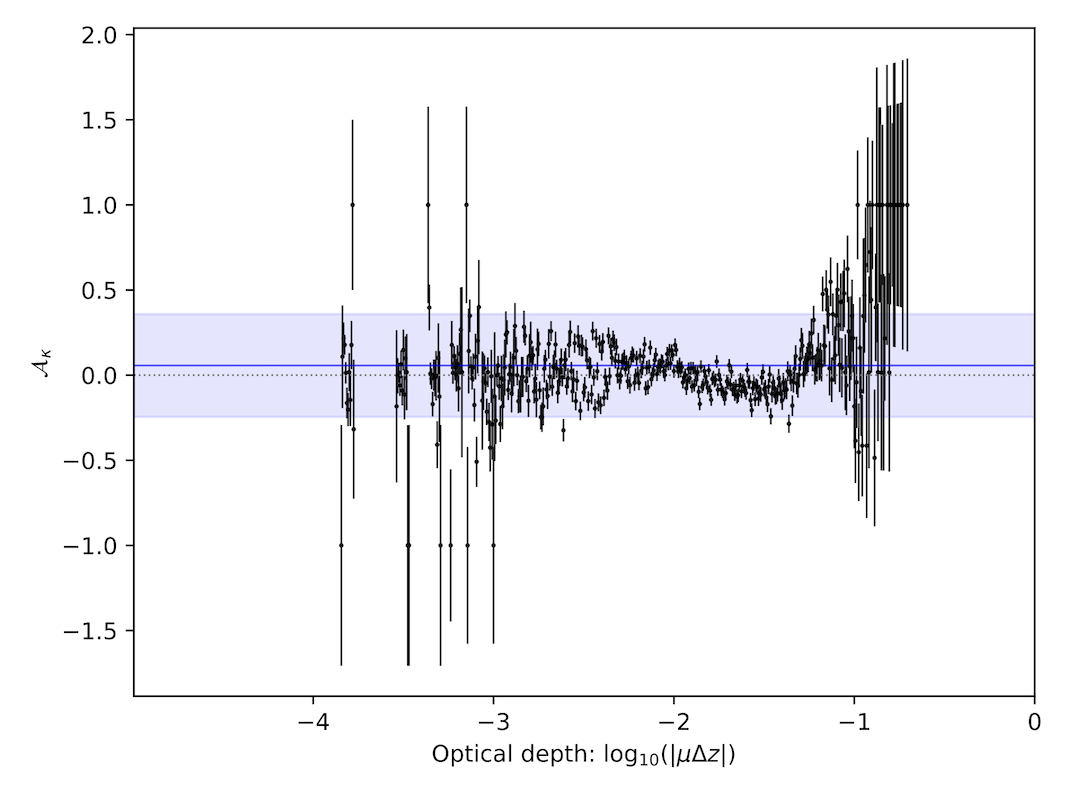}
\includegraphics[width=0.32\linewidth]{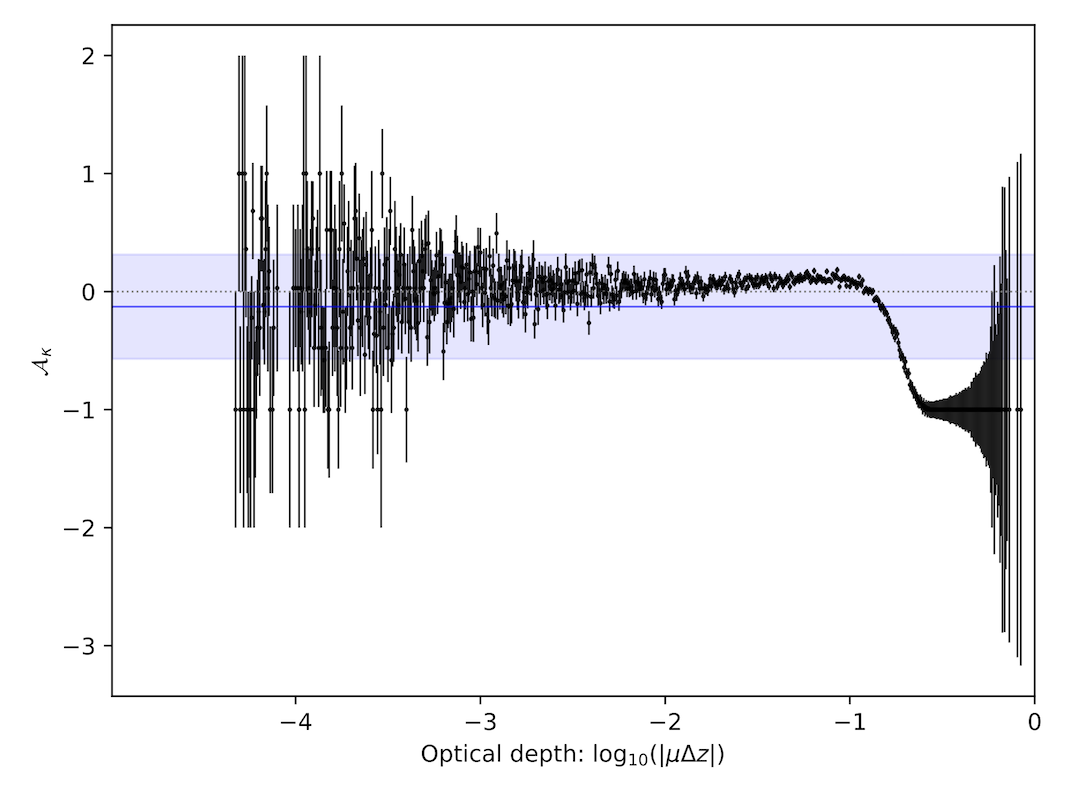}

\leftline{\bf \hspace{0.15\linewidth} MLA \hspace{0.24\linewidth} HeLa cells \hspace{0.18\linewidth} HeLa cell membrane.}

\caption{Results of symmetry analysis. (a) Histograms of propagation probabilities as a function of optical depth. Red and blue lines in the histograms denote the propagation probabilities in forward and backward directions, respectively. (b) Asymmetry property as a function of optical depth. Blue line and bands in the asymmetry property plots are the statistical average and the root mean squared (RMS) value, respectively. Dashed line represents $\mathcal{A}_{\kappa} = 0$.}
\label{figD3;results}
\end{figure}


\newpage

\subsection{D.4 Systematic uncertainties from finite numerical aperture}
High numerical aperture (NA) objectives are essential for achieving high lateral resolution in microscopy, particularly when imaging subcellular structures. However, the use of high-NA optics introduces systematic deviations from the paraxial approximation that underlies the TIE theory. While the transverse ($x$-$y$) plane benefits from improved resolution through larger collection angles (following the Abbe limit $\lambda/2\text{NA}$), the axial ($z$-direction) propagation experiences significant departures from paraxial wave theory.

The fundamental challenge arises because TIE formulation assumes paraxial wave propagation with small angles relative to the optical axis. High-NA objectives, however, collect rays at angles up to $\theta_{\text{max}} = \arcsin(\text{NA}/m)$, which can exceed 60$^{\circ}$ for NA $> 1.2$ in aqueous media. $m$ denotes the refractive index of the objective lens immersion medium. For instance, with NA = 1.20 in water ($m = 1.33$), the maximum collection angle reaches approximately 64.5$^{\circ}$, where $\sin\theta \approx 0.90$ significantly deviates from the paraxial approximation $\sin\theta \approx \theta$.

\subsubsection{D.4.1 Mathematical framework for NA-effect quantification}
To quantify the systematic errors introduced by finite NA, we develop an effective model that accounts for the angular distribution of rays within the collection cone, and derive how this modification propagates through the TIE reconstruction.

\begin{itemize}
\item[--] {\bf Effective model}: In high-NA systems, light rays propagate at various angles relative to the optical axis, up to a maximum angle $\theta_{\text{max}} = \arcsin(\text{NA}/m)$. For a ray at angle $\theta$, the $z$-component of the wave vector is $k_z = kn_0\cos\theta$, where $k = 2\pi/\lambda$ is the vacuum wave number and $n_0$ is the medium refractive index. To account for the contribution of all rays within the NA cone, we need to calculate the average $z$-component of the wave vector.

Assuming uniform intensity distribution within the collection cone (a first-order approximation), the effective wave vector along $z$ becomes $\langle k_z \rangle = kn_0\langle\cos\theta\rangle$. In the paraxial approximation, $\cos\theta \approx 1 - \theta^2/2$ for small angles. Averaging over the NA cone, we obtain $\langle\cos\theta\rangle \approx 1 - \langle\theta^2\rangle/2$. For uniform distribution up to $\theta_{\text{max}}$, this gives $\langle\cos\theta\rangle \approx 1 - \theta_{\text{max}}^2/4$.

Using the relation $\sin\theta_{\text{max}} = \text{NA}/m$, and for moderate NA values where $\theta_{\text{max}} \approx \sin\theta_{\text{max}}$, we find $\theta_{\text{max}}^2 \approx \text{NA}^2/m^2$. This leads to the effective refractive index:
\begin{equation}
n_{\text{eff}} = n_0\langle\cos\theta\rangle = n_0\left(1 - \frac{\text{NA}^2}{4m^2}\right) = n_0(1 - \epsilon)
\end{equation}
where we define the dimensionless parameter $\epsilon = \text{NA}^2/4m^2$. This parameter directly quantifies the deviation from the paraxial regime and determines the magnitude of systematic biases in the reconstructed optical parameters.

\item[--] {\bf Modified TIE}: Under finite NA conditions, the standard TIE is modified by replacing $n_0$ with $n_{\text{eff}}$ in the phase gradient term:
\begin{equation}
\frac{\partial I}{\partial z} = -\frac{1}{kn_{\text{eff}}}\nabla_\perp \cdot (I\nabla_\perp\phi) - 2\frac{\partial\alpha}{\partial z}I
\end{equation}
where $k n_{\text{eff}} = k n_0(1 - \epsilon)$. This modification leads to systematic biases in the reconstructed optical parameters, as detailed in the following error propagation analysis.
\end{itemize}

\subsubsection{D.4.2 Error propagation analysis}
We now analyze how the effective refractive index modification propagates through the TIE reconstruction process to produce systematic errors in each optical parameter. Starting from the modified TIE with $n_{\text{eff}}$, we derive the relative errors for the phase ($\phi$), intensity derivatives ($\partial\alpha/\partial z$, $\partial\beta/\partial z$), and the final reconstructed quantities: refractive index fluctuations ($\Delta n$), attenuation coefficient ($\mu$), and attenuation asymmetry ($\mathcal{A}_\kappa$).

\begin{itemize}
\item[--]  {\bf Phase recovery ($\phi$)}: The phase recovered from TIE inversion exhibits a systematic bias. When solving the modified TIE with $n_{\text{eff}}$ instead of $n_0$, the reconstructed phase gradient is amplified by the factor $1/(1-\epsilon)$ due to the reduced effective refractive index in the denominator of the phase term. This leads to:
\begin{equation}
\phi_{\text{measured}} = \frac{\phi_{\text{true}}}{1-\epsilon}
\end{equation}
Relative error: $\delta_\phi = \epsilon/(1-\epsilon)$

\item[--] {\bf Intensity attenuation derivative ($\partial\alpha/\partial z$)}: The intensity reduction parameter, extracted from symmetric intensity differences according to Eq.~(7) in the main text, is independent of the effective refractive index to first order. This is because the phase contributions cancel in the symmetric combination ($I_+ + I_-$), leaving only the intensity attenuation effect. The NA effect enters only through higher-order corrections related to the spatial extent of the intensity distribution:
\begin{equation}
\delta_{\partial\alpha/\partial z} \approx \mathcal{O}(\epsilon^2) \approx 0
\end{equation}

\item[--] {\bf Phase-coupling derivative ($\partial\beta/\partial z$)}: The phase-coupling parameter is obtained from the Transport of Intensity Equation through antisymmetric intensity differences. Since it is directly proportional to the phase gradient, and the phase gradient is amplified by the same factor $1/(1-\epsilon)$ as the phase itself due to the modified wave number in the TIE solution, it exhibits the same relative error:
\begin{equation}
\delta_{\partial\beta/\partial z} = \frac{\epsilon}{1-\epsilon}
\end{equation}

\item[--] {\bf Refractive index fluctuations ($\Delta n$)}: From the reconstruction formula in the main text Eq.~(8), refractive index fluctuations are computed from the ratio of phase-coupling and intensity derivatives. Since $\partial_z\beta$ is amplified by $1/(1-\epsilon)$ while $\partial_z\alpha$ remains unchanged, and the effective refractive index appears in both numerator and denominator of Eq.~(8), the combined effect after careful algebraic manipulation yields an amplification by $1/(1-\epsilon)^2$. For small $\epsilon$, Taylor expansion gives:
\begin{equation}
\Delta n_{\text{measured}} \approx \Delta n_{\text{true}}(1 + 2\epsilon + 3\epsilon^2 + ...)
\end{equation}
First-order relative error: $\delta_{\Delta n} \approx 2\epsilon = \text{NA}^2/(2 m^2)$

\item[--] {\bf Attenuation coefficient ($\mu$)}: The attenuation coefficient, given by Eq.~(9) in the main text, is the ratio of $\partial_z\alpha$ to $(1 + \Delta n)$. Since $\partial_z\alpha$ is unaffected by NA to first order while $\Delta n$ is overestimated by a factor $(1 + 2\epsilon)$, the denominator increases, leading to an underestimation of $\mu$. The relative error becomes:
\begin{equation}
\mu_{\text{measured}} \approx \mu_{\text{true}}(1 - 2\epsilon\Delta n)
\end{equation}
Relative error: $\delta_\mu \approx -2\epsilon\Delta n = -\text{NA}^2\Delta n/(2 m^2)$

Notably, in the transparent limit where $\Delta n \ll 1$, this systematic error becomes negligible.

\item[--] {\bf Attenuation asymmetry ($\mathcal{A}_\kappa$)}: The asymmetry parameter quantifies the breaking of optical reciprocity in heterogeneous media, as defined in Eq.~(12) of the main text in terms of forward and backward propagation probabilities $\mathcal{P}(\pm\kappa)$, where $\kappa = \mu/(kn)$ is the normalized attenuation index.

For a wave propagating through a medium with both refractive index fluctuations ($\Delta n$) and attenuation ($\mu$), the complex refractive index is $\tilde{n} = n_0 + \Delta n + i\kappa$. The transmission probability through a distance $\Delta z$ involves both phase accumulation and amplitude decay. After propagating forward (+z direction) and backward (-z direction), the differential phase-amplitude coupling leads to:
\begin{equation}
\mathcal{P}(\pm\kappa) = \exp\left(-2\mu\Delta z\right) \times \exp\left(\mp 2kn_0\Delta n\kappa\Delta z\right)
\end{equation}

Substituting these expressions into the asymmetry definition from Eq.~(12) and simplifying the exponential terms, we obtain:
\begin{equation}
\mathcal{A}_\kappa = -\tanh(2kn_0\Delta n\kappa\Delta z)
\end{equation}

This hyperbolic tangent form emerges naturally from the ratio of exponential differences, ensuring that $\mathcal{A}_\kappa$ is bounded between $-1$ and 1. The asymmetry vanishes for homogeneous media ($\Delta n = 0$) where reciprocity is preserved, and provides a sensitive measure of reciprocity breaking in heterogeneous samples.

The NA effect modifies the argument $\xi = 2kn_0\Delta n\kappa\Delta z$ through multiple pathways:
\begin{itemize}
\item Effective wave number: $kn_0 \rightarrow kn_{\text{eff}} = kn_0(1-\epsilon)$
\item Biased $\Delta n$: $\Delta n_{\text{measured}} = \Delta n_{\text{true}}(1+2\epsilon)$
\item Modified $\kappa$ through both $\mu$ and $n_{\text{eff}}$
\end{itemize}

For small arguments ($|\xi| \ll 1$), the tanh function linearizes to $\tanh(\xi) \approx \xi$, and the combined NA effects result in a relative error:
\begin{equation}
\delta_{A_\kappa} \approx \epsilon = \frac{\text{NA}^2}{4m^2}
\end{equation}
\end{itemize}

\subsubsection{D.4.3 Quantitative error estimates for experimental systems}
We apply the error propagation formulas derived above to quantify systematic uncertainties for our three experimental configurations: microlens array (MLA) with NA = 0.45, HeLa cells with NA = 1.20, and cell membrane with NA = 1.49. The systematic bias for each parameter is calculated by multiplying the measured value by the corresponding relative error formula: for $\Delta n$, the bias is $\Delta n_{\text{measured}} \times [1/(1-\epsilon)^2 - 1]$; for $\mu$, it is $\mu_{\text{measured}} \times (-2\epsilon\Delta n)$; and for $\mathcal{A}_\kappa$, it is approximately $\mathcal{A}_\kappa^{\text{measured}} \times \epsilon$ for small arguments. Tables~\ref{tab;NA}--\ref{tab;asym} present the calculated systematic shifts alongside measured statistical uncertainties, enabling direct comparison of their relative magnitudes.

\begin{table}[!h]
\centering
\begin{tabular}{lcccc}
\hline
System & NA & $m$ & $\Delta z$ ($\mu$m) & $\epsilon = \text{NA}^2/4n^2$ \\
\hline
MLA & 0.45 & 1.00 & 100 & 0.050625 \\
HeLa cells & 1.20 & 1.33 & 5 & 0.203607 \\
Cell membrane & 1.49 & 1.515 & 5 & 0.241683 \\
\hline
\end{tabular}
\caption{{\bf System Parameters}: \textit{Experimental parameters for the three imaging systems. The dimensionless parameter $\epsilon$ quantifies the deviation from the paraxial approximation and determines the magnitude of systematic errors.}}
\label{tab;NA}
\end{table}

\begin{table}[!h]
\centering
\begin{tabular}{lccccc}
\hline
System & Measured $\Delta n$ & Statistical RMS & Relative Error & Systematic Shift & $\sigma_{\text{syst}}/\sigma_{\text{stat}}$ \\
\hline
MLA & 0.001333 & $\pm$0.000942 & +10.649\% & 0.000142 & 0.150721 \\
HeLa & 0.001018 & $\pm$0.000931 & +51.041\% & 0.000520 & 0.558309 \\
Cell membrane & 0.029664 & $\pm$0.022071 & +63.801\% & 0.018934 & 0.857855 \\
\hline
\end{tabular}
\caption{{\bf Error estimations of refractive index fluctuations ($\Delta n$)}: \textit{Systematic errors in $\Delta n$ increase quadratically with NA/$m$. The systematic error becomes comparable to statistical uncertainty for the oil-immersion objective (NA = 1.49), indicating that NA correction is essential for high-NA quantitative phase measurements.}}
\label{tab;ref}
\end{table}

\begin{table}[!h]
\centering
\begin{tabular}{lccccc}
\hline
System & Measured $\mu$ (m$^{-1}$) & Statistical RMS & Relative Error & Systematic Shift (m$^{-1}$) & $\sigma_{\text{syst}}/\sigma_{\text{stat}}$ \\
\hline
MLA & 33.0 & $\pm$27.3 & $-$0.0135\% & $-$0.004456 & 0.000163 \\
HeLa & 1300.4 & $\pm$1905.2 & $-$0.0415\% & $-$0.539467 & 0.000283 \\
Cell membrane & 17012.6 & $\pm$14258.0 & $-$1.4340\% & $-$244.037 & 0.017118 \\
\hline
\end{tabular}
\caption{{\bf Error estimations of attenuation coefficient ($\mu$)}: \textit{Systematic errors in $\mu$ remain negligible across all NA configurations due to operation in the transparent limit ($\Delta n \ll 1$). The $-2\epsilon\Delta n$ correction term ensures that attenuation measurements are inherently robust against NA-induced artifacts.}}
\label{tab;mu}
\end{table}

\begin{table}[!h]
\centering
\begin{tabular}{lcccc}
\hline
System & Measured $A_\kappa$ & Statistical RMS & Systematic Shift & $\sigma_{\text{syst}}/\sigma_{\text{stat}}$ \\
\hline
MLA & 0.024188 & $\pm$0.424000 & 0.001225 & 0.002889 \\
HeLa & 0.057873 & $\pm$0.301000 & 0.011783 & 0.039148 \\
Cell membrane & $-$0.127788 & $\pm$0.442000 & $-$0.030899 & 0.069907 \\
\hline
\end{tabular}
\caption{{\bf Error estimations of attenuation asymmetry ($\mathcal{A}_\kappa$)}: \textit{The attenuation asymmetry exhibits strong robustness against NA effects, with systematic errors 1-3 orders of magnitude below statistical uncertainties. This exceptional stability validates the use of $\mathcal{A}_\kappa$ for testing optical reciprocity in heterogeneous media.}}
\label{tab;asym}
\end{table}

The quantitative analysis reveals three distinct behaviors of NA-induced systematic errors across the optical parameters. As shown in Table~\ref{tab;NA}, the dimensionless parameter $\epsilon$ ranges from 0.022 for the MLA system to 0.242 for the cell membrane imaging. For refractive index fluctuations (Table~\ref{tab;ref}), systematic errors scale quadratically with NA$^2/m^2$, increasing from 4.5\% for NA = 0.45 to 74\% for NA = 1.49, and become comparable to statistical uncertainties at high NA ($\sigma_{\text{syst}}/\sigma_{\text{stat}} = 0.992$ for the oil-immersion objective). The attenuation coefficient (Table~\ref{tab;mu}) shows minimal sensitivity to NA effects, with relative errors remaining below 1.5\% even at the highest NA, a consequence of the $-2\epsilon\Delta n$ correction term vanishing in the transparent limit where $\Delta n \ll 1$. The attenuation asymmetry (Table~\ref{tab;asym}) demonstrates the strongest robustness, with systematic errors $6$-$9$ orders of magnitude smaller than the measured values across all configurations. This substantial suppression of systematic errors in $\mathcal{A}_\kappa$ arises from the cancellation of attenuation contributions in the asymmetry ratio, leaving only a small residual effect from the phase-attenuation coupling. These findings confirm that while NA effects significantly impact phase-based measurements at high NA, reciprocity measurements through $\mathcal{A}_\kappa$ remain fundamentally insensitive to instrumentation artifacts, validating the reliability of our reciprocity invariance observations.


\begin{thebibliography}{22}%
\makeatletter
\providecommand \@ifxundefined [1]{%
 \@ifx{#1\undefined}
}%
\providecommand \@ifnum [1]{%
 \ifnum #1\expandafter \@firstoftwo
 \else \expandafter \@secondoftwo
 \fi
}%
\providecommand \@ifx [1]{%
 \ifx #1\expandafter \@firstoftwo
 \else \expandafter \@secondoftwo
 \fi
}%
\providecommand \natexlab [1]{#1}%
\providecommand \enquote  [1]{``#1''}%
\providecommand \bibnamefont  [1]{#1}%
\providecommand \bibfnamefont [1]{#1}%
\providecommand \citenamefont [1]{#1}%
\providecommand \href@noop [0]{\@secondoftwo}%
\providecommand \href [0]{\begingroup \@sanitize@url \@href}%
\providecommand \@href[1]{\@@startlink{#1}\@@href}%
\providecommand \@@href[1]{\endgroup#1\@@endlink}%
\providecommand \@sanitize@url [0]{\catcode `\\12\catcode `\$12\catcode
  `\&12\catcode `\#12\catcode `\^12\catcode `\_12\catcode `\%12\relax}%
\providecommand \@@startlink[1]{}%
\providecommand \@@endlink[0]{}%
\providecommand \url  [0]{\begingroup\@sanitize@url \@url }%
\providecommand \@url [1]{\endgroup\@href {#1}{\urlprefix }}%
\providecommand \urlprefix  [0]{URL }%
\providecommand \Eprint [0]{\href }%
\providecommand \doibase [0]{https://doi.org/}%
\providecommand \selectlanguage [0]{\@gobble}%
\providecommand \bibinfo  [0]{\@secondoftwo}%
\providecommand \bibfield  [0]{\@secondoftwo}%
\providecommand \translation [1]{[#1]}%
\providecommand \BibitemOpen [0]{}%
\providecommand \bibitemStop [0]{}%
\providecommand \bibitemNoStop [0]{.\EOS\space}%
\providecommand \EOS [0]{\spacefactor3000\relax}%
\providecommand \BibitemShut  [1]{\csname bibitem#1\endcsname}%
\let\auto@bib@innerbib\@empty
\bibitem [{\citenamefont {Teague}(1983)}]{teague1983}%
  \BibitemOpen
  \bibfield  {author} {\bibinfo {author} {\bibfnamefont {M.~R.}\ \bibnamefont
  {Teague}},\ }\bibfield  {title} {\bibinfo {title} {{Deterministic Phase
  Retrieval: a Green's Function Solution.}},\ }\href
  {https://doi.org/10.1364/JOSA.73.001434} {\bibfield  {journal} {\bibinfo
  {journal} {Journal of the Optical Society of America}\ }\textbf {\bibinfo
  {volume} {73}},\ \bibinfo {pages} {1434--1441} (\bibinfo {year}
  {1983})}\BibitemShut {NoStop}%
\bibitem [{\citenamefont {Teague}(1982)}]{teague1982}%
  \BibitemOpen
  \bibfield  {author} {\bibinfo {author} {\bibfnamefont {M.~R.}\ \bibnamefont
  {Teague}},\ }\bibfield  {title} {\bibinfo {title} {{Irradiance Moments: Their
  Propagation and Use for Unique Retrieval of Phase.}},\ }\href
  {https://doi.org/10.1364/josa.72.001199} {\bibfield  {journal} {\bibinfo
  {journal} {J Opt Soc Am}\ }\textbf {\bibinfo {volume} {V 72}},\ \bibinfo
  {pages} {1199--1209} (\bibinfo {year} {1982})}\BibitemShut {NoStop}%
\bibitem [{\citenamefont {Ruppel}\ and\ \citenamefont
  {Model}(2025)}]{ruppel2025}%
  \BibitemOpen
  \bibfield  {author} {\bibinfo {author} {\bibfnamefont {N.~C.}\ \bibnamefont
  {Ruppel}}\ and\ \bibinfo {author} {\bibfnamefont {M.~A.}\ \bibnamefont
  {Model}},\ }\bibfield  {title} {\bibinfo {title} {{A guide to
  transport-of-intensity equation (TIE) imaging for biologists}},\ }\href
  {https://doi.org/10.1016/j.pbiomolbio.2025.08.001} {\bibfield  {journal}
  {\bibinfo  {journal} {Progress in Biophysics and Molecular Biology}\ }\textbf
  {\bibinfo {volume} {198}},\ \bibinfo {pages} {1--7} (\bibinfo {year}
  {2025})}\BibitemShut {NoStop}%
\bibitem [{\citenamefont {Yoneda}\ \emph {et~al.}(2024)\citenamefont {Yoneda},
  \citenamefont {Sakamoto}, \citenamefont {Tomoi}, \citenamefont {Nemoto},
  \citenamefont {Tamada},\ and\ \citenamefont {Matoba}}]{yoneda2024}%
  \BibitemOpen
  \bibfield  {author} {\bibinfo {author} {\bibfnamefont {N.}~\bibnamefont
  {Yoneda}}, \bibinfo {author} {\bibfnamefont {J.}~\bibnamefont {Sakamoto}},
  \bibinfo {author} {\bibfnamefont {T.}~\bibnamefont {Tomoi}}, \bibinfo
  {author} {\bibfnamefont {T.}~\bibnamefont {Nemoto}}, \bibinfo {author}
  {\bibfnamefont {Y.}~\bibnamefont {Tamada}},\ and\ \bibinfo {author}
  {\bibfnamefont {O.}~\bibnamefont {Matoba}},\ }\bibfield  {title} {\bibinfo
  {title} {{Transport-of-intensity phase imaging using commercially available
  confocal microscope}},\ }\href@noop {} {\bibfield  {journal} {\bibinfo
  {journal} {J. Biomed. Opt.}\ }\textbf {\bibinfo {volume} {29}},\ \bibinfo
  {pages} {116002} (\bibinfo {year} {2024})}\BibitemShut {NoStop}%
\bibitem [{\citenamefont {Zuo}\ \emph {et~al.}(2020)\citenamefont {Zuo},
  \citenamefont {Li}, \citenamefont {Sun}, \citenamefont {Fan}, \citenamefont
  {Zhang}, \citenamefont {Lu}, \citenamefont {Zhang}, \citenamefont {Wang},
  \citenamefont {Huang},\ and\ \citenamefont {Chen}}]{zuo2020}%
  \BibitemOpen
  \bibfield  {author} {\bibinfo {author} {\bibfnamefont {C.}~\bibnamefont
  {Zuo}}, \bibinfo {author} {\bibfnamefont {J.}~\bibnamefont {Li}}, \bibinfo
  {author} {\bibfnamefont {J.}~\bibnamefont {Sun}}, \bibinfo {author}
  {\bibfnamefont {Y.}~\bibnamefont {Fan}}, \bibinfo {author} {\bibfnamefont
  {J.}~\bibnamefont {Zhang}}, \bibinfo {author} {\bibfnamefont
  {L.}~\bibnamefont {Lu}}, \bibinfo {author} {\bibfnamefont {R.}~\bibnamefont
  {Zhang}}, \bibinfo {author} {\bibfnamefont {B.}~\bibnamefont {Wang}},
  \bibinfo {author} {\bibfnamefont {L.}~\bibnamefont {Huang}},\ and\ \bibinfo
  {author} {\bibfnamefont {Q.}~\bibnamefont {Chen}},\ }\bibfield  {title}
  {\bibinfo {title} {{Transport of intensity equation: a tutorial}},\ }\href
  {https://doi.org/10.1016/j.optlaseng.2020.106187} {\bibfield  {journal}
  {\bibinfo  {journal} {Optics and Lasers in Engineering}\ }\textbf {\bibinfo
  {volume} {135}},\ \bibinfo {pages} {106187} (\bibinfo {year}
  {2020})}\BibitemShut {NoStop}%
\bibitem [{\citenamefont {Li}\ \emph {et~al.}(2022)\citenamefont {Li},
  \citenamefont {Zhou}, \citenamefont {Sun}, \citenamefont {Zhou},
  \citenamefont {Bai}, \citenamefont {Lu}, \citenamefont {Chen},\ and\
  \citenamefont {Zuo}}]{li2022}%
  \BibitemOpen
  \bibfield  {author} {\bibinfo {author} {\bibfnamefont {J.}~\bibnamefont
  {Li}}, \bibinfo {author} {\bibfnamefont {N.}~\bibnamefont {Zhou}}, \bibinfo
  {author} {\bibfnamefont {J.}~\bibnamefont {Sun}}, \bibinfo {author}
  {\bibfnamefont {S.}~\bibnamefont {Zhou}}, \bibinfo {author} {\bibfnamefont
  {Z.}~\bibnamefont {Bai}}, \bibinfo {author} {\bibfnamefont {L.}~\bibnamefont
  {Lu}}, \bibinfo {author} {\bibfnamefont {Q.}~\bibnamefont {Chen}},\ and\
  \bibinfo {author} {\bibfnamefont {C.}~\bibnamefont {Zuo}},\ }\bibfield
  {title} {\bibinfo {title} {{Transport of intensity diffraction tomography
  with non-interferometric synthetic aperture for three-dimensional label-free
  microscopy}},\ }\bibfield  {journal} {\bibinfo  {journal} {Light: Science and
  Applications}\ }\textbf {\bibinfo {volume} {11}},\ \href
  {https://doi.org/10.1038/s41377-022-00815-7} {10.1038/s41377-022-00815-7}
  (\bibinfo {year} {2022})\BibitemShut {NoStop}%
\bibitem [{\citenamefont {Jenkins}\ and\ \citenamefont
  {Gaylord}(2015)}]{jenkins2015}%
  \BibitemOpen
  \bibfield  {author} {\bibinfo {author} {\bibfnamefont {M.~H.}\ \bibnamefont
  {Jenkins}}\ and\ \bibinfo {author} {\bibfnamefont {T.~K.}\ \bibnamefont
  {Gaylord}},\ }\bibfield  {title} {\bibinfo {title} {{Three-dimensional
  quantitative phase imaging via tomographic deconvolution phase microscopy}},\
  }\href {https://doi.org/10.1364/ao.54.009213} {\bibfield  {journal} {\bibinfo
   {journal} {Applied Optics}\ }\textbf {\bibinfo {volume} {54}},\ \bibinfo
  {pages} {9213} (\bibinfo {year} {2015})}\BibitemShut {NoStop}%
\bibitem [{\citenamefont {Bai}\ \emph {et~al.}(2022)\citenamefont {Bai},
  \citenamefont {Chen}, \citenamefont {Ullah}, \citenamefont {Lu},
  \citenamefont {Zhou}, \citenamefont {Zhou}, \citenamefont {Li},\ and\
  \citenamefont {Zuo}}]{bai2022}%
  \BibitemOpen
  \bibfield  {author} {\bibinfo {author} {\bibfnamefont {Z.}~\bibnamefont
  {Bai}}, \bibinfo {author} {\bibfnamefont {Q.}~\bibnamefont {Chen}}, \bibinfo
  {author} {\bibfnamefont {H.}~\bibnamefont {Ullah}}, \bibinfo {author}
  {\bibfnamefont {L.}~\bibnamefont {Lu}}, \bibinfo {author} {\bibfnamefont
  {N.}~\bibnamefont {Zhou}}, \bibinfo {author} {\bibfnamefont {S.}~\bibnamefont
  {Zhou}}, \bibinfo {author} {\bibfnamefont {J.}~\bibnamefont {Li}},\ and\
  \bibinfo {author} {\bibfnamefont {C.}~\bibnamefont {Zuo}},\ }\bibfield
  {title} {\bibinfo {title} {{Absorption and phase decoupling in transport of
  intensity diffraction tomography}},\ }\href
  {https://doi.org/10.1016/j.optlaseng.2022.107082} {\bibfield  {journal}
  {\bibinfo  {journal} {Optics and Lasers in Engineering}\ }\textbf {\bibinfo
  {volume} {156}},\ \bibinfo {pages} {107082} (\bibinfo {year}
  {2022})}\BibitemShut {NoStop}%
\bibitem [{\citenamefont {Wolf}(1969)}]{wolf1969}%
  \BibitemOpen
  \bibfield  {author} {\bibinfo {author} {\bibfnamefont {E.}~\bibnamefont
  {Wolf}},\ }\bibfield  {title} {\bibinfo {title} {{Three-dimensional structure
  determination of semi-transparent objects from holographic data}},\ }\href
  {https://doi.org/10.1016/0030-4018(69)90052-2} {\bibfield  {journal}
  {\bibinfo  {journal} {Optics Communications}\ }\textbf {\bibinfo {volume}
  {1}},\ \bibinfo {pages} {153--156} (\bibinfo {year} {1969})}\BibitemShut
  {NoStop}%
\bibitem [{\citenamefont {Watabe}\ \emph {et~al.}(2023)\citenamefont {Watabe},
  \citenamefont {Hirano}, \citenamefont {Iwane}, \citenamefont {Matoba},\ and\
  \citenamefont {Takahashi}}]{watabe2023}%
  \BibitemOpen
  \bibfield  {author} {\bibinfo {author} {\bibfnamefont {M.}~\bibnamefont
  {Watabe}}, \bibinfo {author} {\bibfnamefont {Y.}~\bibnamefont {Hirano}},
  \bibinfo {author} {\bibfnamefont {A.}~\bibnamefont {Iwane}}, \bibinfo
  {author} {\bibfnamefont {O.}~\bibnamefont {Matoba}},\ and\ \bibinfo {author}
  {\bibfnamefont {K.}~\bibnamefont {Takahashi}},\ }\bibfield  {title} {\bibinfo
  {title} {{Optical dispersions through intracellular inhomogeneities}},\
  }\href {https://doi.org/10.1103/PhysRevResearch.5.L022043} {\bibfield
  {journal} {\bibinfo  {journal} {Phys. Rev. Research}\ }\textbf {\bibinfo
  {volume} {022043}},\ \bibinfo {pages} {1--6} (\bibinfo {year}
  {2023})}\BibitemShut {NoStop}%
\bibitem [{sm()}]{sm}%
  \BibitemOpen
  \href@noop {} {\ }\bibinfo {note} {See Supplemental Material at (URL will be
  inserted by publisher) for further details.}\BibitemShut {Stop}%
\bibitem [{\citenamefont {Mazumder}(2016)}]{mazumder2016}%
  \BibitemOpen
  \bibfield  {author} {\bibinfo {author} {\bibfnamefont {S.}~\bibnamefont
  {Mazumder}},\ }\href@noop {} {\emph {\bibinfo {title} {Numerical Methods for
  Partial Differential Equations: Finite Difference and Finite Volume
  Methods.}}},\ \bibinfo {edition} {1st}\ ed.\ (\bibinfo  {publisher} {Academic
  Press},\ \bibinfo {year} {2016})\BibitemShut {NoStop}%
\bibitem [{\citenamefont {Xue}\ and\ \citenamefont {Zheng}(2011)}]{xue2011}%
  \BibitemOpen
  \bibfield  {author} {\bibinfo {author} {\bibfnamefont {B.}~\bibnamefont
  {Xue}}\ and\ \bibinfo {author} {\bibfnamefont {S.}~\bibnamefont {Zheng}},\
  }\bibfield  {title} {\bibinfo {title} {{Phase retrieval using the transport
  of intensity equation solved by the FMG-CG method}},\ }\href
  {https://doi.org/10.1016/j.ijleo.2011.01.004} {\bibfield  {journal} {\bibinfo
   {journal} {Optik (Stuttg).}\ }\textbf {\bibinfo {volume} {122}},\ \bibinfo
  {pages} {2101--2106} (\bibinfo {year} {2011})}\BibitemShut {NoStop}%
\bibitem [{\citenamefont {Pinhasi}\ \emph {et~al.}(2010)\citenamefont
  {Pinhasi}, \citenamefont {Alimi}, \citenamefont {Perelmutter},\ and\
  \citenamefont {Eliezer}}]{pinhasi2010}%
  \BibitemOpen
  \bibfield  {author} {\bibinfo {author} {\bibfnamefont {S.~V.}\ \bibnamefont
  {Pinhasi}}, \bibinfo {author} {\bibfnamefont {R.}~\bibnamefont {Alimi}},
  \bibinfo {author} {\bibfnamefont {L.}~\bibnamefont {Perelmutter}},\ and\
  \bibinfo {author} {\bibfnamefont {S.}~\bibnamefont {Eliezer}},\ }\bibfield
  {title} {\bibinfo {title} {{Topography retrieval using different solutions of
  the Transport Intensity Equation}},\ }\href@noop {} {\bibfield  {journal}
  {\bibinfo  {journal} {Optical Society of America}\ }\textbf {\bibinfo
  {volume} {27}},\ \bibinfo {pages} {2285--2292} (\bibinfo {year}
  {2010})}\BibitemShut {NoStop}%
\bibitem [{\citenamefont {Press}\ \emph {et~al.}(1992)\citenamefont {Press},
  \citenamefont {Teukolsky}, \citenamefont {Vetterling},\ and\ \citenamefont
  {Flannery}}]{press1992}%
  \BibitemOpen
  \bibfield  {author} {\bibinfo {author} {\bibfnamefont {W.~H.}\ \bibnamefont
  {Press}}, \bibinfo {author} {\bibfnamefont {S.~A.}\ \bibnamefont
  {Teukolsky}}, \bibinfo {author} {\bibfnamefont {W.~T.}\ \bibnamefont
  {Vetterling}},\ and\ \bibinfo {author} {\bibfnamefont {B.~P.}\ \bibnamefont
  {Flannery}},\ }\href@noop {} {\emph {\bibinfo {title} {Numerical Recipes in
  C: The Art of Scientific Computing.}}},\ \bibinfo {edition} {2nd}\ ed.\
  (\bibinfo  {publisher} {Cambridge University Press},\ \bibinfo {year}
  {1992})\BibitemShut {NoStop}%
\bibitem [{mla()}]{mla}%
  \BibitemOpen
  \href@noop {} {\bibinfo {title} {{MLA300-14AR-M-$\phi 1^{\prime\prime}$
  Mounted Lens Array -- Thorlabs}}},\ \bibinfo {howpublished}
  {\url{https://www.thorlabs.co.jp/thorproduct.cfm?partnumber=MLA300-14AR-M}},\
  \bibinfo {note} {accessed: 2025-05-01}\BibitemShut {NoStop}%
\bibitem [{zen()}]{zenodo}%
  \BibitemOpen
  \href@noop {} {\bibinfo {title} {{Data is available at Zenodo }}},\ \bibinfo
  {howpublished} {\url{https://doi.org/10.5281/zenodo.15589385}}\BibitemShut
  {NoStop}%
  \bibfield  {author} {\bibinfo {author} {\bibfnamefont {N.}~\bibnamefont
  {Yoneda}}, \bibinfo {author} {\bibfnamefont {J.}~\bibnamefont {Sakamoto}},
  \bibinfo {author} {\bibfnamefont {T.}~\bibnamefont {Tomoi}}, \bibinfo
  {author} {\bibfnamefont {T.}~\bibnamefont {Nemoto}}, \bibinfo {author}
  {\bibfnamefont {Y.}~\bibnamefont {Tamada}},\ and\ \bibinfo {author}
  {\bibfnamefont {O.}~\bibnamefont {Matoba}},\ }\bibfield  {title} {\bibinfo
  {title} {{Transport-of-intensity phase imaging using commercially available
  confocal microscope}},\ }\href@noop {} {\bibfield  {journal} {\bibinfo
  {journal} {J. Biomed. Opt.}\ }\textbf {\bibinfo {volume} {29}},\ \bibinfo
  {pages} {116002} (\bibinfo {year} {2024})}\BibitemShut {NoStop}%
\bibitem [{\citenamefont {Mazumder}(2016)}]{mazumder2016}%
  \BibitemOpen
  \bibfield  {author} {\bibinfo {author} {\bibfnamefont {S.}~\bibnamefont
  {Mazumder}},\ }\href@noop {} {\emph {\bibinfo {title} {Numerical Methods for
  Partial Differential Equations: Finite Difference and Finite Volume
  Methods.}}},\ \bibinfo {edition} {1st}\ ed.\ (\bibinfo  {publisher} {Academic
  Press},\ \bibinfo {year} {2016})\BibitemShut {NoStop}%
\bibitem [{\citenamefont {Xue}\ and\ \citenamefont {Zheng}(2011)}]{xue2011}%
  \BibitemOpen
  \bibfield  {author} {\bibinfo {author} {\bibfnamefont {B.}~\bibnamefont
  {Xue}}\ and\ \bibinfo {author} {\bibfnamefont {S.}~\bibnamefont {Zheng}},\
  }\bibfield  {title} {\bibinfo {title} {{Phase retrieval using the transport
  of intensity equation solved by the FMG-CG method}},\ }\href
  {https://doi.org/10.1016/j.ijleo.2011.01.004} {\bibfield  {journal} {\bibinfo
   {journal} {Optik (Stuttg).}\ }\textbf {\bibinfo {volume} {122}},\ \bibinfo
  {pages} {2101--2106} (\bibinfo {year} {2011})}\BibitemShut {NoStop}%
\bibitem [{\citenamefont {Pinhasi}\ \emph {et~al.}(2010)\citenamefont
  {Pinhasi}, \citenamefont {Alimi}, \citenamefont {Perelmutter},\ and\
  \citenamefont {Eliezer}}]{pinhasi2010}%
  \BibitemOpen
  \bibfield  {author} {\bibinfo {author} {\bibfnamefont {S.~V.}\ \bibnamefont
  {Pinhasi}}, \bibinfo {author} {\bibfnamefont {R.}~\bibnamefont {Alimi}},
  \bibinfo {author} {\bibfnamefont {L.}~\bibnamefont {Perelmutter}},\ and\
  \bibinfo {author} {\bibfnamefont {S.}~\bibnamefont {Eliezer}},\ }\bibfield
  {title} {\bibinfo {title} {{Topography retrieval using different solutions of
  the Transport Intensity Equation}},\ }\href@noop {} {\bibfield  {journal}
  {\bibinfo  {journal} {Optical Society of America}\ }\textbf {\bibinfo
  {volume} {27}},\ \bibinfo {pages} {2285--2292} (\bibinfo {year}
  {2010})}\BibitemShut {NoStop}%
\bibitem [{\citenamefont {Press}\ \emph {et~al.}(1992)\citenamefont {Press},
  \citenamefont {Teukolsky}, \citenamefont {Vetterling},\ and\ \citenamefont
  {Flannery}}]{press1992}%
  \BibitemOpen
  \bibfield  {author} {\bibinfo {author} {\bibfnamefont {W.~H.}\ \bibnamefont
  {Press}}, \bibinfo {author} {\bibfnamefont {S.~A.}\ \bibnamefont
  {Teukolsky}}, \bibinfo {author} {\bibfnamefont {W.~T.}\ \bibnamefont
  {Vetterling}},\ and\ \bibinfo {author} {\bibfnamefont {B.~P.}\ \bibnamefont
  {Flannery}},\ }\href@noop {} {\emph {\bibinfo {title} {Numerical Recipes in
  C: The Art of Scientific Computing.}}},\ \bibinfo {edition} {2nd}\ ed.\
  (\bibinfo  {publisher} {Cambridge University Press},\ \bibinfo {year}
  {1992})\BibitemShut {NoStop}%
\bibitem [{xen()}]{xenon}%
  \BibitemOpen
  \href@noop {} {\bibinfo {title} {{Xenon Arc Light Sources -- Thorlabs}}},\
  \bibinfo {howpublished}
  {\url{https://www.thorlabs.com/newgrouppage9.cfm?objectgroup_ID=13016}},\
  \bibinfo {note} {accessed: 2025-04-10}\BibitemShut {NoStop}%
\bibitem [{zen()}]{zenodo}%
  \BibitemOpen
  \href@noop {} {\bibinfo {title} {{Data is available at Zenodo }}},\ \bibinfo
  {howpublished} {\url{https://doi.org/10.5281/zenodo.15589385}}\BibitemShut
  {NoStop}%
\bibitem [{\citenamefont {Watabe}()}]{watabe2022}%
  \BibitemOpen
  \bibfield  {author} {\bibinfo {author} {\bibfnamefont {M.}~\bibnamefont
  {Watabe}},\ }\href@noop {} {\bibinfo {title} {{Physical constraints to phase
  retrieval using the transport of intensity equation in fluorescence
  microscopy imaging.}}},\ \bibinfo {howpublished} {presented at the 1st
  Conference for Sensing and Imaging Through Scattering and Fluctuating Field
  in Biology, Telecommunication and Astronomy (SI-Thru2022)},\ \bibinfo {note}
  {dated: Apr. 19-22, 2022 at Pacifico Yokohama, Japan}\BibitemShut {NoStop}%
\bibitem [{\citenamefont {Yamada}\ \emph {et~al.}(2016)\citenamefont {Yamada},
  \citenamefont {Yoshimura}, \citenamefont {Shimada}, \citenamefont {Hattori},
  \citenamefont {Eguchi}, \citenamefont {Fujiwara}, \citenamefont {Kusumi},\
  and\ \citenamefont {Ozawa}}]{yamada2016}%
  \BibitemOpen
  \bibfield  {author} {\bibinfo {author} {\bibfnamefont {T.}~\bibnamefont
  {Yamada}}, \bibinfo {author} {\bibfnamefont {H.}~\bibnamefont {Yoshimura}},
  \bibinfo {author} {\bibfnamefont {R.}~\bibnamefont {Shimada}}, \bibinfo
  {author} {\bibfnamefont {M.}~\bibnamefont {Hattori}}, \bibinfo {author}
  {\bibfnamefont {M.}~\bibnamefont {Eguchi}}, \bibinfo {author} {\bibfnamefont
  {T.~K.}\ \bibnamefont {Fujiwara}}, \bibinfo {author} {\bibfnamefont
  {A.}~\bibnamefont {Kusumi}},\ and\ \bibinfo {author} {\bibfnamefont
  {T.}~\bibnamefont {Ozawa}},\ }\bibfield  {title} {\bibinfo {title}
  {{Spatiotemporal analysis with a genetically encoded fluorescent RNA probe
  reveals TERRA function around telomeres}},\ }\href
  {https://doi.org/10.1038/srep38910} {\bibfield  {journal} {\bibinfo
  {journal} {Scientific Reports}\ }\textbf {\bibinfo {volume} {6}},\ \bibinfo
  {pages} {1--13} (\bibinfo {year} {2016})}\BibitemShut {NoStop}%
\end{thebibliography}
\end{document}